\documentclass[a4paper,11pt, arrowdel]{article}
\usepackage{jcappub} % for details on the use of the package, please see the JINST-author-manual
\RequirePackage[top=12mm,bottom=11.5mm,left=22mm,right=22mm,head=12mm,includeheadfoot]{geometry}

\usepackage{enumitem}
\usepackage{fancyhdr}
\usepackage{mathtools}
\usepackage{amssymb}
\usepackage{amsmath}
\usepackage{hyperref}
\usepackage{amsfonts}
\usepackage{amssymb}
\usepackage{caption}
\usepackage{float}
\usepackage{bm}
\usepackage{graphicx}
\usepackage{color}
\usepackage{verbatim}
\usepackage{blindtext, multicol}
\usepackage{multirow}
\usepackage{epigraph}
\usepackage[dvipsnames]{xcolor}
\usepackage{physics}

\usepackage[normalem]{ulem}

\usepackage{tensor}
\usepackage{physics}
\usepackage{romannum}
\usepackage{subfigure}

\usepackage[capitalise]{cleveref}
\usepackage[activate={true,nocompatibility},final,kerning=true,factor=1100,stretch=10,shrink=10]{microtype}

\usepackage[sorting=none,style=numeric-comp]{biblatex}
\usepackage[english]{babel}
\usepackage{csquotes}
\DeclareUnicodeCharacter{0301}{\'{i}}
\usepackage[nottoc]{tocbibind}

\usepackage{academicons}
\usepackage{xcolor}
\usepackage{fontawesome5}
\definecolor{orcidlogocol}{rgb}{0.65, 0.807, 0.223}

\newcommand{\orcid}[1]{$\,$\href{https://orcid.org/#1}{\textcolor{orcidlogocol}{\faOrcid}}}

\usepackage[T1]{fontenc} % if needed

\numberwithin{equation}{section}

\newcommand\num{\addtocounter{equation}{1}\tag{\theequation}}

\def\beq{\begin{equation}}
\def\eeq{\end{equation}}
\def\ber{\begin{eqnarray}}
\def\eer{\end{eqnarray}}

\def\wre{w_{_{\rm re}}}

\def\l{\left}
\def\r{\right}
\def\d{\mathrm{d}}
\def\p{\partial}
\newcommand{\sq}{\lower.25ex\hbox{\large$\Box$}}

\def\f{\frac}
\def\mpl{m_{p}}

\def\mpc{\rm Mpc^{-1}}

\def\ng{n_{_{\rm GW}}}
\def\Og{\Omega_{_{\rm GW}}}
\def\Ogo{\Omega_{_{\rm GW}}^{0}}

\def\xvec{\vec{x}}
\def\kvec{\vec{k}}

\def\lleq{\lower0.9ex\hbox{ $\buildrel < \over \sim$} ~}
\def\ggeq{\lower0.9ex\hbox{ $\buildrel > \over \sim$} ~}

\def\half{\frac{1}{2}}
\def\thalf{\frac{3}{2}}
\def\fhalf{\frac{5}{2}}
\def\t{\times}

\hypersetup{
    colorlinks = true,
    linkcolor = blue,
    citecolor = blue,
    urlcolor = Blue
}

\title{Inflationary Gravitational Waves as a probe of the unknown post-inflationary primordial Universe}

\author[a,b]{Athul K. Soman \orcid{0009-0002-0947-6892},}
\author[c,d]{Swagat S. Mishra \orcid{0000-0003-4057-145X},}
\author[b,c]{Mohammed Shafi \orcid{0000-0003-0438-4155},}
\author[b]{Soumen Basak}
\affiliation[a]{International School for Advanced Studies (SISSA), via Bonomea 265, 34136 Trieste, Italy.}
\affiliation[b]{School of Physics, Indian Institute of Science Education and Research, Thiruvananthapuram, 695551, India.}
\affiliation[c]{School of Physics and Astronomy,  University of Nottingham, Nottingham NG7 2RD, UK.}
\affiliation[d]{Inter-University Centre for Astronomy and Astrophysics (IUCAA),
Post Bag 4, Ganeshkhind, Pune 411~007, India.}
\emailAdd{akuruvai@sissa.it}
\emailAdd{swagat.mishra@nottingham.ac.uk}
\emailAdd{mohammedshafi.hashimnazeer@nottingham.ac.uk}
\emailAdd{sbasak@iisertvm.ac.in}

\abstract{One of the key predictions of the standard inflationary paradigm is the quantum mechanical generation of  the transverse and traceless tensor fluctuations due to  the rapid accelerated expansion of space, which  later constitute  a stochastic background of primordial gravitational waves (GWs). The amplitude of the (nearly) scale-invariant inflationary tensor power spectrum at large scales provides us with crucial information about the energy scale of inflation in the case of the minimal inflaton coupling to gravity. Furthermore, the spectral energy density,  $\Omega_{_{\rm GW}}(f)$, of the GWs at sufficiently small scales (or, large frequencies $f$) serves as an important observational probe of post-inflationary primordial dynamics. In fact, the small-scale spectral tilt, $n_{_{\rm GW}} = \frac{{\rm d}\log{\Omega_{_{\rm GW}}}}{{\rm d}\log{f}}$, of the  spectral energy density of  GWs is  sensitive to the (unknown) post-inflationary equation of state (EoS), $w$, of the universe; with a softer EoS ($w < 1/3$) leading to a red tilt: $n_{_{\rm GW}} < 0$, while a stiffer EoS ($w > 1/3$) resulting in  a blue tilt: $n_{_{\rm GW}} > 0$. The post-inflationary dynamics, however,  is generically  expected to be quite complex, potentially involving a number of distinct  phases. Hence,  in this work, we discuss the possibility of multiple sharp transitions, namely $w_1 \to w_2 \to w_3 \to ... \to w_n$, in the EoS of the post-inflationary universe and compute the corresponding spectral energy density of the inflationary GWs. We explicitly determine the region of the parameter space $\lbrace{ w_1, \, w_2, \, w_3, ..., w_n\rbrace}$ which leads to a potentially detectable signal in the upcoming GW detectors, without violating the current constraints.} 
\begin{document}
\maketitle

\section{Introduction}\label{sec:intro}
Cosmic inflation~\cite{Starobinsky:1980te,Guth:1980zm,Linde:1981mu,Albrecht:1982wi,Linde:1983gd,Linde:1990flp,kinney2009tasi,Martin:2013tda,Baumann_TASI,Baumann:2018muz,Kodama1984_CPT,Riotto2002_CPT,Ellis:2023wic,Mishra:2024axb} is currently the leading paradigm in providing natural initial conditions for the hot Big Bang  phase~\cite{Linde:1990flp,Kolb:1990vq,Dodelson:2003ft,Rubakov:2017xzr}. In the simplest scenario, inflation is assumed to be sourced by a single (real) scalar field $\phi$, called the \textit{inflaton field}, which rolls slowly down its potential $V(\phi)$, and  couples minimally to the Ricci scalar~\cite{Linde:1990flp,Martin:2013tda,Baumann_TASI,Mishra:2024axb}.  Of particular importance is the spectrum of (nearly) scale-invariant scalar perturbations on super-Hubble scales, generated \textit{via} quantum fluctuations due to  the rapid accelerated expansion of space during inflation~\cite{Mukhanov:1981xt,Hawking:1982cz,Starobinsky:1982ee,Guth:1982ec,Linde:1990flp}. Upon their Hubble-entry,  the scalar curvature fluctuations induce the temperature and density inhomogeneities in the primordial plasma, which later grow \textit{via} gravitational instability, leading to the formation of the \textit{large-scale structure} (LSS) in the universe~\cite{Linde:1990flp,Dodelson:2003ft,Mukhanov:2005sc,Baumann:2022mni,Gorbunov:2011zzc}. The latest \textit{Cosmic Microwave Background} (CMB) observations~\cite{Planck:2018vyg, Planck_overview, Planck_inflation,2021PhRvL.127o1301A} provide strong support for the single field slow-roll inflationary paradigm.

Another important aspect of the inflationary dynamics is the quantum mechanical generation of a nearly scale-invariant spectrum of transverse and traceless tensor fluctuations~\cite{Grishchuk:1974ny} on super-Hubble scales. The inflationary tensor fluctuations, upon their Hubble-entry in the post-inflationary epochs,  constitute a stochastic background of primordial gravitational waves (GWs)~\cite{Starobinsky:1979ty, Sahni:1990tx, Allen:1987bk}. Both scalar and tensor primordial fluctuations leave distinct imprints on the CMB, which can be parameterised in terms of two sets of inflationary observables,  namely, the amplitudes and the spectral indices of the primordial power spectra. In particular,  the parameter space comprising of  the scalar spectral index $n_{_S}-1$ and the tensor-to-scalar ratio $r$ is quite illustrative in ruling out a large class of inflationary models~\cite{Dodelson:1997hr,Martin:2013tda,Planck_overview,Planck_inflation, Mishra:2021wkm,Mishra:2022ijb}.   The latest  CMB observations~\cite{Planck_overview,2021PhRvL.127o1301A} impose constraints on the amplitude of primordial GWs,  expressed as an upper bound on the tensor-to-scalar ratio, namely, $r < 0.036$ at $95\%$ confidence, which already disfavours a large class of single field inflationary models, see Ref.~\cite{Mishra:2022ijb}. Moreover, the inflationary tensor fluctuations may potentially be important in  distinguishing between different competing frameworks of the pre-hot Big Bang universe~\cite{Brandenberger:2009jq,Brandenberger:2018wbg,Creminelli:2007aq,Ijjas:2019pyf,Ijjas:2018qbo,Graham:2017hfr,Boyle:2003km, Baumann:2007zm,Finelli:2001sr,Brandenberger:2016vhg,Wands:1998yp,Gasperini:1992em,Lidsey:1999mc,Gasperini:2002bn,Khoury:2001wf,Raveendran:2018yyh,Raveendran:2017vfx,Levy:2015awa,Ijjas:2014fja,Brandenberger:1988aj,Brandenberger:2011et,Brandenberger:2008nx} observationally. 

Since these GWs originate during the pre-hot Big Bang universe and propagate towards the present epoch, they encode important information about the physical processes during inflation as well as the post-inflationary dynamics of the universe. In particular,  given that they are generated due to the exponential expansion during inflation, the amplitude of the spectral energy density $\Og(f)$ of the primordial GWs at large length scales (or small frequencies $f$) is related to the energy scale  (Hubble scale $H$) of inflation~\cite{Starobinsky:1979ty,Sahni:1990tx,Giovannini:1999hx,Mishra:2021wkm,Haque:2021dha}.   Additionally, the spectral index of $\Og(f)$, defined by $n_{_{\rm GW}} = \frac{\d \log{\Omega_{_{\rm GW}}}}{\d \log{f}}$, is sensitive to the equation of state (EoS) of  the  post-inflationary evolution~\cite{Sahni:1990tx}.   In particular, the spectral tilt $\ng$ at small scales (or high frequencies $f$)  is determined by the (unknown) post-inflationary EoS $w$ of the primordial universe prior to the commencement of the Big Bang Nucleosynthesis (BBN)~\cite{Sahni:1990tx,Mishra:2021wkm}. A softer EoS ($w < 1/3$) leads to a red tilt: $n_{_{\rm GW}} < 0$, while a stiffer EoS ($w > 1/3$) results in  a blue tilt: $n_{_{\rm GW}} > 0$.  A combination of the amplitude and tilt of $\Og(f)$ across a broad range of frequency bands would then enable us to probe the dynamics of the early universe at energy scales far beyond the reach of the most powerful terrestrial particle accelerators~\cite{Sahni:1990tx, Giovannini:1999hx, Giovannini:2019oii, Caprini:2018mtu,Guzzetti:2016mkm, Wang:2016tbj, Mishra:2021wkm, Haque:2021dha}. Hence the search for the primordial  GWs is often regarded as one of the key future targets in Cosmology.

A number of GW detectors are already operational at present, while more sensitive and advanced future detectors, both ground-based and space-based, at different frequency ranges are under construction.   The  existing (and upcoming) GW detectors can be primarily classified into three categories, namely, \textit{(i)} CMB B-mode polarisation probes operating in the  ultra-low frequency range ($10^{-18} \, {\rm Hz} \lesssim f \lesssim 10^{-16} \, {\rm Hz}$), \textit{(ii)} pulsar timing arrays (PTAs) sensitive to GWs in the low frequency range ($10^{-9} \, {\rm Hz} \lesssim f \lesssim 10^{-7} \, {\rm Hz} $), and \textit{(iii)} atomic and laser interferometers operating in the relatively high frequency range $(10^{-5} \, {\rm Hz} \lesssim f \lesssim 10^3 \, {\rm Hz})$~\cite{Campeti:2020xwn}. Amongst the ultra-low frequency GW detectors, there are CMB probes such as  the BICEP/Keck Array~\cite{BICEP:2018czh}, LiteBIRD~\cite{Hazumi:2019lys}, and POLARBEAR/Simons Array~\cite{POLARBEAR:2015ixw}, which are designed to measure the B-mode polarisation of the CMB photons, induced by the quadrupolar primordial GWs~\cite{Seljak:1996gy, Kuroyanagi:2014qaa}. The recent data release by the PTA observations, such as the NANOGrav collaboration~\cite{NANOGrav:2023gor,, NANOGrav:2023tcn, NANOGrav:2023hde}, the EPTA/InPTA~\cite{EPTA:2023fyk, EPTA:2023sfo, EPTA:2023gyr} and IPTA~\cite{InternationalPulsarTimingArray:2023mzf} provide tentative evidence for the detection of  a background of stochastic GWs at low frequencies  ($\sim {\cal O}\l( 10^1\r)\, {\rm nHz}$ range), whose primary origin, whether astrophysical or primordial, is yet to be established~\cite{NANOGrav:2023hvm,EPTA:2023xxk}. The currently operational interferometers such as the advanced Laser Interferometer Gravitational-wave Observatory (aLIGO)~\cite{LIGOScientific:2014pky, Bode:2020dge}, Virgo~\cite{VIRGO:2014yos}, KAGRA~\cite{KAGRA:2018plz} and the upcoming/proposed interferometers, such as the Laser Interferometer Space Antenna (LISA)~\cite{LISA:2017pwj, Bayle:2022hvs, LISACosmologyWorkingGroup:2022jok}, Cosmic Explorer (CE)~\cite{Reitze:2019iox}, Einstein Telescope (ET)~\cite{Coccia:2023wag}, Big Bang Observer (BBO)~\cite{Harry:2006fi}, and the DECi-hertz Interferometer Gravitational wave Observatory (DECIGO)~\cite{Sato:2009zzb, Kawamura:2019jqt},  will be able to probe  the (relatively) large frequency GWs from the early universe associated with the small-scale inflationary dynamics, as well as the post-inflationary expansion history of the universe, prior to the epoch of Quantum Chromodynamics (QCD) phase transition.
 
During the period of single field slow-roll inflation, the energy density of the universe is primarily comprised of  the homogeneous inflaton condensate $\phi$, which rolls down its potential $V(\phi)$ in a slow terminal speed, thereby driving the near-exponential expansion of space. If the inflaton couplings to other external fields are low enough, then particle production during inflation can be neglected, since the rapid accelerated expansion of space quickly dilutes away the decay products. After the end of inflation, however,  the inflaton field begins to oscillate around the minimum of its potential~\cite{Turner:1983he} and begins to transfer its energy to the particles of external offspring fields  that it is coupled to via parametric resonance~\cite{Kofman_1994, Shtanov:1994ce, Kofman:1997yn, Kofman:1996mv, Lozanov:2019jxc}, and/or to the inflaton inhomogeneities $\delta\varphi$ \textit{via} self-resonance (when attractive self-interactions are present in $V(\phi)$)~\cite{Amin:2011hj, Lozanov:2017hjm, Lozanov:2019jxc, Mahbub:2023faw, Shafi:2024jig}. In certain classes of inflation, where the inflaton potential does not feature a stable minimum, rather it decays to zero at large field values, particle production \textit{via} resonance is absent. However, the inflaton can still decay either perturbatively~\cite{Parker:1969au} or \textit{via} the process of {\em instant preheating}~\cite{Felder:1998vq}. Nevertheless, the offspring fields (and other decay products) undergo further interaction, scattering and  thermalization; as a result, the universe  eventually transitions to the  radiation-dominated thermal plasma phase (or, the hot Big bang phase) before the commencement of BBN. This transient phase between the end of inflation and  the beginning of the thermal radiation domination is referred to as the epoch of {\em reheating}, which is supposed to be the origin of all primordial matter and energy in the universe~\cite{Kofman:1997yn, Lozanov:2019jxc, Mishra:2024axb}. 

A number of comprehensive studies  conduced in the past three decades~\cite{Kofman_1994, Kofman:1997yn, Shtanov:1994ce, Kofman:1996mv, Lozanov:2019jxc, Antusch:2020iyq, Antusch:2021aiw, Antusch:2022mqv, Figueroa:2022iho, Allahverdi:2020bys, Lozanov:2016hid, Amin:2014eta}  demonstrate that the physical processes taking place during reheating is highly complex, and potentially non-linear. However, in most cases, the reheating dynamics can be broadly  divided into three stages: \textit{(i)} preheating, \textit{(ii)} backreaction, and \textit{(iii)} thermalization~\cite{Kofman_1994, Kofman:1997yn, lozanov2019lectures, Mishra:2024axb}.  The initial stage of preheating usually exhibits rapid and efficient particle production \textit{via} broad parametric resonance~\cite{Kofman_1994, Kofman:1997yn, Amin:2014eta}. Eventually, the backreaction of the produced particles leads to the complete/partial fragmentation of the coherent inflaton condensate. Additionally, cosmological redshifting drives  the momenta of produced particles  away from the broad resonance band, thereby shutting the resonance down and  quenching the resonant particle production. This phase is succeeded by a long and gradual perturbative decay of the inflaton, as well as the re-scattering of decay products. Hence, the final stage of  thermalization  is usually the longest of the three stages, which eventually leads to the hot Big Bang phase, as discussed before.  If strong attractive self-interaction of the inflaton field is present, then it can  lead to the formation of quasi-solitonic  structures like oscillons, resulting in the production of GWs~\cite{Lozanov_2018, Lozanov:2019ylm, Lozanov:2022yoy, Amin:2011hj, Lozanov:2017hjm, Lozanov:2019jxc, Mahbub:2023faw, Shafi:2024jig}. Depending upon the type of inflaton potential and the inflaton couplings to external fields, the duration and EoS of each of these phases can be different.

Despite the profusion of theoretical and numerical work in the field, the epoch of reheating  remains observationally inaccessible at present. In general, there exist two distinct types of observational probes of the early universe physics, namely \textit{(1)} primordial relics such as ultra long-lived solitons (\textit{e.g.} strings, domain walls, oscillons), and \textit{(2)} propagating primordial messengers (signals) such as primordial GWs. Furthermore, these primordial GWs may either originate from the physical processes during reheating or from an earlier epoch, such as inflation. In this work, we focus on the GW signal which gets 
 generated during inflation and propagates towards us, encoding the unknown primordial dynamics of reheating after inflation.
 In particular, the duration of and the equation of state (EoS) during reheating can be inferred from the spectral energy density of primordial GWs at higher frequencies~\cite{Mishra:2021wkm, lozanov2019lectures, Haque:2021dha}.  Furthermore, the GW spectrum is also influenced by a number of other factors,  such as the anisotropic stress from the free-streaming relativistic neutrinos~\cite{Weinberg:2003ur} and the variation in the effective number of relativistic degrees of freedom throughout the history of the universe~\cite{Schwarz:1997gv,Watanabe:2006qe,Kuroyanagi:2008ye,Watanabe:2006qe}. Since our interest lies with in the gross features of the GW spectrum, we will not consider these factors in our computation.

There has been a number of interesting papers in the recent literature~\cite{Martin:2014nya, Figueroa:2019paj, Mishra:2021wkm, Giovannini:2022eue, Giovannini:2022vha, Haque:2021dha,Vagnozzi:2020gtf,Benetti:2021uea,Vagnozzi:2023lwo,Barman:2022qgt, Barman:2023ymn, Barman:2023ktz, Barman:2023rpg, Barman:2024mqo} investigating the possibility  of probing  the epoch of reheating \textit{via} inflationary GWs, where the kinematics during reheating is parameterised by the expansion rate during ($\sim$ at the end of) inflation $\, H_{\rm inf}$, the temperature achieved at the end of reheating $T_{\rm r*}$,  the duration and the EoS $\wre$ during reheating. However, a majority of the papers has primarily  assumed  a single (average) EoS, $\wre$, throughout the reheating history\footnote{ However, see refs.~\cite{Giovannini:2022vha, Haque:2021dha} for earlier work involving multiple primordial equation of states.}. As mentioned earlier, the dynamics of reheating features a number of complex non-linear phases, and hence, the universe is likely to have undergone  multiple transitions through different equations of state during reheating~\cite{Ng:1993pv,Antusch:2021aiw}. Additionally, the universe might have undergone multiple phase-transitions~\cite{Gouttenoire:2021jhk,Gouttenoire:2021wzu,Gouttenoire:2019kij,Gouttenoire:2019rtn} after the completion of reheating in the radiation dominated epoch, leading to a significant deviation of the EoS from $w=1/3$. 

Consequently, in this paper, we carry out a thorough investigation of the spectrum of primordial stochastic GWs by considering the unknown post-inflationary history of the universe to have undergone multiple phases of piece-wise (nearly) constant EoS, namely, $w_1 \to w_2 \to w_3 \to ... \to w_n$. For the sake of simplicity in calculation, we further assume that  the transition between any two successive epochs is  sharp enough to be modelled as  an {\em instantaneous transition}. Under the aforementioned assumptions, we first solve the evolution equation for  the Fourier mode functions of the tensor fluctuations in different epochs of constant $w_i$,  and  then use the Israel junction matching conditions~\cite{Deruelle:1995kd, Sahni:1990tx, Mishra:2023lhe} to determine the full tensor mode functions. Using the thus obtained analytical expression for the mode functions, we finally compute  the present-day spectral energy density of the (first-order) inflationary GWs.  We highlight the subspace of the parameter space of EoS $\lbrace w_1, \, w_2, \,  ... \, w_n\rbrace$, and the corresponding duration of each epoch of constant $w_i$, satisfying two conditions: \textit{(i)} yielding  a potentially detectable background of stochastic GWs in the upcoming detectors, such as  LISA, BBO, DECIGO, CE, and ET; while at the same time, \textit{(ii)} satisfying the constraints imposed by the already existing GW, CMB, and BBN observations. To be specific, we provide forecasts for GWs corresponding to the unknown primordial history of the universe comprising of (up to) three different piece-wise constant EoS phases $\lbrace w_1, \, w_2, \, w_3\rbrace$. As a concrete example, we apply our method to determine the parameter space of the duration of different piece-wise constant $w_i$ epochs before the commencement of BBN  in the framework of the recently proposed String Theory inspired model of the post-inflationary universe in Ref.~\cite{Apers:2024ffe}, which leads to a detectable GW background in the aforementioned detectors.

\medskip

 Our paper is organised as follows:  In Sec.~\ref{sec:GWs_theory}, we first derive  analytical expressions for the tensor mode functions in the post-inflationary epochs using junction matching conditions, and  then move on to compute the corresponding spectral energy density of GWs.  Secs.~\ref{sec:probing_primodial_EOS} and \ref{sec:effect_GW_Trh} are dedicated to determining the parameter space of the multiple equations of states (and their corresponding duration) which leads to a potentially detectable signal in the upcoming GW detectors. In Sec.~\ref{sec:String_phenomenology}, we focus on  the post-inflationary evolution of the universe in a specific (String-inspired) model of the primordial universe, as proposed in Ref.~\cite{Apers:2024ffe}. Finally, we spell out the primary conclusions from our results, and provide further discussions on our work in Sec.~\ref{sec; discussion}. Various appendices provide supplementary material to the analysis carried out in the main sections. For example, we provide a brief review of the single field slow-roll  inflationary dynamics in App.~\ref{sec:Inf_dyn}, including the spectra of scalar and tensor  quantum fluctuations generated during inflation.  We also provide a succinct discussion about the latest CMB constraints on the inflationary observables in App.~\ref{sec:Inf_dyn}, which serve as the initial conditions for the post-inflationary evolution. App.~\ref{appendix; soln to MS eqn} discusses analytical solutions of the Mukhanov-Sasaki equation, while App.~\ref{appendix; aH} shows the evolution of the scale factor and Hubble parameter in terms of conformal time. Similarly, App.~\ref{App:Om_GWs} is dedicated to a discussion on the energy density of inflationary GWs, while App.~\ref{app; energy denisty param of CMB photons and neutrinos} discusses the energy density of photons and neutrinos. 

\medskip

We work with  natural units $\hbar = c = 1$ throughout this paper, and define the reduced Planck mass to be $m_p = 1/\sqrt{8\pi G} = 2.44 \t 10^{18}$ GeV. The background universe is considered to be described by the flat Friedmann-Lemaitre-Robertson-Walker (FLRW) line element, with metric  signature $(-, \,+, \,+, \,+)$. Derivative with respect to cosmic time `$t$' is denoted with an overdot $\l( \; \dot{} \; \r)$, while derivative with respect to conformal time `$\tau$' is denoted with an overprime $\l( \; ' \; \r)$.

\section{Primordial gravitational waves in the post-inflationary universe}
\label{sec:GWs_theory}

The rapid accelerated expansion during inflation leads to the generation of almost scale-invariant primordial tensor fluctuations on super-Hubble scales, as discussed in Sec.~\ref{sec:inf_dyn_tensor_QF}. 
These tensor fluctuations, upon making their Hubble-entry in the post-inflationary (decelerating) epochs, then propagate as waves and thus, they constitute a stochastic gravitational wave (GW) background. In this section, we carry out a detailed study of the evolution of these primordial tensor fluctuations in the post-inflationary universe. There has been a number of important papers in the literature in this direction, \textit{e.g.} Refs.~\cite{Figueroa:2019paj, Mishra:2021wkm, Giovannini:2022eue, Giovannini:2022vha, Haque:2021dha,Gouttenoire:2021jhk}, to quote a few. In particular, we would like to emphasize the work carried out in Refs.~\cite{Figueroa:2019paj} where the authors determined the parameter space of the post-inflationary epoch which leads to a detectable signal in the aLIGO and LISA GW observatories. To be specific, their analysis primarily focused on the estimation of the post-inflationary parameter space, such as the expansion rate during (at the end of) inflation ($H_{\rm inf}$), the duration of reheating (characterised by the temperature achieved at the end of reheating, $T_{\rm r*}$), and the EoS during reheating ($\wre)$.

It is important to note that the authors of Ref.~\cite{Figueroa:2019paj} modelled the post-inflationary reheating  epoch  by assuming a single EoS. However, as mentioned earlier, since the dynamics of reheating is quite complex,  the universe is likely to have undergone  multiple transitions through different equations of state during reheating~\cite{Ng:1993pv,Antusch:2021aiw}. Furthermore, the early universe might have undergone a number of phase-transitions after the completion of reheating in the radiation dominated epoch,  resulting in  a significant deviation of the EoS from $w=1/3$.  Therefore, it is important to extend the analysis of Ref.~\cite{Figueroa:2019paj} by allowing for the possibility of multiple phases of piece-wise constant EoS.

Before proceeding to determine the post-inflationary evolution of the tensor fluctuations, let us lay out the primary assumptions under which we base our analysis of multiple primordial EoS. Our technique is similar to the one carried out in  Ref.~\cite{Mishra:2023lhe} for multiple transitions during inflation in the context of primordial black hole formation.

\begin{enumerate}
    \item We assume the universe to  have  undergone multiple sharp (instantaneous) transitions in the EoS, $\wre$, \textit{i.e.}  $w_1 \rightarrow w_2$ at $\tau = \tau_1$\, ;  $w_2 \rightarrow w_3$ at $\tau = \tau_2$\,; $w_3 \rightarrow w_4$ at $\tau = \tau_3$\, ;  and so on,   in the post-inflationary epoch before the commencement of BBN. Hence, the EoS can be written as
\beq
\wre (\tau) = w_1 \, + \, \l( w_2 - w_1 \r) \Theta\l( \tau-\tau_1\r) \, + \, \l( w_3 - w_2 \r) \Theta\l( \tau-\tau_2\r) \, + \, \l( w_4 - w_3 \r) \Theta\l( \tau-\tau_3\r) \, + \, ... \, ,
\label{eq:w_inst_theta}
\eeq
where $\Theta(\tau-\tau_n)$ is the Heaviside step function. 
    \item The inflationary tensor modes  are frozen on the super-Hubble scales, with amplitude $h_{k, \, {\rm inf}}^{\lambda}$, before their Hubble-entry. Hence they serve as  the initial conditions for the post-inflationary evolution as,
    \begin{equation}\label{eq; h_k initial condition}
        h_k ^{\lambda} (\tau_k) =   h_{k, \, {\rm inf}} ^{\lambda} \quad ; \quad h_k ^{' \lambda} (\tau_k) = 0 \, ,
    \end{equation}
    where $\tau_k$ is the conformal time when the tensor mode of interest with the (comoving) wavenumber $k$ enters the comoving Hubble radius. The expression for $h_{k, \, {\rm inf}} ^{\lambda}$ is given by  (see Eq.~\eqref{eq; h_k, inf})
    \begin{equation}
        h_{k, \, {\rm inf}} ^{\lambda} (\tau) = i \, \sqrt{\frac{2}{k^3}} \,  \f{H}{m_p} 
      \label{eq:tensor_inf_sup_Hubble}
      \end{equation}
    \item We match the solutions of the equation of motion of tensor fluctuations  across two adjacent epochs with different (constant) EoS at the time of the instantaneous transition,  $\tau = \tau_n$, by using the Israel Junction conditions~\cite{Deruelle:1995kd}, \textit{i.e.} 
    \begin{align}
        h_{k,\, b} ^{\lambda} (\tau_n) &= h_{k,\, a} ^{\lambda} (\tau_n) \,, ~~~~~~~~~(\text{Continuity}), \label{eq; continuity of hk}\\
        h_{k,\, b} ^{' \lambda} (\tau)  \Big|_{\tau = \tau_n} &= h_{k,\, a} ^{' \lambda} (\tau)  \Big|_{\tau = \tau_n}  \, , \quad(\text{Differentiability}) \label{eq; differentaibility of hk},
    \end{align}
    where $h_{k,\, b} ^{\lambda}$ and $h_{k,\, a} ^{\lambda}$ correspond to the tensor amplitude before and after the transition respectively.
\end{enumerate}

\subsection{Evolution of gravitational waves}\label{sub sec; Evolution of gravitational waves }

First, we solve the equation of motion for the tensor mode functions in the post-inflationary epoch. The action in Eq.~\eqref{eq;action in terms of h_lambda} leads to the field equation 
\beq
h_{ij}^{''\lambda}  + 2 \, \frac{a'}{a} \, h_{ij}^{'\lambda}  - \vec{\nabla}^2  \, h_{ij}^{\lambda}= 0 \, ,
\label{eq:tensor_field_Eq}
\eeq
and the corresponding  equation for the evolution of Fourier mode functions (for each polarisation) becomes
\begin{equation}\label{eq; EOM in hk}
    h_k ^{'' \lambda}  + 2 \, \frac{a'}{a} \, h_k ^{' \lambda}  + k^2  \, h_k ^{\lambda}= 0 \, .
\end{equation}
For an epoch with a constant EoS, say $w$, the conformal Hubble parameter can be written as (see App.~\ref{appendix; aH})
\begin{equation}
    {\cal H} = \frac{a'}{a} = aH =  a_i H_i \l[1+ \frac{ a_i H_i  (\tau-\tau_i)}{\alpha}\r]^{-1} \, , \label{eq; conformal Hubble}
\end{equation}
where `$i$' denotes the beginning of the epoch and the parameter $\alpha = 2/(1+ 3 w)$. We introduce the variable $y(\tau) = k/[a(\tau) H(\tau)]$ as the new independent variable in order to rewrite Eq.~\eqref{eq; EOM in hk} in the desired form
\begin{equation}\label{eq; EOM in y}
    \frac{\d ^2 h_k ^{\lambda}}{\d y^2} + 2 \, \frac{\alpha}{y} \, \frac{\d h_k ^{\lambda}}{\d y} + \alpha^2 \, h_k ^{\lambda} = 0 \, .
\end{equation}
Note that $y < 1$ for super-Hubble modes and $y > 1$ for sub-Hubble modes. Defining $f_k^{\lambda} (x) = x^{(\alpha - 1/2)} \, h_k ^{\lambda}$ with $x = \alpha \, y$, we can convert Eq.~\eqref{eq; EOM in y} to the following expression
\begin{equation}\label{eq; Bessel eqn}
    \frac{\d ^2 f_k ^{\lambda}}{\d x^2} + \frac{1}{x} \, \frac{\d f_k ^{\lambda}}{\d x} + \l[ 1 - \frac{\l(\alpha - \half \r)^2}{x^2}\r]\, f_k ^{\lambda} = 0 \, ,
\end{equation}
which is of the form of a Bessel equation. Hence, the general solution of this equation can be written in terms of a linear combination of two independent Bessel functions of the first-kind, $J_{ \l(\alpha - \half \r)} (x)$ and $J_{- \l(\alpha - \half \r)} (x)$~\cite{NIST:DLMF},
\begin{align}
     h_k ^{\lambda} (y) &= \frac{1}{(\alpha \, y)^{\alpha - \half}} \l[A_k \, J_{ \l(\alpha - \half \r)} (\alpha \, y) +  B_k \, J_{- \l(\alpha - \half \r)} (\alpha \, y) \r] \label{eq; General exp for hk} \; , \\
     h_k ^{' \lambda} (y) &= \frac{-k}{(\alpha \, y)^{\alpha - \half}} \l[A_k \, J_{ \l(\alpha + \half \r)} (\alpha \, y) -  B_k \, J_{- \l(\alpha + \half \r)} (\alpha \, y) \r] \label{eq; General exp for (hk)'} \, .
\end{align}
The coefficients $A_k$ and $B_k$ in the solution are obtained by junction matching conditions at the transitions of two adjacent epochs as specified in Eqs.~\eqref{eq; continuity of hk} and  \eqref{eq; differentaibility of hk}. Specifically, consider a sharp transition from the $(n-1)^{\rm th}$ epoch to the $n^{\rm th}$ epoch with the equations of state $w_{(n-1)}$ and $w_n$ respectively, at time $\tau = \tau_{(n-1)}$. The expression for the corresponding tensor amplitude can be written as
\begin{equation}
    h_k ^{\lambda} (y) = 
    \begin{cases}
        \cfrac{\l[A_{k, \, m} \, J_{ \l(\alpha_{m} - \half \r)} (\alpha_{m} \, y) +  B_{k, \, m} \, J_{- \l(\alpha_{m} - \half \r)} (\alpha_{m} \, y) \r]}{(\alpha_{m} \, y)^{\alpha_{m} - \half}} , & y_{(m-1)} \leq y \leq y_{m} \, ,\\
        \\
        \cfrac{\l[A_{k, \, n} \, J_{ \l(\alpha_n - \half \r)} (\alpha_n \, y) +  B_{k, \, n} \, J_{- \l(\alpha_n - \half \r)} (\alpha_n \, y) \r]}{(\alpha_n \, y)^{\alpha_n - \half}} , & y_{m} \leq y \leq y_{n} \, ,
    \end{cases}
     \label{eq; h_k for two consecutive epoch}
\end{equation}
where $m = n-1$ and $y_n = k/[a(\tau_n) H(\tau_n)]$. Similarly
the derivatives of the tensor amplitude is given by
\begin{equation}
    h_k ^{' \lambda} (y)  = 
    \begin{cases}
        -k \, \cfrac{\l[A_{k, \, m} \, J_{ \l(\alpha_{m} + \half \r)} (\alpha_{m} \, y) -  B_{k, \, m} \, J_{- \l(\alpha_{m} + \half \r)} (\alpha_{m} \, y) \r]}{(\alpha_{m} \, y)^{\alpha_{m} - \half}} , & y_{(m-1)} \leq y \leq y_{m} \, ,\\
        \\
        -k \, \cfrac{\l[A_{k, \, n} \, J_{ \l(\alpha_n + \half \r)} (\alpha_n \, y) -  B_{k, \, n} \, J_{- \l(\alpha_n + \half \r)} (\alpha_n \, y) \r]}{(\alpha_n \, y)^{\alpha_n - \half}} , & y_{m} \leq y \leq y_{n} \, .
    \end{cases}
    \label{eq; h_k' for two consecutive epochs}
\end{equation}
Imposing junction matching conditions~(\ref{eq; continuity of hk}) and (\ref{eq; differentaibility of hk}) on  Eqs.~\eqref{eq; h_k for two consecutive epoch} and  \eqref{eq; h_k' for two consecutive epochs},  we obtain
\begin{align}
    \frac{A_{k, \, n} \, g_1 + B_{k, \, n} \, f_1}{\l( \alpha_n \, y_m \r)^{\l(\alpha_n - \half \r)}} = \frac{A_{k, \, m} \, g_2 + B_{k, \, m} \, f_2}{\l( \alpha_m \, y_m \r)^{\l(\alpha_m - \half \r)}}, \label{eq; Israel_cont_general}\\
    \frac{A_{k, \, n} \, g_3 - B_{k, \, n} \, f_3}{\l( \alpha_n \, y_m \r)^{\l(\alpha_n - \half \r)}} = \frac{A_{k, \, m} \, g_4 - B_{k, \, m} \, f_4}{\l( \alpha_m \, y_m \r)^{\l(\alpha_m - \half \r)}}, \label{eq; Israel_diff_general}
\end{align}
where
\begin{align}
    g_1 &= J_{\l(\alpha_{n} - \half\r)} (\alpha_{n} \, y_{m}) , &f_1 &= J_{-\l(\alpha_{n} - \half\r)} (\alpha_{n} \, y_{m}) \, , \label{eq; Bessel fun simple notations_order alpha_n -1/2}\\
    g_2 &= J_{\l(\alpha_{m} - \half\r)} (\alpha_{m} \, y_{m}) , &f_2 &= J_{-\l(\alpha_{m} - \half\r)} (\alpha_{m} \, y_{m}) \, , \label{eq; Bessel fun simple notations_order alpha_m -1/2}\\
    g_3 &= J_{\l(\alpha_{n} + \half\r)} (\alpha_{n} \, y_{m}) , &f_3 &= J_{-\l(\alpha_{n} + \half\r)} (\alpha_{n} \, y_{m}) \, ,\label{eq; Bessel fun simple notations_order alpha_n + 1/2}\\
    g_4 &= J_{\l(\alpha_{m} + \half\r)} (\alpha_{m} \, y_{m}) , &f_4 &= J_{-\l(\alpha_{m} + \half\r)} (\alpha_{m} \, y_{m}) \, \label{eq; Bessel fun simple notations_order alpha_m +1/2} \, .
\end{align}
A slight rearrangement of the terms yields the following relations between the coefficients before and after transition
\begin{align}
    A_{k, \, n} = \frac{\l( \alpha_n \, y_m \r)^{\l(\alpha_n - \half \r)}}{\l( \alpha_m \, y_m \r)^{\l(\alpha_m - \half \r)}}  \, \frac{\l[ A_{k, \, m} \l( g_2 \, f_3 + g_4 \, f_1 \r) + B_{k, \, m} \l( f_2 \, f_3 - f_4 \, f_1 \r) \r]}{f_1 \, g_3 + g_1 \, f_3} \, , \label{eq; coeff A_k,n}\\
    B_{k, \, n} = \frac{\l( \alpha_n \, y_m \r)^{\l(\alpha_n - \half \r)}}{\l( \alpha_m \, y_m \r)^{\l(\alpha_m - \half \r)}}  \, \frac{\l[ A_{k, \, m} \l( g_2 \, g_3 - g_4 \, g_1 \r) + B_{k, \, m} \l( f_2 \, g_3 + f_4 \, g_1 \r) \r]}{f_1 \, g_3 + g_1 \, f_3}\label{eq; coeff B_k,n} \, .
\end{align}
Utilizing Eqs.~\eqref{eq; coeff A_k,n} and \eqref{eq; coeff B_k,n}, we can find the coefficients $A_k$ and $B_k$ for any epoch in the post-inflationary universe, except for the first epoch right after the end of inflation as they have to be calculated using Eq.~\eqref{eq; h_k initial condition}. We stress that the expression for mode functions given in Eq.~\eqref{eq; General exp for hk}  is applicable at all scales in the post-inflationary universe, irrespective of whether the mode functions  are sub-Hubble\footnote{Note that we do not consider the deep-UV modes that never exited the Hubble-radius during inflation. In this sense, we have a UV cutoff around $k_{\rm UV} = k_{e}$. For a thorough discussion on regularization and renormalization of tensor modes, see Refs.~\cite{Negro:2024bbf,Negro:2024iwy}.} or super-Hubble, with the junction matching conditions having been imposed at the time of sharp transition.

\subsubsection{Universe with a single equation of state during reheating}\label{subsub sec; single EOS during reheating}

We begin by providing expressions for the coefficients  of  the tensor mode functions in Eq.~(\ref{eq; General exp for hk}) for the simplest case, where the post-inflationary reheating epoch is characterized by a single constant EoS. 

\begin{itemize}
    \item \textbf{Reheating epoch}

    Assume that the reheating epoch is characterised by a single effective EoS, $\wre$. Then, the resultant tensor amplitude  Eq.~\eqref{eq; General exp for hk} for reheating epoch can be written as
    \begin{equation}\label{eq; h_k for reheating}
         h_{k, \, {\rm re}} ^{\lambda} (y) = \frac{1}{(\alpha_{\rm re} \, y)^{\alpha_{\rm re} - \half}} \l[A_{k, \, {\rm re}} \, J_{ \l(\alpha_{\rm re} - \half \r)} (\alpha_{\rm re} \, y) +  B_{k, \, {\rm re}} \, J_{- \l(\alpha_{\rm re} - \half \r)} (\alpha_{\rm re} \, y) \r] \, ,
    \end{equation}
    where $\alpha_{\rm re} = 2/ (1+ 3 \, w_{\rm re})$. In order to use the initial conditions, we need the super-Hubble limits of the tensor mode and its derivative, for which we make use of the small argument limit of the Bessel function, $J_{\nu}(y \ll 1) \approx (y/2)^{\nu} / \Gamma(\nu+1)$, in the super-Hubble regime. The tensor mode function and its derivative  on super-Hubble scales take the form
    \begin{align}
        \lim_{y \rightarrow 0} \, h_{k, \, {\rm re}} ^{\lambda} (y) &= A_{k, \, {\rm re}} \, \frac{1}{2^{\l(\alpha_{\rm re} - \half \r)}} \frac{1}{\Gamma \l(\alpha_{\rm re} + \half \r)} +  B_{k, \, {\rm re}} \, \frac{2^{\l(\alpha_{\rm re} - \half \r)}}{\l(\alpha_{\rm re} \, y\r)^{2\l(\alpha_{\rm re} - \half \r)}} \frac{1}{\Gamma \l(-\alpha_{\rm re} + \thalf \r)} \, , \label{eq; h_k_re_super_Hubble}\\
        \lim_{y \rightarrow 0} \, h_{k, \, {\rm re}} ^{' \lambda} (y) &= -k \, \l[A_{k, \, {\rm re}} \, \frac{\alpha_{\rm re} \, y}{2^{\l(\alpha_{\rm re} + \half \r)}} \frac{1}{\Gamma \l(\alpha_{\rm re} + \thalf \r)} -  B_{k, \, {\rm re}} \, \frac{2^{\l(\alpha_{\rm re} + \half \r)}}{\l(\alpha_{\rm re} \, y\r)^{2\alpha_{\rm re} }} \frac{1}{\Gamma \l(-\alpha_{\rm re} + \half \r)}\r] \, . \label{eq; h_k'_re_super_Hubble}
    \end{align}
    Incorporating this into Eq.~\eqref{eq; h_k initial condition}, we obtain
    \begin{equation}\label{eq; coeff A_k and B_k for reheating}
        A_{k, \, {\rm re}} = 2^{\l(\alpha_{\rm re} - \half \r)} \,   \Gamma \l(\alpha_{\rm re} + 1/2 \r) \,  h_{k, \, {\rm inf}} ^{\lambda}, \quad B_{k, \, {\rm re}} = 0 \, .
    \end{equation}
    
    \item \textbf{Radiation-dominated epoch}

    The  tensor mode functions in the RD epoch ($w = 1/3, \, \alpha = 1$) take the form
    \begin{equation}
         h_{k, \, {\rm RD}}^{\lambda} (y) = \frac{1}{y^{\half}} \l[ A_{k, \, {\rm RD}} \, J_{ \half } ( y) +   B_{k, \, {\rm RD}}\, J_{- \half } (y) \r] \label{eq; h_k for RD epoch} \, .
    \end{equation}
    The coefficients for this epoch are  determined using Eqs.~\eqref{eq; coeff A_k,n} and \eqref{eq; coeff B_k,n} at the end of reheating (denoted by `${\rm r*}$')
    \begin{align}
        A_{k, \, {\rm RD}} &= \frac{y_{\rm r*}^{(1 - \alpha_{\rm re})}}{\alpha_{\rm re}^{\l(\alpha_{\rm re} - \half \r)}} \t A_{k, \, {\rm re}} \l[ \frac{J_{\l(\alpha_{\rm re} - \half \r)} (\alpha_{\rm re} \, y_{\rm r*}) \, J_{-\thalf } (y_{\rm r*}) + J_{\l(\alpha_{\rm re} + \half \r)} (\alpha_{\rm re} \, y_{\rm r*}) \, J_{-\half } (y_{\rm r*}) }{ J_{-\half } (y_{\rm r*}) \, J_{\thalf } (y_{\rm r*}) + J_{\half } (y_{\rm r*}) \, J_{-\thalf } (y_{\rm r*}) } \r] \, , \label{eq; coeff A_k RD} \\
        B_{k, \, {\rm RD}} &= \frac{y_{\rm r*}^{(1 - \alpha_{\rm re})}}{\alpha_{\rm re} ^{\l(\alpha_{\rm re} - \half \r)}} \t A_{k, {\rm re} } \l[ \; \frac{J_{\l(\alpha_{\rm re} - \half \r)} (\alpha_{\rm re} \, y_{\rm r*}) \, J_{\thalf } (y_{\rm r*}) \;  - \; J_{\l(\alpha_{\rm re} + \half \r)} (\alpha_{\rm re} \, y_{\rm r*}) \, J_{\half } (y_{\rm r*}) }{ J_{-\half } (y_{\rm r*}) \, J_{\thalf } (y_{\rm r*}) + J_{\half } (y_{\rm r*}) \, J_{-\thalf } (y_{\rm r*}) } \; \r] \, , \label{eq; coeff B_k RD}
    \end{align}
    where $y_{\rm r*} = k/k_{\rm r*}$ with $k_{\rm r*}$ being the comoving wavenumber that made its Hubble-entry at the end of reheating (beginning of radiation domination).
    
    \item \textbf{Matter-dominated epoch}
    
    The tensor mode functions in  the MD epoch ($w = 0, \, \alpha = 2$) take the form
    \begin{equation}\label{eq; h_k for MD epoch}
        h_{k, \, {\rm MD}}^{\lambda} (y) = \frac{1}{(2 \, y)^{ \thalf}} \l[A_{k, \, {\rm MD}} \, J_{  \thalf } (2 \, y) +   B_{k, \, {\rm MD}} \, J_{- \thalf } (2 \, y) \r] \, .
    \end{equation}
    Again, Eqs.~\eqref{eq; coeff A_k,n} and \eqref{eq; coeff B_k,n} at matter-radiation equality (denoted by `${\rm eq}$') lead to
    \begin{align*}
        A_{k, \, {\rm MD}} &= 2^{\thalf} \, y_{\rm eq} \t \Bigg\{ A_{k, \, {\rm RD}} \l[ \frac{J_{\half } (y_{\rm eq}) \, J_{-\fhalf} (2 \, y_{\rm eq}) + J_{\thalf } (y_{\rm eq}) \, J_{-\thalf} (2 \, y_{\rm eq}) }{ J_{-\thalf} (2 \, y_{\rm eq}) \, J_{\fhalf} (2 \, y_{\rm eq}) + J_{\thalf} (2 \, y_{\rm eq}) \, J_{-\fhalf} (2 \, y_{\rm eq}) } \r]\\
        &\hspace{30mm} + B_{k, \, {\rm RD}} \l[ \frac{J_{-\half } (y_{\rm eq}) \, J_{-\fhalf} (2 \, y_{\rm eq}) - J_{-\thalf } (y_{\rm eq}) \, J_{-\thalf} (2 \, y_{\rm eq}) }{ J_{-\thalf} (2 \, y_{\rm eq}) \, J_{\fhalf} (2 \, y_{\rm eq}) + J_{\thalf} (2 \, y_{\rm eq}) \, J_{-\fhalf} (2 \, y_{\rm eq})}\r]\Bigg\} \num \label{eq; coeff A_k MD} \, ,
    \end{align*}
    \begin{align*}
        B_{k, \, {\rm MD}} &= 2^{\thalf} \, y_{\rm eq} \t \Bigg\{ A_{k, \, {\rm RD}} \l[ \frac{J_{\half } (y_{\rm eq}) \, J_{\fhalf} (2 \, y_{\rm eq}) - J_{\thalf } (y_{\rm eq}) \, J_{\thalf} (2 \, y_{\rm eq}) }{ J_{-\thalf} (2 \, y_{\rm eq}) \, J_{\fhalf} (2 \, y_{\rm eq}) + J_{\thalf} (2 \, y_{\rm eq}) \, J_{-\fhalf} (2 \, y_{\rm eq}) } \r]\\
        &\hspace{30mm} + B_{k, \, {\rm RD}} \l[ \frac{J_{-\half } (y_{\rm eq}) \, J_{\fhalf} (2 \, y_{\rm eq}) + J_{-\thalf } (y_{\rm eq}) \, J_{\thalf} (2 \, y_{\rm eq}) }{ J_{-\thalf} (2 \, y_{\rm eq}) \, J_{\fhalf} (2 \, y_{\rm eq}) + J_{\thalf} (2 \, y_{\rm eq}) \, J_{-\fhalf} (2 \, y_{\rm eq})}\r]\Bigg\}\num \label{eq; coeff B_k MD} \, , 
    \end{align*}
    where $y_{\rm eq} = k/k_{\rm eq}$ with $k_{\rm eq}$ being the comoving wavenumber that made its Hubble-entry at the time of matter-radiation equality.

\end{itemize}

\subsubsection{Universe with multiple equations of state during reheating}
\label{subsub sec; multiple EOS during reheating}

\begin{figure}[hbt]
    \centering
    \includegraphics[width = \textwidth]{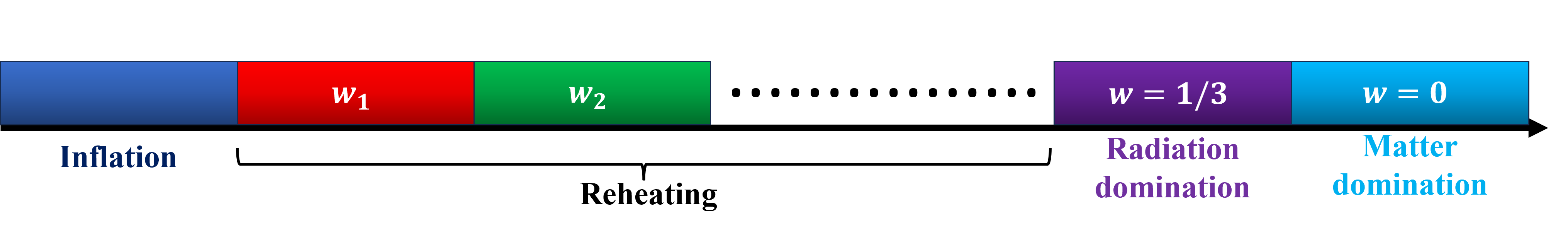}
    \caption{A schematic depiction of the timeline of the universe with a post-inflationary reheating phase consisting of multiple equations of state. Note, the horizontal length is not a representation of the actual duration.}
    \label{fig: timeline of multiple epochs}
\end{figure}

As stressed before, since reheating is a complex phase involving non-linear  processes, it is natural to represent it by multiple  equation of state parameters, as illustrated in Fig.~\ref{fig: timeline of multiple epochs}. We model reheating by multiple epochs of piece-wise constant EoS parameters with sharp (instantaneous) transitions in between. The procedure to obtain the coefficients  appearing in the solution for the tensor mode functions is therefore the same as that given in Sec.~\ref{subsub sec; single EOS during reheating}, with some additional (minor) changes that one need to take into account when considering multiple equations of state parameters, as detailed below.

The coefficients for the first epoch following the end of inflation are the same as that given in Eq.~\eqref{eq; coeff A_k and B_k for reheating}, for reheating with single EoS. The main difference comes while labelling them. For example, if the equation of state of the first epoch is $w_1$, then the coefficients of the mode functions are
\begin{equation}\label{eq; coeff A_k and B_k for first epoch}
    A_{k, \, 1} = 2^{\l(\alpha_1 - \frac{1}{2}\r)} \,   \Gamma \l(\alpha_1 + 1/2\r) \,  h_{k, \, {\rm inf}} ^{\lambda}, \quad B_{k, \, 1} = 0 \, ,
\end{equation}
where $\alpha_1 = 2/(1+3 \, w_1)$. Suppose, there are $n$ epochs during reheating, each with a constant EoS ($w_i$ being the EoS of the $i^{\rm th}$ epoch, with $i \in \mathbb{Z^{+}}$). Then,  the coefficients $A_{k, \, i}$ and $B_{k, \, i}$ corresponding to  all subsequent epochs ($i > 1$) during reheating need to be computed using Eqs.~\eqref{eq; coeff A_k,n} and \eqref{eq; coeff B_k,n}. For the sake of completeness, we  provide closed-form expressions for the coefficients at the onset of radiation domination following the final transition ($i = n$), which are different from the ones corresponding to a single transition as given in Eqs.~\eqref{eq; coeff A_k RD} and \eqref{eq; coeff B_k RD}, to be
\begin{align}
    A_{k, \, {\rm RD}} &= \frac{y_{\rm r*}^{(1 - \alpha_n)}}{\alpha_n^{(\alpha_n - \half )}} \Bigg\{   A_{k, \, n} \l[ \frac{J_{\l(\alpha_n - \half \r)} (\alpha_n \, y_{\rm r*}) \, J_{-\thalf } (y_{\rm r*}) + J_{\l(\alpha_n + \half \r)} (\alpha_n \, y_{\rm r*}) \, J_{-\half } (y_{\rm r*}) }{ J_{-\half } (y_{\rm r*}) \, J_{\thalf } (y_{\rm r*}) + J_{\half } (y_{\rm r*}) \, J_{-\thalf } (y_{\rm r*}) } \r] \nonumber\\
    &\hspace{20mm} +   B_{k, \, n} \l[ \frac{J_{-\l(\alpha_n - \half \r)} (\alpha_n \, y_{\rm r*}) \, J_{-\thalf } (y_{\rm r*}) - J_{-\l(\alpha_n + \half \r)} (\alpha_n \, y_{\rm r*}) \, J_{-\half } (y_{\rm r*}) }{ J_{-\half } (y_{\rm r*}) \, J_{\thalf } (y_{\rm r*}) + J_{\half } (y_{\rm r*}) \, J_{-\thalf } (y_{\rm r*}) } \r] \Bigg\} \, , \label{eq; coeff A_k RD multiple case}
\end{align}
\begin{align}
    B_{k, \, {\rm RD}} &= \frac{y_{\rm r*}^{(1 - \alpha_n)}}{\alpha_n ^{(\alpha_n - \half )}}  \Bigg\{ A_{k, n } \l[ \; \frac{J_{\l(\alpha_n - \half \r)} (\alpha_n \, y_{\rm r*}) \, J_{\thalf } (y_{\rm r*}) \;  - \; J_{\l(\alpha_n + \half \r)} (\alpha_n \, y_{\rm r*}) \, J_{\half } (y_{\rm r*}) }{ J_{-\half } (y_{\rm r*}) \, J_{\thalf } (y_{\rm r*}) + J_{\half } (y_{\rm r*}) \, J_{-\thalf } (y_{\rm r*}) } \; \r] \nonumber\\
    &\hspace{20mm} +  B_{k, \, n} \l[ \; \frac{J_{-\l(\alpha_n - \half \r)} (\alpha_n \, y_{\rm r*}) \, J_{\thalf } (y_{\rm r*}) \;  + \; J_{-\l(\alpha_n + \half \r)} (\alpha_n \, y_{\rm r*}) \, J_{\half } (y_{\rm r*}) }{ J_{-\half } (y_{\rm r*}) \, J_{\thalf } (y_{\rm r*}) + J_{\half } (y_{\rm r*}) \, J_{-\thalf } (y_{\rm r*}) } \; \r]  \Bigg\} \, , \label{eq; coeff B_k RD multiple case}
\end{align}
where $\alpha_n = 2/(1+3 \, w_n)$. While, the analytical expression for $A_{k, \, {\rm MD}}$ and $B_{k, \, {\rm MD}}$ remains the same as in Eqs.~\eqref{eq; coeff A_k MD} and \eqref{eq; coeff B_k MD}.

\subsection{Spectral energy density of gravitational waves}\label{sub sec; Spectral energy density of gravitational waves}

A cosmological background of stochastic gravitational waves at a given time is characterized in terms of the spectral energy density of GWs, defined as the energy density of GWs per unit logarithmic wavenumber interval relative to the critical energy density of the universe, $\rho_{c} (\tau) = 3 \, \mpl^2 \, H^2 (\tau)$ at that epoch~\cite{Kohri:2018awv, Giovannini:2019ioo}, \textit{i.e.,}
\begin{equation}
    \Omega_{_{\text{GW}}}(\tau, k) =  \frac{1}{\rho_c}\, \frac{\d \rho_{_{\rm GW}}}{\d \ln k} \, .
    \label{eqn; defn of Omega_GW}
\end{equation}
The energy density of GWs obtained from the time-time component of the energy-momentum tensor $\tensor{T}{^{\mu}_{\nu}} $ of tensor fluctuations in the FLRW background (see App.~\ref{appendix; energy momentum tensor}) is 
\begin{equation}\label{eq; rho_GW}
    \hat{\rho}_{_{\rm GW}}(\tau,\vec{x}) = -\tensor{\hat{T}}{^0_0}(\tau,\vec{x}) = \frac{\mpl^2}{8a^2 (\tau)} \l[(\hat{h}'_{ij}(\tau,\vec{x}))^2 + (\grad \hat{h}_{ij}(\tau,\vec{x}))^2\r] \, .
\end{equation} 
The vacuum expectation value of $\rho_{_{\rm GW}}$ (with summation over the two polarisation states) is given by
\begin{equation}\label{eq; expectation of rho_GW}
    \rho_{_{\rm GW}}(\tau) = \l<0\l|\hat{\rho}_{_{\rm GW}}(\tau,\vec{x})\r|0\r>= \frac{\mpl^2}{8a^2 (\tau)} \int \d \ln k\; \frac{ k^3}{\pi^2}\Big[\l|h_k ^{' \lambda} (\tau) \r|^2 + k^2 \l|h_k ^{\lambda} (\tau) \r|^2 \Big] \, ,
\end{equation}
where we used the following Fourier representation of the tensor operator $\hat{h}_{ij}$
\begin{equation}\label{eq; operator h_ij}
    \hat{h}_{ij}(\tau,\vec{x}) = \sum_{\lambda=+,\times} \int \frac{\d^3\vec{k}}{(2\pi)^3} \, \epsilon^{\lambda}_{ij} \, \l[h_k^{\lambda} (\tau) \, \hat{a}_{\vec{k}} \, e^{i\vec{k} \cdot \vec{x}} + \l(h_k^{\lambda} (\tau) \r)^*  \, \hat{a}_{\vec{k}}^{\dag} \, e^{-i\vec{k} \cdot \vec{x}}\r] \, ,
\end{equation}
and the vacuum state is defined by $\hat{a}_{\vec{k}} \l|0 \r> = 0$. Using Eq.~\eqref{eq; expectation of rho_GW},  expression for the GW spectral energy density becomes
\begin{equation}\label{eq; equation of pre-Omega_GW }
    \Omega_{_{\text{GW}}}(\tau, k) = \frac{k^2 }{24 \, a^2 (\tau) H^2 (\tau)} \; {\cal P}_h(\tau,k) \, \l[ 1 + \frac{ 1}{k^2} \, \frac{\l|h_k ^{' \lambda} (\tau) \r|^2}{\l|h_k ^{\lambda} (\tau) \r|^2}\r] \, .
\end{equation}
where ${\cal P}_h(\tau,k)$ is the post-inflationary tensor power spectrum (at some time $\tau$), defined by
\begin{equation}
    \label{eq:Tensor_power_spectrum_tau}
    {\cal P}_h(\tau, k) =  \frac{k^3}{2 \pi^2} \left(\l|h_k^+(\tau)\r|^2 + \l|h_k^{\times}(\tau)\r|^2\right) \, .
\end{equation} 
Considering modes that are deep in the Hubble radius so that the frequency of GWs is substantially larger than the expansion rate of the universe, \textit{i.e.}, $k \gg aH$, the second term inside the square bracket becomes~\cite{Boyle:2005se, Giovannini:2023itq}
\begin{equation}
    \frac{ 1}{k^2} \, \frac{\l|h_k ^{' \lambda} (\tau) \r|^2}{\l|h_k ^{\lambda} (\tau) \r|^2} = 1 + {\cal O} \l( \frac{a^2 H^2}{k^2} \r) \simeq 1 \,. \label{eq; high freq limit of power spectrums}
\end{equation}
Therefore, in the sub-Hubble regime, expression for  the GW spectral energy density becomes
\begin{equation}\label{eq; equation of Omega_GW }
    \Omega_{_{\text{GW}}}(\tau, k)=\frac{k^2 }{12\, a^2 (\tau) H^2 (\tau)} \; {\cal P}_h(\tau,k) \, ,
\end{equation}
which coincides with the standard Isaacson approximation~\cite{Brill:1964zz, Isaacson:1968zza}. We again emphasize that this approximation is valid as long as the wavelength of GWs are much shorter than the Hubble radius of the universe\footnote{In the cosmological context,  there are long wavelength tensor modes comparable to the Hubble scale that might be of observational relevance. Hence, in general, one needs to derive the energy-momentum tensor without invoking the frequently used approximation of high frequency GWs propagating in flat background~\cite{Giovannini:2019ioo, Negro:2024bbf}. Nevertheless, in this work we will continue using Eq.~\eqref{eq; equation of Omega_GW } since we are primarily interested in high frequency GWs  that are deep inside the comoving Hubble radius in the MD epoch.}.

In order to calculate the spectral energy density at the present epoch, we compute tensor mode functions deep inside the Hubble radius, in the limit $y \gg 1$, during the MD epoch by making use of the large argument expansion of  Bessel function, namely, $J_{\nu} (y) \approx  \sqrt{2/(\pi y)} \, \cos(y -  \nu \frac{\pi}{2} - \frac{\pi}{4})$. {Accordingly,} Eq.~\eqref{eq; h_k for MD epoch} reduces to
\begin{equation}\label{eq; h_k MD_ sub Hubble limit}
    h_{k, \, {\rm MD}}^{\lambda} (y \gg 1) = - \frac{1}{(2 \, y)^{ \thalf}} \, \frac{1}{\sqrt{\pi y}} \l[A_{k, \, {\rm MD}} \, \cos(2 \, y) +   B_{k, \, {\rm MD}} \, \sin(2 \, y) \r] \, .
\end{equation}
Consequently, the average of the modulus squared value of tensor amplitude becomes
\begin{equation}\label{eq; h_k in MD sq_avg value}
     \overline{|h_{k, \, {\rm MD}}^{\lambda} (y \gg 1)|^2} = \frac{1}{(2 \, y)^{4}} \, \frac{1}{\pi} \l[|A_{k, \, {\rm MD}}|^2 +  |B_{k, \, {\rm MD}}|^2\r] \, .
\end{equation}
Upon substituting $y(\tau)  = y_{\rm eq} \, \frac{a_{\rm eq} H_{\rm eq}}{a(\tau) H(\tau)} = y_{\rm eq} \,  \l( \frac{a(\tau)}{a_{\rm eq}}\r)^{1/2}$, for $\tau > \tau_{\rm eq}$, into the above expression, we can observe that the term $\overline{|h_{k, \, {\rm MD}}^{\lambda} (y \gg 1)|^2} \propto a^{-2}$. In fact, this is true in general for any epoch, \textit{i.e.,} $\overline{|h_{k} ^{\lambda} (y \gg 1)|^2} \propto a^{-2}$. This implies that the tensor amplitude dampens as $h_{k} ^{\lambda} \propto a^{-1}$ due to the expansion of the universe, irrespective of {the EoS}~\cite{Figueroa:2019paj}. At late times, as the universe begins to accelerate again,  no new additional tensor modes become sub-Hubble in this epoch. Instead, a narrow range of long-wavelength tensor modes, that had freshly become sub-Hubble, begin to exit the  Hubble radius during this accelerated epoch. We therefore consider these tensor modes inaccessible, and only focus on  sub-Hubble modes deep inside the Hubble radius at the onset of the MD epoch. Substituting Eq.~\eqref{eq; h_k in MD sq_avg value} and $k = y_{\rm eq} \, a_{\rm eq} \, H_{\rm eq}$ in Eq.~\eqref{eq; equation of Omega_GW } we obtain
\begin{align}\label{eq; final exp for Omega_GW}
    \overline{\Omega_{_{\rm GW}}(\tau_0, k)} = \frac{1}{96 \pi^3} \,\frac{g_{*, \, {\rm r*}}}{g_{*, \, 0}} \l(\frac{g_{s, \, 0}}{g_{s, \, {\rm r*}}}\r)^{4/3} \, \Omega_{\rm rad, \, 0} \,    \frac{1}{y_{\rm eq}^2} \l[\tilde{A}^2_{k, \, {\rm MD}} +  \tilde{B}^2_{k, \, {\rm MD}}\r]  \,\frac{H^2_{\rm inf}}{\mpl^2} \l(\frac{k}{k_*}\r)^{n_{_T}} \, ,
\end{align}
where $g_*, \, g_s$ are the effective number of relativistic degrees of freedom of the universe in energy density, entropy density respectively, with the second subscript in $g_*$ and $g_s$ denoting the time of estimation of its value (with `0' denoting the present epoch). $H_{{\rm inf}}$  corresponds to the Hubble parameter at the Hubble-exit of the CMB pivot scale during inflation, \textit{i.e.}, $H_{{\rm inf}} = H_*$. Note that in  Eq.~\eqref{eq; final exp for Omega_GW}, the  coefficients $\tilde{A}_{k, \, {\rm MD}}$ and $\tilde{B}_{k, \, {\rm MD}}$ differ from  $A_{k, \, {\rm MD}}$ and $B_{k, \, {\rm MD}}$ in Eq.~\eqref{eq; h_k for MD epoch} by the factor $h_{k, \, {\rm inf}} ^{\lambda}$ as discussed in App.~\ref{appendix; Omega_GW}.

\bigskip

From detection perspective, it is important to characterize the GW spectral energy density in terms of the present-day (physical, as opposed to comoving) frequency of GWs~\cite{NANOGrav:2023hvm, Figueroa:2019paj, Campeti:2020xwn, Kuroyanagi:2008ye, Boyle:2005se}. The present-day frequency $f_k$ of a mode of comoving wavenumber $k$ is given by
\begin{equation}\label{eq; rel bw present freq f and wave no k}
    f_k = \frac{k}{2 \pi a_0} = \frac{1}{2 \pi} \frac{a_k}{a_0} H_k \, ,
\end{equation}
where $a_k$ and $H_k$ are the scale factor and Hubble parameter at the time when the  mode with comoving wavenumber $k$ made its Hubble-entry.

In order for us to mark the beginning and the end of each post-inflationary epoch in our computation, we relate the present-day frequency of a mode to the temperature at the time of its Hubble-entry.
In a radiation-dominated universe, the Hubble parameter can be expressed in terms of its temperature as
\begin{equation}\label{eq; Hubble H for univ. with rel. species}
    H^2(T) = \frac{\rho_{\rm rad} (T)}{3 \, \mpl^2} = \frac{1}{3 \, \mpl^2} \,\frac{\pi^2}{30}\, g_{*,\, T} \, T^4 \, . 
\end{equation}
From the conservation of entropy, we have
\begin{equation}\label{eq; entropy conservation}
    \l(\frac{g_{s, \, 0}}{g_{s,T}} \r)^{1/3} \, \frac{T_0}{T} = \frac{a(T)}{a_0} \, .
\end{equation}
By incorporating Eqs.~\eqref{eq; Hubble H for univ. with rel. species} and \eqref{eq; entropy conservation} to Eq.~\eqref{eq; rel bw present freq f and wave no k}, one can  relate the temperature $T_k$ of the universe (in GeV) at the Hubble-entry of a mode to its present-day frequency $f_k$ (in Hz) as~\cite{Mishra:2021wkm}
\begin{align}\label{eq; present-day freq of GW}
    \frac{f_k}{{\rm Hz}} = 7.43 \t 10^{-8} \l(\frac{g_{s, \, 0}}{g_{s,T_k}} \r)^{1/3} \, \l(\frac{g_{*,T_k}}{90} \r)^{1/2} \,  \l(\frac{T_k}{ {\rm GeV}} \r) \, .
\end{align}
Note that the above expression is valid for a radiation-dominated universe in thermal equilibrium. With reheating being primarily a non-thermal process, achieving thermal equilibrium only at the end of it, Eq.~\eqref{eq; present-day freq of GW} cannot be used during reheating. However, we shall be using Eq.~\eqref{eq; present-day freq of GW} to relate the energy scale of the universe during reheating to the physical frequency of GWs; where $T_k$ is treated as an \textit{effective temperature} (related to $f_k$ via Eq.~(\ref{eq; present-day freq of GW}), which serves as a proxy for the energy-scale at the Hubble-entry time of the mode. In fact, to be precise,  we can define the corresponding  energy scale of the universe to be 
\begin{equation}\label{eq; energy and effective temp}
    E_k = \rho_c^{1/4}(T_k) = \l(\frac{\pi^2}{30} \, g_{*,\, T_k} \r)^{1/4} \, T_k \, .
\end{equation}
 Therefore, the present-day frequency of a mode making its Hubble-entry at an energy scale $E_k$ is given by
\begin{equation}\label{eq; f_k and E_k}
    \frac{f_k}{{\rm Hz}} = 1.03 \t 10^{-8} \l(\frac{g_{s, \, 0}}{g_{s,T_k}} \r)^{1/3} \, g_{*,T_k} ^{1/4} \,  \l(\frac{E_k}{ {\rm GeV}} \r) \, .
\end{equation}
Table \ref{tab:temperature and frequencies} lists the present-day frequency of GWs generated at the Hubble-entry of some of the important cosmological events (and their corresponding energy scales).

While computing the spectral energy density of GWs, in order to be concrete we have fixed the energy scale during inflation to be $E_{\rm inf} = 5.76 \t 10^{15}$ GeV corresponding to the tensor-to-scalar ratio $r= 0.001$, which is a target of next generation CMB experiments. Similarly, in order to be consistent with the BBN predictions for the abundance of light nuclei, reheating must end  before the onset of BBN, \textit{i.e.,} $T_{\rm r*} \gg T_{\rm BBN} \sim 1$ MeV. Therefore, we fix the energy scale of the universe at the end of reheating to be $E_{\rm r*} = 1 \, {\rm GeV} \; (T_{\rm r*} \simeq 0.45 \, {\rm GeV}) $, which is safely above the BBN abound. Note that for a given reheating history, by decreasing the tensor-to-scalar ratio, the probability of detection of GWs will become lower. Therefore, our analysis in this work must be treated as a best case scenario for the detection. By fixing the reheating temperature, we primarily focus on illustrating how to probe the duration and EoS of successive epochs during reheating.

\begin{table}[htb]
    \centering
    \begin{tabular}{|c|c|c|}
    \hline
    & & \\[-1ex]
      Event & Energy scale ($E_k$)  & Present-day frequency $f_k$ (Hz) \\[1.5ex]
    \hline
    \hline
     & & \\[-1ex]
    Matter-radiation equality   & $\sim 1$ eV  & $1.4 \t 10^{-17}$\\[1.5ex]
    Hubble-entry time of CMB pivot scale   & $\sim 5$ eV  & $7.2 \t 10^{-17}$\\[1.5ex]
    Onset of BBN   & $\sim 1.4$ MeV & $1.8 \t 10^{-11}$\\[1.5ex]
    QCD Phase Transition   & $\sim 320$ MeV & $3.7 \t 10^{-9}$ \\[1.5ex]
    Electro-weak symmetry breaking   & $\sim 240$ GeV & $2.7 \t 10^{-6}$\\[1.5ex]
    \hline
    \end{tabular}
    \caption{ Present-day frequencies of primordial GWs generated at different (Hubble-entry) energy scale corresponding to some of the important events in the thermal history of our universe.}
    \label{tab:temperature and frequencies}
\end{table}

By  replacing the comoving wavenumber $k$ with the present-day frequency $f_k$, and  denoting the oscillation-averaged spectral energy density as $\Og ^{(0)} (f)$ instead of $\overline{\Og ^{(0)} (f)}$, we obtain
\begin{align}
\label{eq; exp for Omega_GW as fun of freq}
    h^2 \, \Og ^{(0)} (f_k) = \frac{1}{96 \pi^3} \,\frac{g_{*, \, {\rm r*}}}{g_{*, \, 0}} \l(\frac{g_{s, \, 0}}{g_{s, \, {\rm r*}}}\r)^{4/3} \, h^2 \Omega_{\rm rad, \, 0} \,    \l(\frac{f_{\rm eq}}{f_k} \r)^2 \l[\tilde{A}^2_{k, \, {\rm MD}} +  \tilde{B}^2_{k, \, {\rm MD}}\r]  \,\frac{H^2_{\rm inf}}{\mpl^2} \, ,
\end{align}
where $h$ is a dimensionless parameter, related to the Hubble parameter at the present epoch by $H_0 = 100 \, h \, {\rm km} \,{\rm  s}^{-1} \, {\rm Mpc}^{-1}$. Values of the relativistic degrees of freedom at our fixed reheating temperature are $g_{*, \, {\rm r*}} = g_{s, \, {\rm r*}} = 73.0, \, g_{s, \, 0} = 3.94, \, {\rm and} \, g_{*, \, 0} = 3.38$. The present-day density parameter of radiation is $ h^2 \Omega_{\rm rad, \, 0} = 4.16 \t 10^{-5}$ (see Eq.~\eqref{eq; present den param of rad}).
The spectral-tilt of relic GWs is defined as~\cite{Giovannini:1999hx}
\begin{equation}
    n_{_{\rm GW}} = \frac{\d \ln \Og ^{(0)} (f_k)}{\d \ln f_k}\,, \label{eq; spectral tilt defn}
\end{equation}
which,  for a range of  frequencies corresponding to modes making their Hubble-entry at the beginning and end of a post-inflationary cosmic epoch with constant EoS $w$, reduces to\footnote{ Note that we have ignored contribution from the tiny red-tilt $n_{_T} \simeq -r/8$ of the inflationary tensor modes, since they are very small for $r\leq 0.001$, see Ref.~\cite{Mishra:2024axb}.}~\cite{Mishra:2021wkm}
\begin{equation}
     n_{_{\rm GW}} = 2 \l( \frac{w - 1/3}{w + 1/3} \r)\,.
\label{eq; spectral tilt}
\end{equation}
as explicitly demonstrated in App.~\ref{App:GW_tilt}.
\begin{figure}[hbt]
    \centering
    \includegraphics[width=0.75\textwidth]{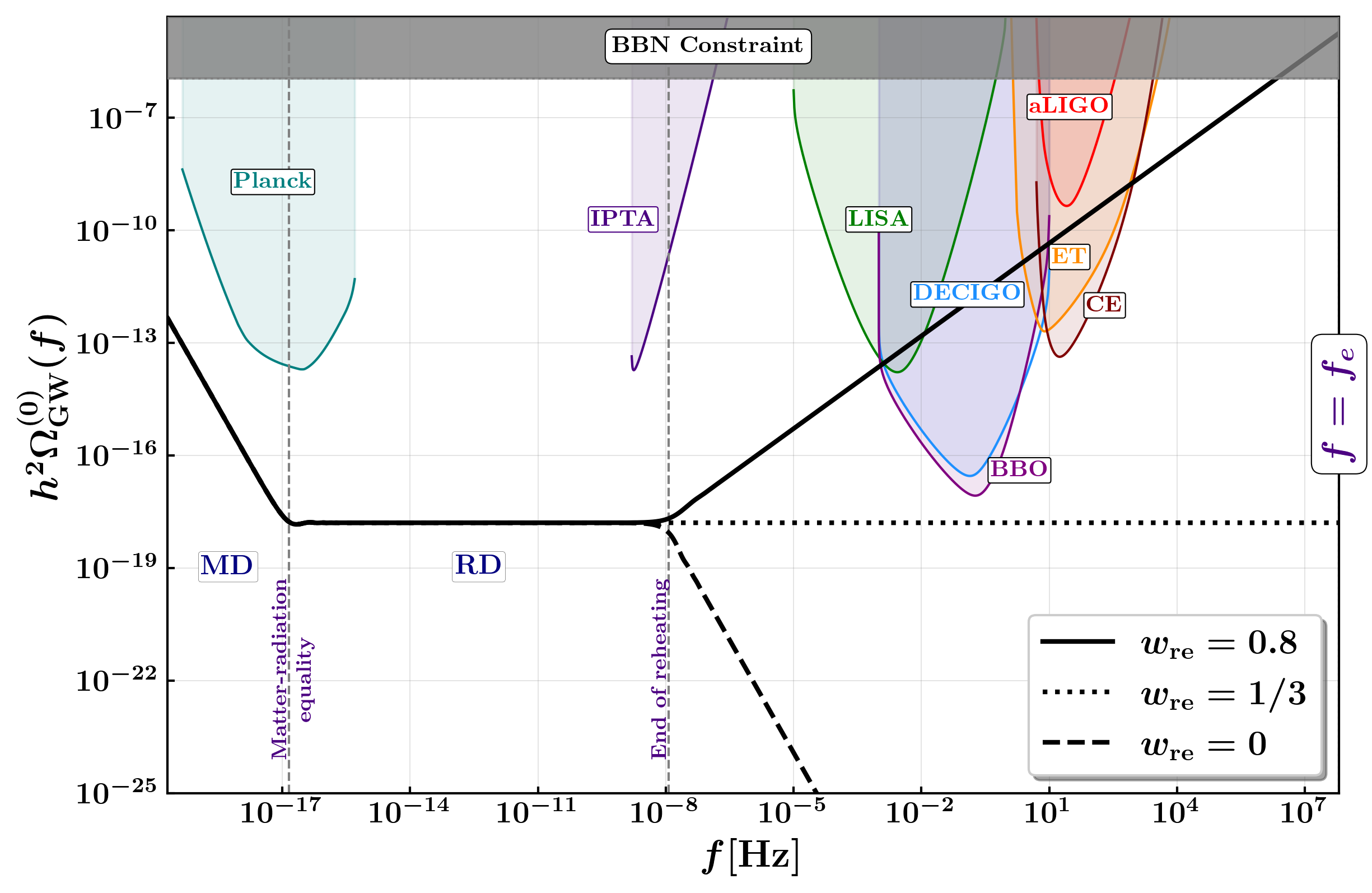}
    \caption{The spectral energy density of first-order inflationary GWs is shown as a function of the present-day GW frequencies for a single constant post-inflationary EoS during reheating; along with the sensitivity curves of various GW detectors~\cite{schmitz_2020_3689582}, as well as the BBN constraint. The spectrum is shown for three different possibilities of reheating EoS above $f \gtrsim 10^{-8}\,{\rm Hz}$, corresponding to a blue-tilted stiff matter  phase with $w_{_{\rm re}} = 0.8$ (solid black), radiative phase with $w_{_{\rm re}} = 1/3$ (dotted black), and softer matter-like phase with $w_{_{\rm re}} = 0$ (dashed black).  The  tensor-to-scalar ratio has been set to  $r = 0.001$ and the  reheating energy scale is fixed at  $E_{\rm r*} = 1$ GeV. Note, the spectrum depicted in solid line ($w_{_{\rm re}}= 0.8$) violates the BBN constraint, in line with Ref.~\cite{Figueroa:2019paj}. The maximum frequency considered here (the rightmost point in the plot) is the UV cut-off frequency, $f_e = 6.37 \t 10^7$ Hz, which corresponds to $k_e = 4.05 \t 10^{22} \, {\rm Mpc}^{-1}$ for $r = 0.001$.}
    \label{fig:EDS single}
\end{figure}

Fig.~\ref{fig:EDS single} shows the GW spectral energy density as a function of the present-day frequency of GWs for the case of a single constant EoS during reheating.  During the MD epoch, the energy density of relic GWs redshifts faster than the background density,  implying that $\Og(f)$ exhibits a red-tilt, with slope $n_{_{\rm GW}} \simeq -2$ as determined from Eq.~(\ref{eq; spectral tilt}). 
Similarly, $\Og$ is (nearly) flat for the range of frequencies corresponding to modes that made their Hubble-entry in the RD epoch, since the GWs redshift at an equal rate in comparison to the background radiation. Finally, the spectrum of those modes that made their Hubble-entry during reheating can be either  blue-tilted, or flat or red-tilted depending upon the EoS during reheating. For example, we see a red-tilt for a softer EoS  $w_{_{\rm re}} < 1/3$ (black dashed curve), and a blue-tilt for a stiff-matter phase  $w_{_{\rm re}} > 1/3$ (solid black curve); while for radiation like EoS $w_{_{\rm re}} = 1/3$ (black dotted curve), the spectrum is flat, as expected. Therefore, we observe that if the post-inflationary reheating EoS is stiff-matter like, then the relic GWs have a greater likelihood of getting  detected by the upcoming GW detectors~\cite{Sahni:2001qp,Figueroa:2019paj, Mishra:2021wkm,Chen:2024roo}. 

Keeping in mind that GW energy density contributes to the total energy density of the universe, the observations on the light nuclei abundance impose an upper bound on the GW energy density during BBN to be~\cite{Caprini:2018mtu} 
\begin{equation}\label{eq; BBN upper bound for GWs}
    \rho_{_{\rm GW}} (T_{\rm BBN}) \leq \Delta \rho_{_{\rm rad}} (T_{\rm BBN}) = \frac{\pi^2}{30} \t \frac{7}{8} \, (2 \, \Delta N_{\nu}) \,  T_{\rm BBN}^4 = \frac{7}{8} \, \rho_{\gamma} (T_{\rm BBN}) \, \Delta N_{\nu} \, , 
\end{equation}
where $\Delta N_{\nu}$ is the extra neutrino (relativistic) degrees of freedom outside the Standard Model from the Standard Model value $N_{\nu} = 3$, and $\rho_{\gamma}$ is the energy density of photons.  Using entropy conversation from Eq.~\eqref{eq; entropy conservation}, we obtain the expression for the total energy density of primordial GWs that are sub-Hubble before the BBN to be
\begin{equation}\label{eq; rho_GW by ratio}
    \frac{\rho_{_{\rm GW}}(T_{\rm BBN})}{\rho_{_{\rm GW}, \, 0}} = \l(\frac{a_0}{a_{\rm BBN}} \r)^4 = \l(\frac{g_{s, \, T_{\rm BBN}}}{g_{s, \, 0}} \r)^{4/3} \l(\frac{T_{\rm BBN}}{T_0} \r)^4 \, .
\end{equation}
\begin{figure}
    \centering
    \subfigure[]{\includegraphics[width=0.47\textwidth]{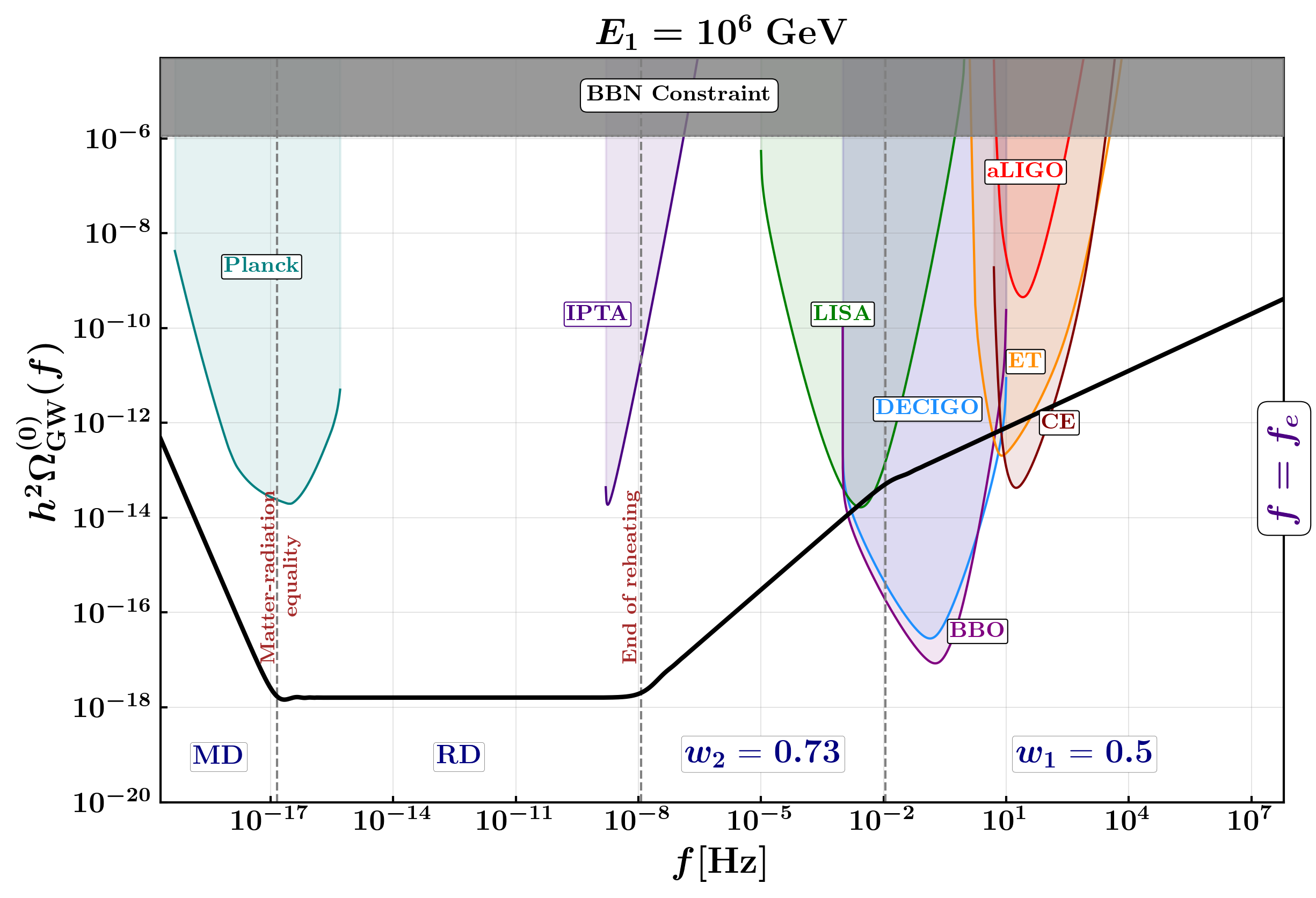}}
    \subfigure[]{\includegraphics[width=0.47\textwidth]{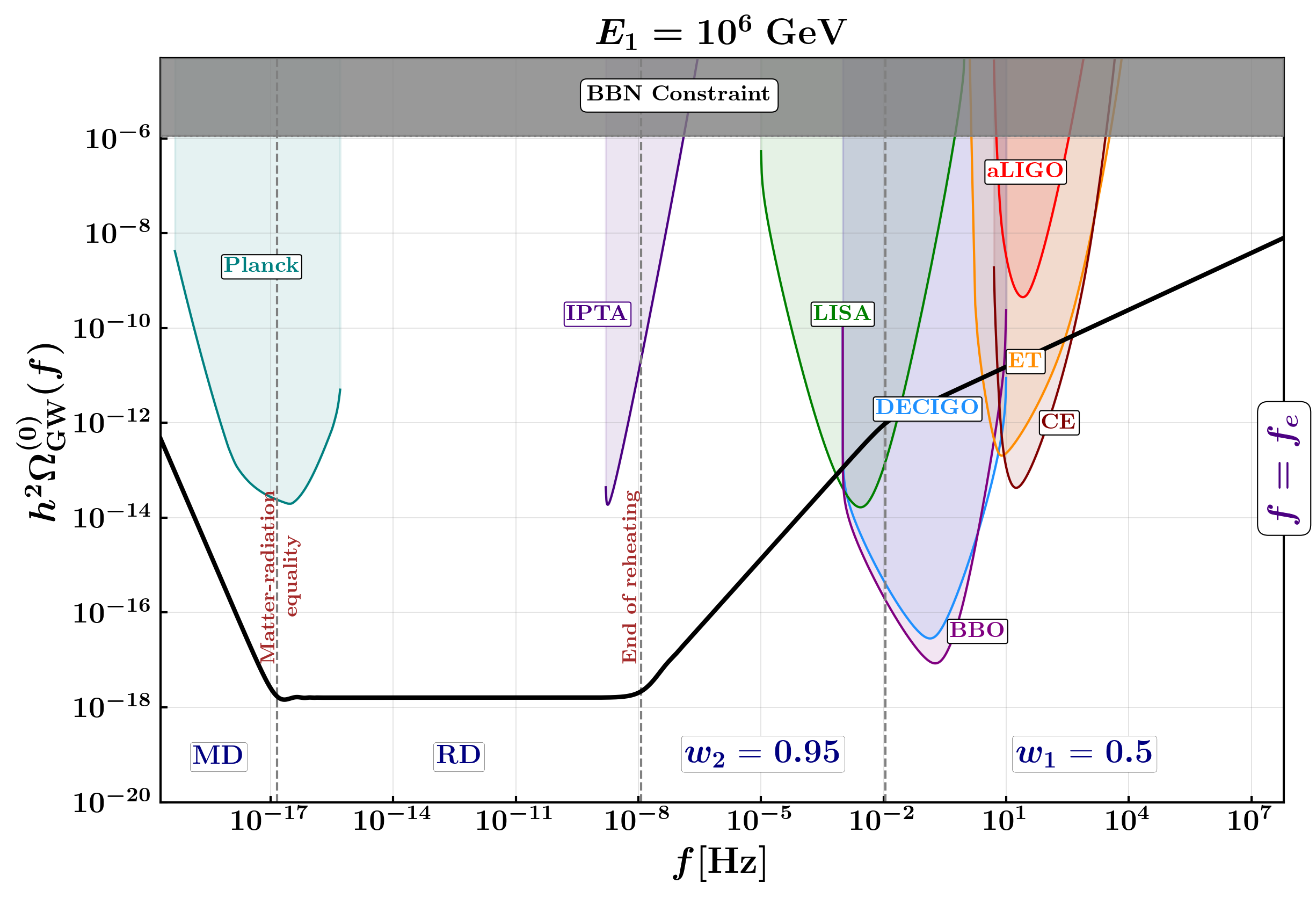}}
    \subfigure[]{\includegraphics[width=0.47\textwidth]{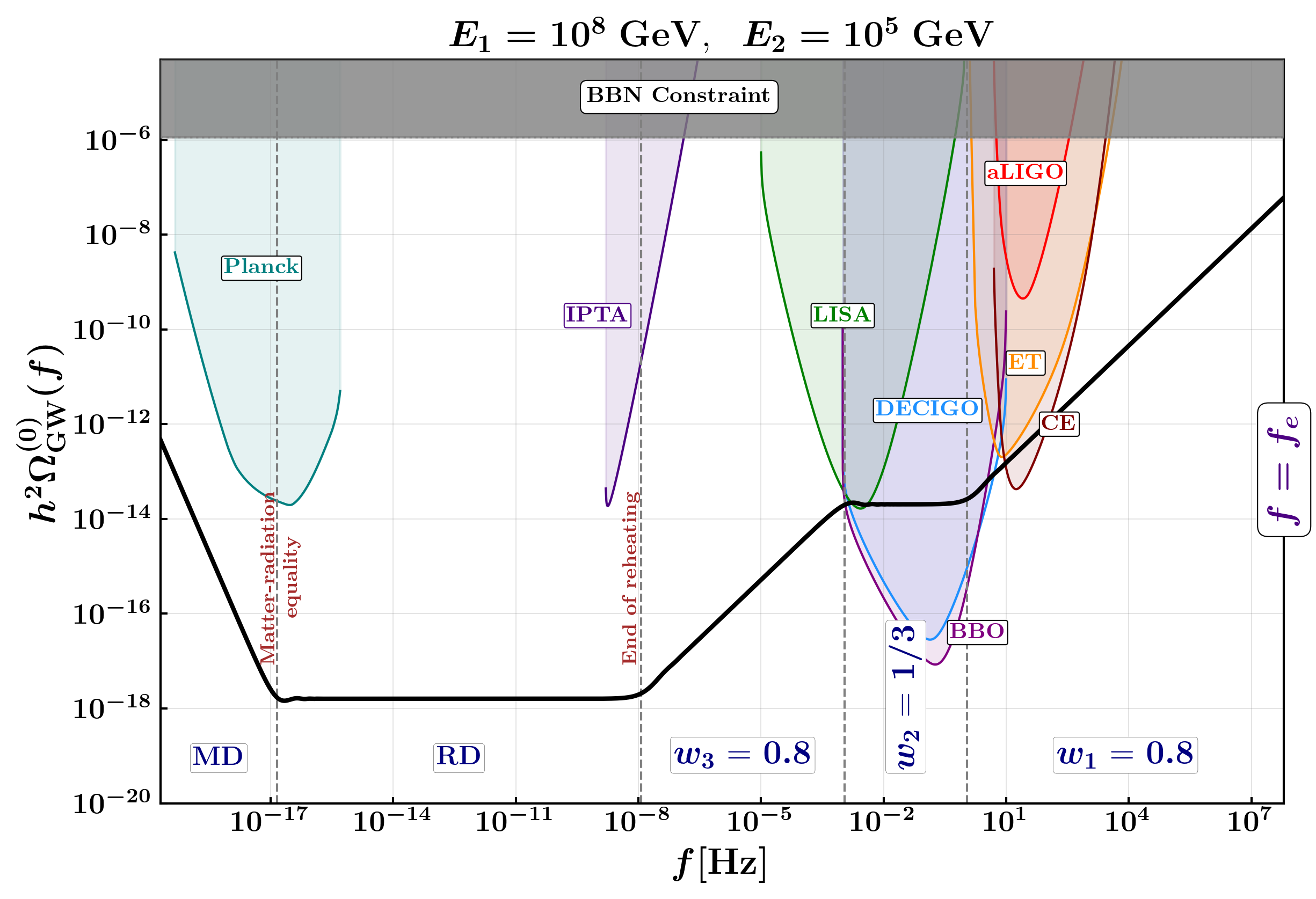}}
    \subfigure[]{\includegraphics[width=0.47\textwidth]{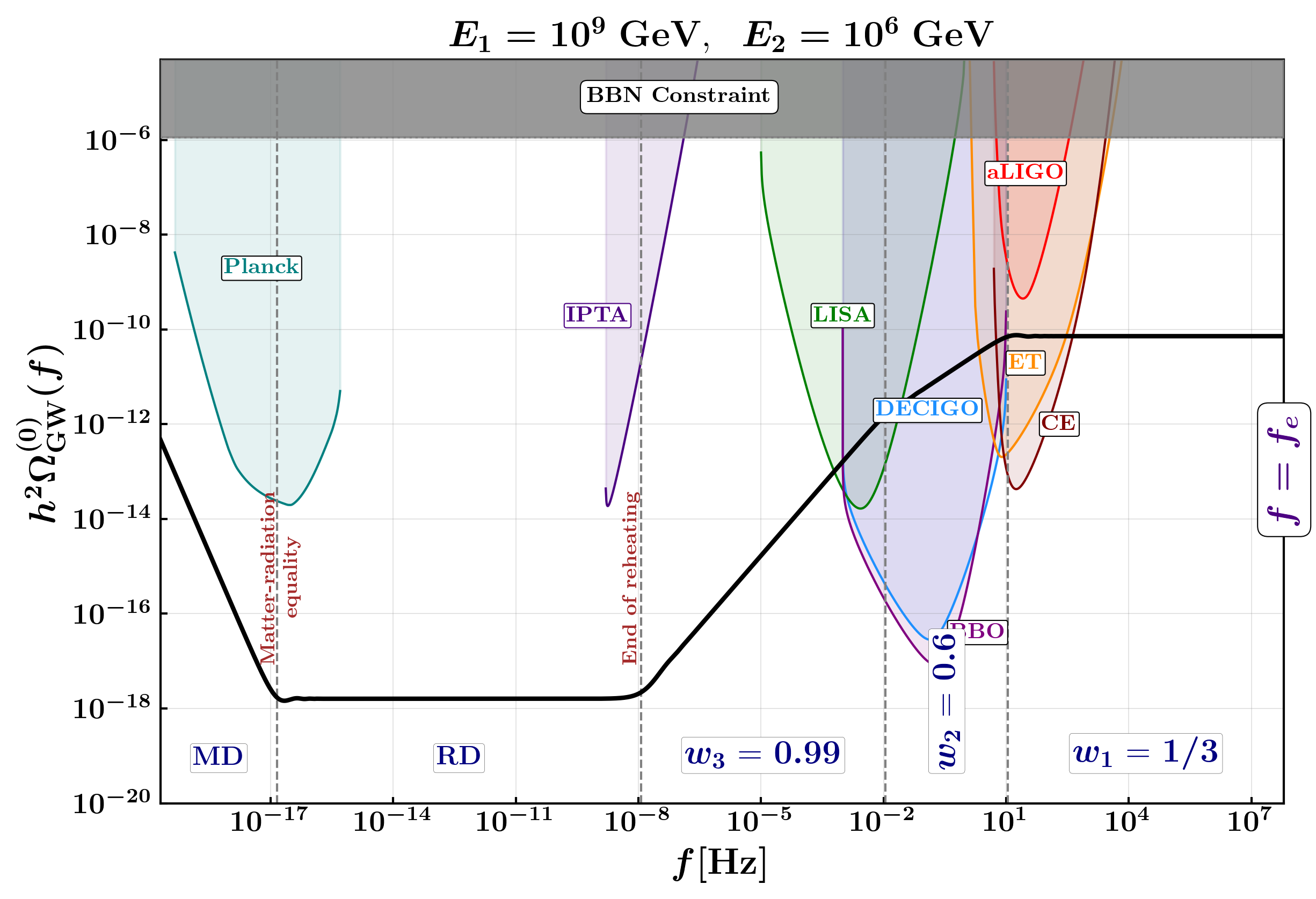}}
    \subfigure[]{\includegraphics[width=0.47\textwidth]{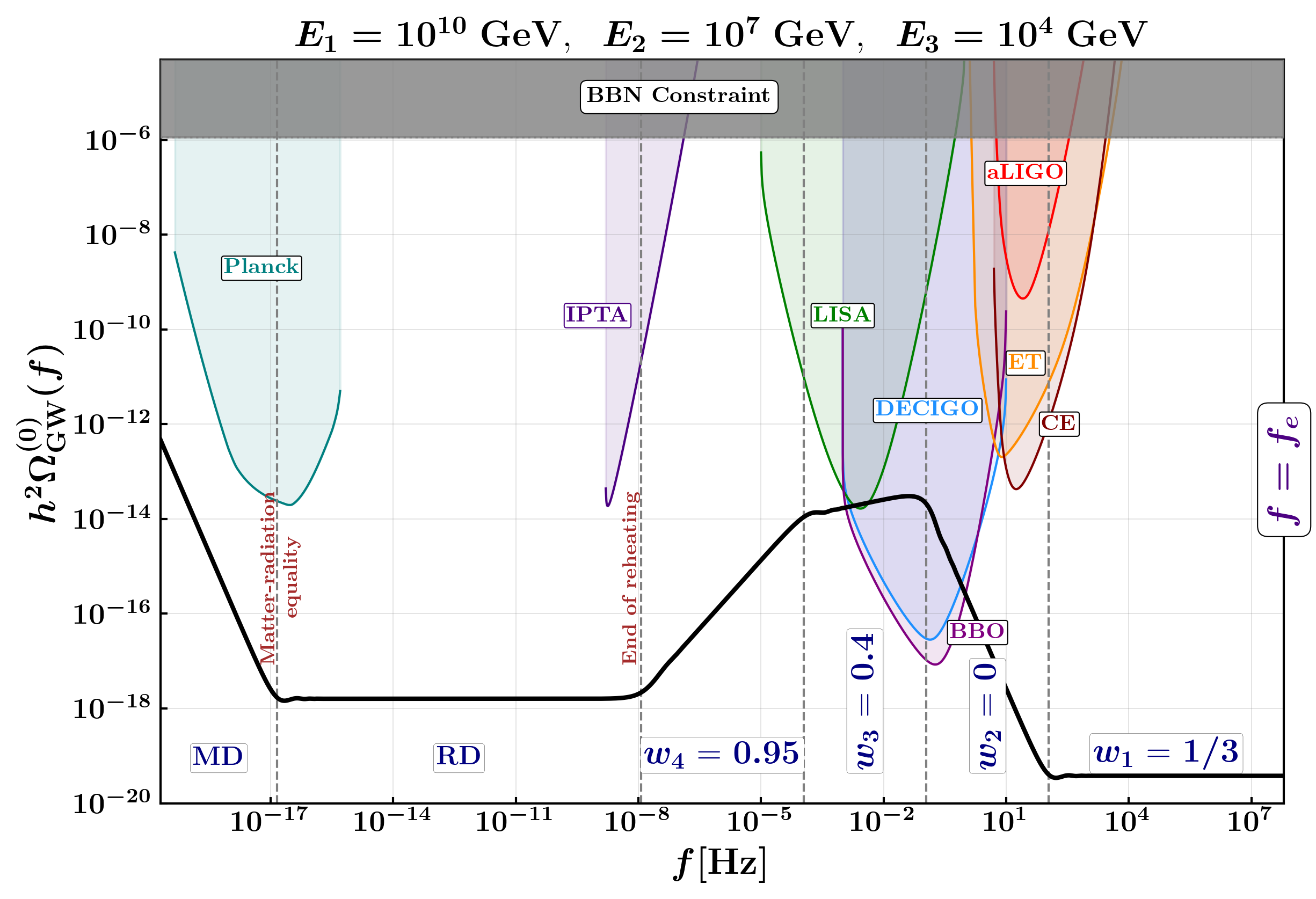}}
    \subfigure[]{\includegraphics[width=0.47\textwidth]{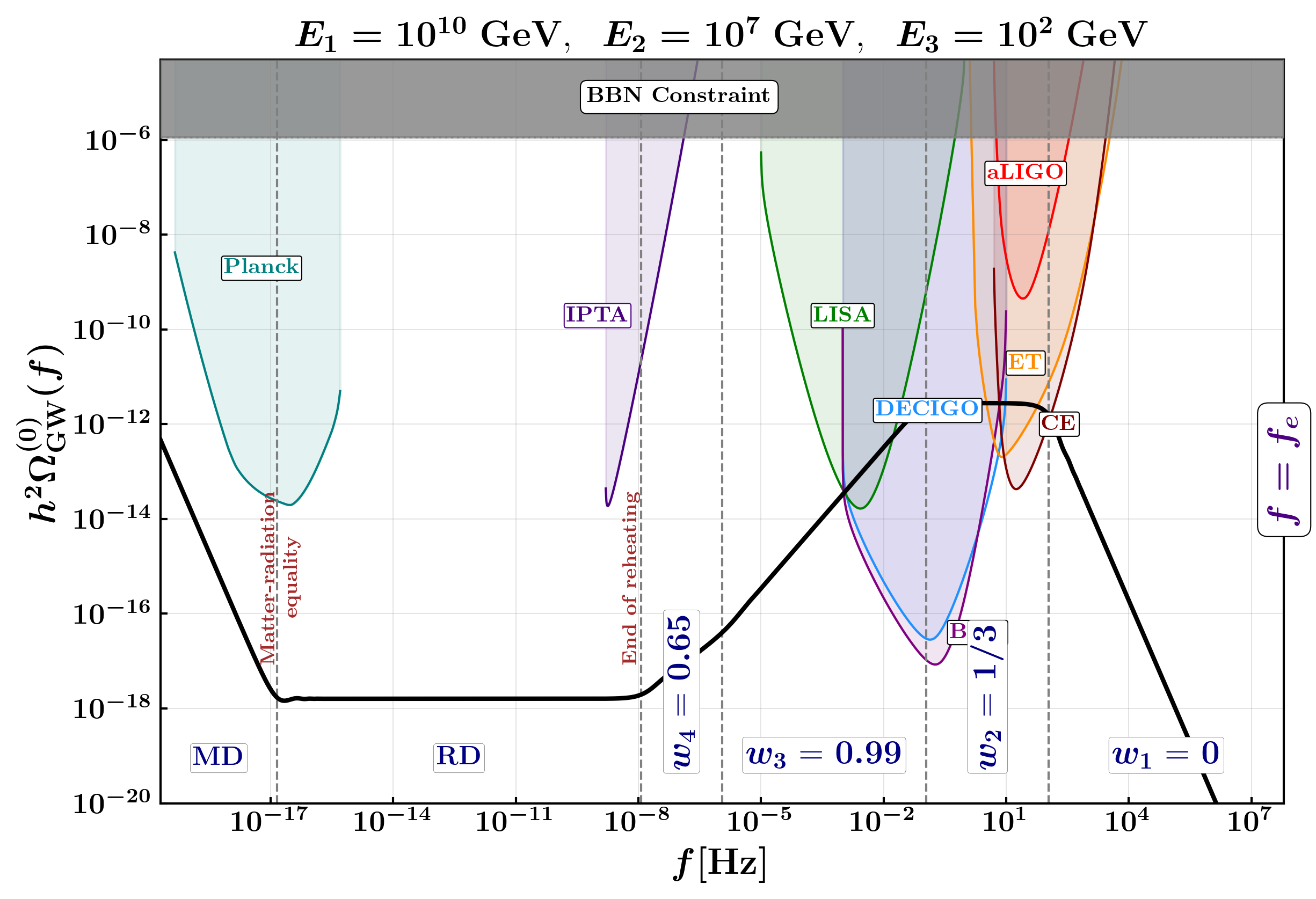}}
    \caption{The spectral energy density ($\Og$) of inflationary GWs associated with two (\textbf{top row}), three (\textbf{middle row}) and four (\textbf{bottom row}) epochs of piece-wise constant EoS parameters during reheating. Plots in the \textbf{left column} correspond to scenarios where $\Og$ has a minimal overlap with the LISA sensitivity curve, while those in the \textbf{right column} belong to  scenarios where $\Og$ has a substantial overlap with LISA sensitivity region, without violating the existing BBN and aLIGO constraints. The energy scales $E_i$ (with $i \in \mathbb{Z}^+$) corresponding to the end of each epoch, are given in the top labels of each plot. The energy scales at the end of reheating and inflation are the same in all plots, namely, $E_{\rm r*} = 1$ GeV and $E_{\rm inf} = 5.76 \t 10^{15}$ GeV respectively, so as the tensor-to-scalar ratio, $r = 0.001$. The maximum frequency considered here (the rightmost point in the plot) is the UV cut-off frequency, $f_e = 6.37 \t 10^7$ Hz, which corresponds to $k_e = 4.05 \t 10^{22} \, {\rm Mpc}^{-1}$ for $r = 0.001$.}
    \label{fig: EDS two, three, four epoch}
\end{figure}

We can further modify Eq.~\eqref{eq; rho_GW by ratio} using the relation involving energy density of photons, $\rho_{\gamma}(T) = (\pi^2 / 15) \,  T^4$,
\begin{equation}\label{eq; BBN bound for present time}
    \frac{\rho_{_{\rm GW}, \, 0}}{\rho_{c, \, 0}} = \frac{\rho_{_{\rm GW}} (T_{\rm BBN})}{\rho_{\gamma} (T_{\rm BBN})} \, \l(\frac{g_{s, \, 0}}{g_{s, \, T_{\rm BBN}}} \r)^{4/3} \, \Omega_{\gamma,\,  0}\, .
\end{equation}
Using Eqs.~\eqref{eqn; defn of Omega_GW}
and \eqref{eq; BBN upper bound for GWs}, we can rewrite Eq.~\eqref{eq; BBN bound for present time} as
\begin{align}\label{eq; BBN constraint integral}
    h^2 \int_{f_{\rm BBN}}^{f_{e}} \d \ln f \;   \Og(\tau_0, \, f)  & \leq \frac{7}{8} \, \Delta N_{\nu} \, \l(\frac{g_{s, \, 0}}{g_{s, \, T_{\rm BBN}}} \r)^{4/3} \, h^2 \, \Omega_{\gamma,\,  0} \nonumber\\
    & = 5.64 \t 10^{-6} \, \Delta N_{\nu} < 1.13 \t 10^{-6} \, ,
\end{align}
where we have used the values $h^2 \, \Omega_{\gamma,\,  0} = 2.46 \t 10^{-5}$ (see Eq.~\eqref{eq; density param of CMB photons})$, \, g_{s, \, 0} = 3.94, \, g_{s, \, T_{\rm BBN}} = 10.75$. The upper bound, $ N_{\nu} < 3.2$, is obtained with a $95\%$ C.L. from combining the deuterium-to-hydrogen ratio along with CMB baryon density~\cite{RevModPhys.88.015004}. Eq.~\eqref{eq; BBN constraint integral} is integrated from the frequencies of those modes that have entered the comoving Hubble radius at the onset of BBN $(f_{\rm BBN})$ to those at the end of inflation $(f_e)$. 
 However, computing the integral each time for a case of different multiple epochs is computationally expensive.  Therefore, one can incorporate an analytical approximation as follows. For tensor modes making their Hubble-entry during a given epoch with $\alpha_i = 2/(1+3w_i)$, the spectral energy density at the present epoch obeys $\Ogo(f) \equiv   \Og(\tau_0, f) \propto f^{2\l(1-\alpha_i\r)}$, as can be inferred from Eq.~(\ref{eq; spectral tilt defn}). Therefore, the spectral energy density corresponding to $n$ number of multiple piece-wise constant EoS parameters prior to the BBN can be represented in terms of a combination of the Heaviside Theta function, as 
\ber
\Ogo(f) = \Omega_{\rm GW}^{0, \rm RD} &+& \l[  \Omega_{\rm GW}^{0,n}(f) - \Omega_{\rm GW}^{0, \rm RD}\r]\,\Theta\l( f-f_{{\rm r}*}\r) +  \l[ \Omega_{\rm GW}^{0,n-1}(f) - \Omega_{\rm GW}^{0,n}(f)\r]\,\Theta\l( f-f_{n-1}\r)  \nonumber \\
&+& \,.\,.\,.\,.\, + \l[ \Omega_{\rm GW}^{0,1}(f) - \Omega_{\rm GW}^{0,2}(f) \r]\,\Theta\l( f-f_1\r) \, ,
\label{eq:Omega_GW_f_multiple_analyt}
\eer
 where
 \beq
\Omega_{\rm GW}^{0,i}(f) = \Omega_{\rm GW}^{0,i+1}(f_i) \l(\f{f}{f_i}\r)^{2\l(1-\alpha_i\r)} = ~\Omega_{\rm GW}^{0,i+1}(f_i) \l(\f{f}{f_i}\r)^{2\l(\f{w_i-1/3}{w_i+1/3}\r)} \, ,
\label{eq:O_GW_i_analyt}
 \eeq
 and 
 $ \Omega_{\rm GW}^{0, \rm RD}$ is a constant with its value given by~\cite{Figueroa:2019paj, Mishra:2021wkm,}
 \begin{equation}
     \Omega_{\rm GW}^{0, \rm RD} \simeq \frac{1}{24} \, \Omega_{\rm rad,\, 0} \times A_{_S} \times r = 3.64 \t 10^{-18} \, \l( \frac{r}{0.001} \r) h^{-2}
     \label{eq; value of Omega_GW^rad} \, .
 \end{equation}
 Also note that $f_i$ is the present-day frequency corresponding to the transition from $i^{\rm th}$ to $(i+1)^{\rm th}$ epoch and $f_{n} = f_{\rm r*}$. Eq.~\eqref{eq:O_GW_i_analyt} can also be written as
 \begin{equation}
     \Omega_{\rm GW}^{0, \, i} = \Omega_{\rm GW}^{0, \rm RD} \t \l[ \prod_{m = i}^{n} \l( \frac{f_m}{f_{m+1}}\r)^{2(1-\alpha_{m+1})} \r] \l( \frac{f}{f_i} \r)^{2(1-\alpha_i)} \, . \label{eq; Omega_GW_i_analyt_as_Omega_RD}
 \end{equation}
 Therefore, for the case of multiple EoS parameters before radiation domination, one can analytically integrate  the BBN constraint Eq.~\eqref{eq; BBN constraint integral}, using Eq.~(\ref{eq; Omega_GW_i_analyt_as_Omega_RD}), to obtain 
% \begin{align*}
%     h^2 \Omega_{\rm GW}^{0, \, \rm RD} \l[ \ln \l( \frac{f_{\rm r*}}{f_{\rm BBN}} \r) 
%     & + \frac{1}{2(1-\alpha_n)} \l\{ \l(\frac{f_{n-1}}{f_n} \r)^{2(1-\alpha_n)} -1 \r\} \\
%     + \frac{(f_{n-1}/f_n)^{2(1-\alpha_n)}}{2(1-\alpha_{n-1})} \l\{ \l(\frac{f_{n-2}}{f_{n-1}} \r)^{2(1-\alpha_{n-1})} -1 \r\} \r]
%     < 1.13 \t 10^{-6} \,, \label{eq; BBN bound general approx} \num
% \end{align*}

\begin{align}
    & h^2 \Omega_{\rm GW}^{0, \, \rm RD} \l[ 
        \ln \l( \frac{f_{\rm r*}}{f_{\rm BBN}} \r) 
        +  \frac{{\cal F}_n}{2(1-\alpha_n)}
          \l\{ \l(\frac{f_{n-1}}{f_n} \r)^{2(1-\alpha_n)} -1 \r\} 
        + \frac{{\cal F}_{n-1}}{2(1-\alpha_{n-1})}
          \l\{ \l(\frac{f_{n-2}}{f_{n-1}} \r)^{2(1-\alpha_{n-1})} -1 \r\} \right. \notag \\
    &\quad \left.
     \hspace{4cm} + ..... +  \frac{{\cal F}_1}{2(1-\alpha_1)}
          \l\{ \l(\frac{f_{e}}{f_{1}} \r)^{2(1-\alpha_{1})} -1 \r\} 
    \r] < 1.13 \t 10^{-6} \,, \label{eq; BBN bound general approx} \num
\end{align}
where ${\cal F}_i$'s are defined as
\begin{equation}
    {\cal F}_i = \prod_{m = i}^n \l( \frac{f_m}{f_{m+1}} \r)^{2(1-\alpha_{m+1})} \label{eq; coeff F_i} \,, i \in \{1, 2, ... , n \}.
\end{equation}
Since we consider $n$ number of EoS parameters before radiation domination, in order to compute the value of ${\cal F}_i$ using Eq.~\eqref{eq; coeff F_i}, one needs to identify $f_{n+1} = f_{\rm eq}$ and $\alpha_{n+1} = 1$, since the $(n+1)^{\rm th}$ epoch is the radiation-dominated epoch of the hot Big Bang phase. Therefore, ${\cal F}_n =1$. Furthermore, one should note that whenever $w_i = 1/3$ (prior to the final radiation domination), the contribution to Eq.~\eqref{eq; BBN bound general approx} from that epoch will contain an additional logarithm term, apart from the already existing logarithm term in Eq.~(\ref{eq; BBN bound general approx}).

Fig.~\ref{fig: EDS two, three, four epoch}, displays the GW spectral energy density for the cases of reheating with two, three, and four epochs of piece-wise constant EoS parameters; with each epoch having a different EoS and duration. The figure helps to highlight post-inflationary scenarios where the predicted signals for the spectral energy density of GWs  have, approximately, either a minimal overlap  (left column) or maximal overlap (right column) with the LISA sensitivity curve,  without violating the existing aLIGO and  BBN constraints. Hence, it is important to analyse the possible combinations of EoS parameters and duration of these epochs resulting in signals that are detectable in the near future, not only by LISA, but also by other upcoming GW detectors.  Note that the energy scale of inflation corresponding to a tensor-to-scalar ratio $r = 0.001$  is  $E_{\rm inf} = 5.76 \t 10^{15} \, {\rm GeV}$, as can be inferred from Eq.~(\ref{eq:E_r_SR_rel}) in App.~\ref{sec:inf_dyn_obs}. Therefore, the value of the UV cut-off frequency used in the spectral energy density plots of this paper can be obtained, by using the relation between $f_k$ and $E_k$ in Eq.~\eqref{eq; f_k and E_k}, to be $f_e = 6.37 \t 10^7 \, {\rm Hz} $.

The case of instantaneously sharp transitions, apart from being computationally simple,  gives a reasonable insight into the  detectable post-inflationary properties  of the universe. However, we must stress that realistic sharp cosmological transitions  are not expected to be instantaneous, but rather smooth~\cite{Haque:2021dha}. Hence, our  $\Og$ curves are not fully accurate for frequencies right around the transitions. 

\section{Inflationary gravitational waves as a probe of the primordial equation of state}
\label{sec:probing_primodial_EOS}
\begin{figure}
    \centering
    \includegraphics[width=0.75\linewidth]{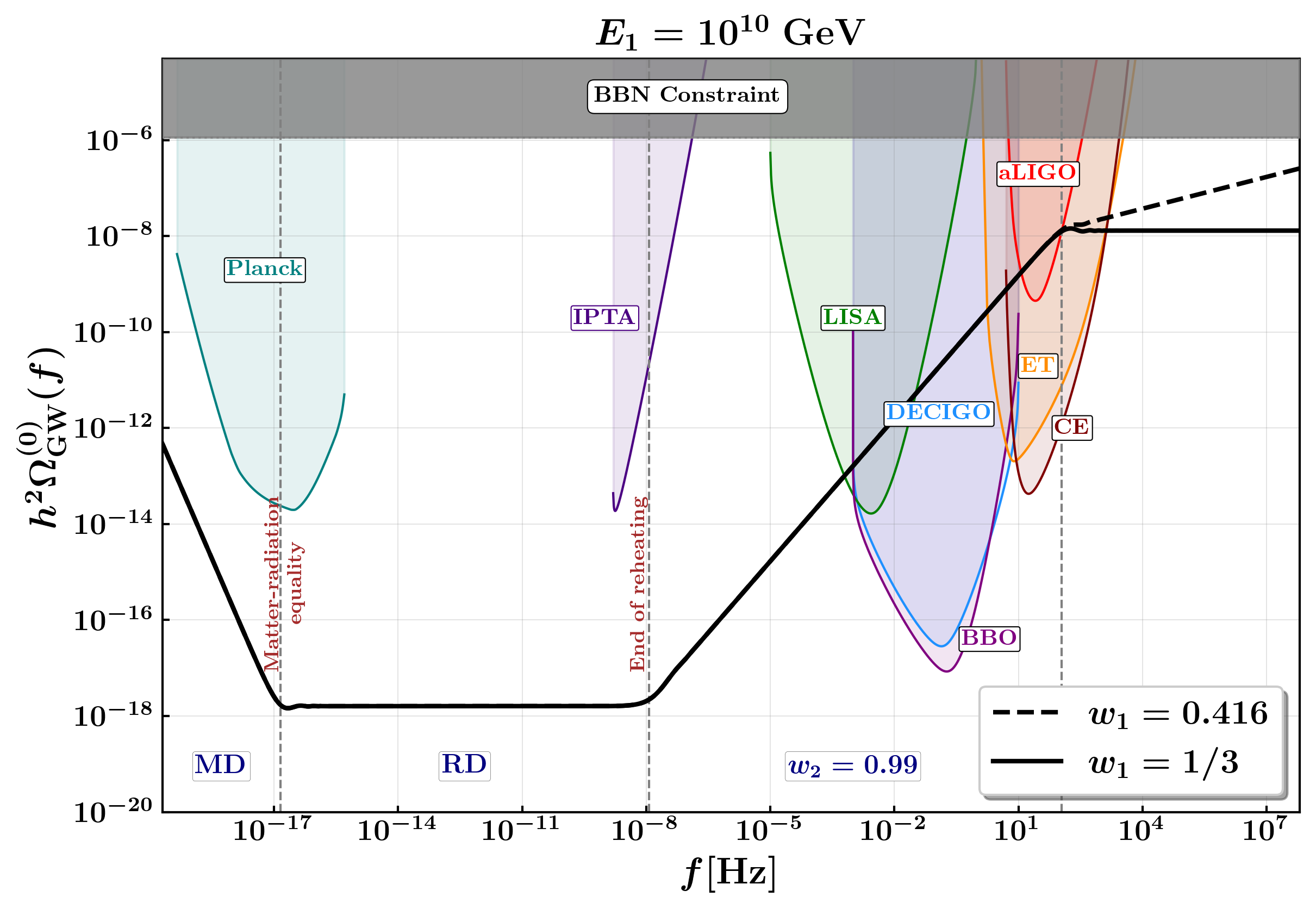}
    \caption{Spectral energy density of inflationary GWs at the present epoch, for the case of a two-epoch reheating scenario, with $r=0.001$, $E_1 =  10^{10}\,{\rm GeV}$ and $E_{{\rm r}*} = 1\,{\rm GeV}$. The solid line corresponds to $w_1 = 1/3$ which easily satisfies the BBN constraint~(\ref{eq; BBN constraint integral}). Therefore,  $w_1 \leq 1/3$, the BBN constraint is guaranteed to be satisfied. However, we find that the BBN constraint is bound to be violated when $w_1 \geq 0.417$,  (the dashed line is plotted for $w_1 =0.416$. The UV cut-off frequency is $f_e = 6.37 \t 10^7$ Hz which corresponds to $k_e = 4.05 \t 10^{22} \, {\rm Mpc}^{-1}$ for $r = 0.001$.}
    \label{fig:limit for BBN}
\end{figure}

\begin{figure}
  \centering
  \subfigure[]{\includegraphics[width=0.385\textwidth]{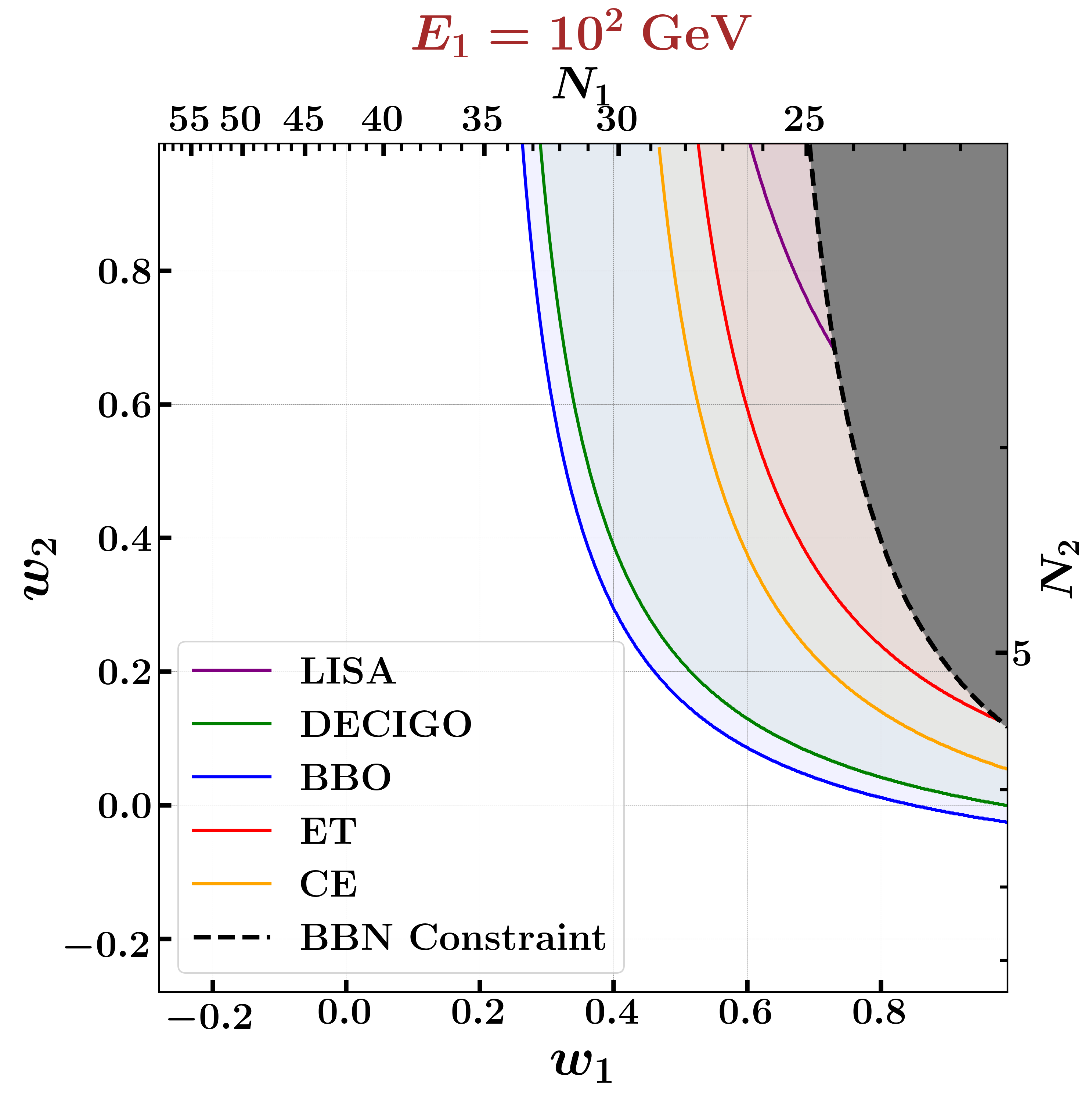}\label{subfig; two epoch 10^2}}
  \subfigure[]{\includegraphics[width=0.385\textwidth]{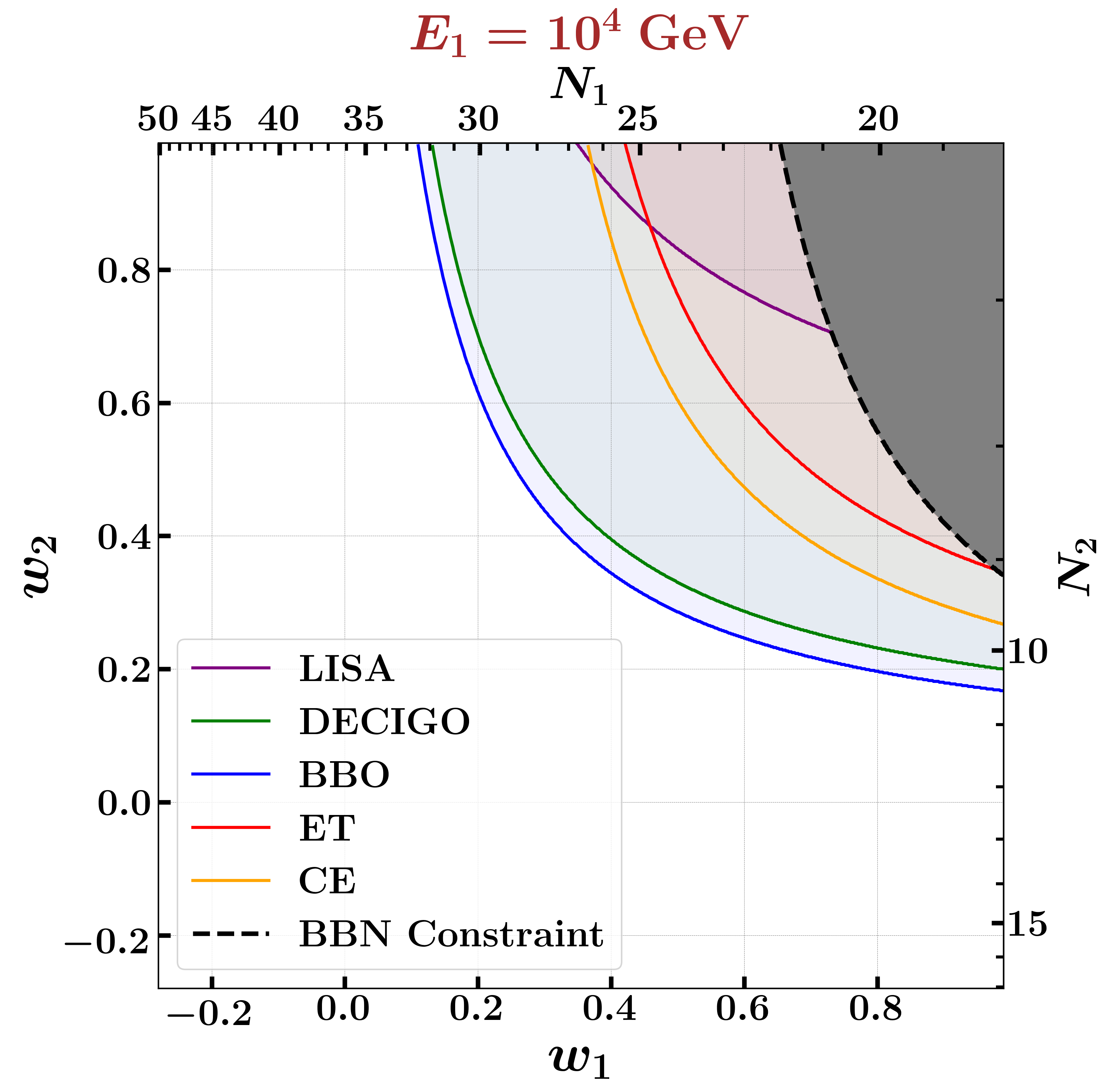}\label{subfig; two epoch 10^4}}
  \subfigure[]{\includegraphics[width=0.385\textwidth]{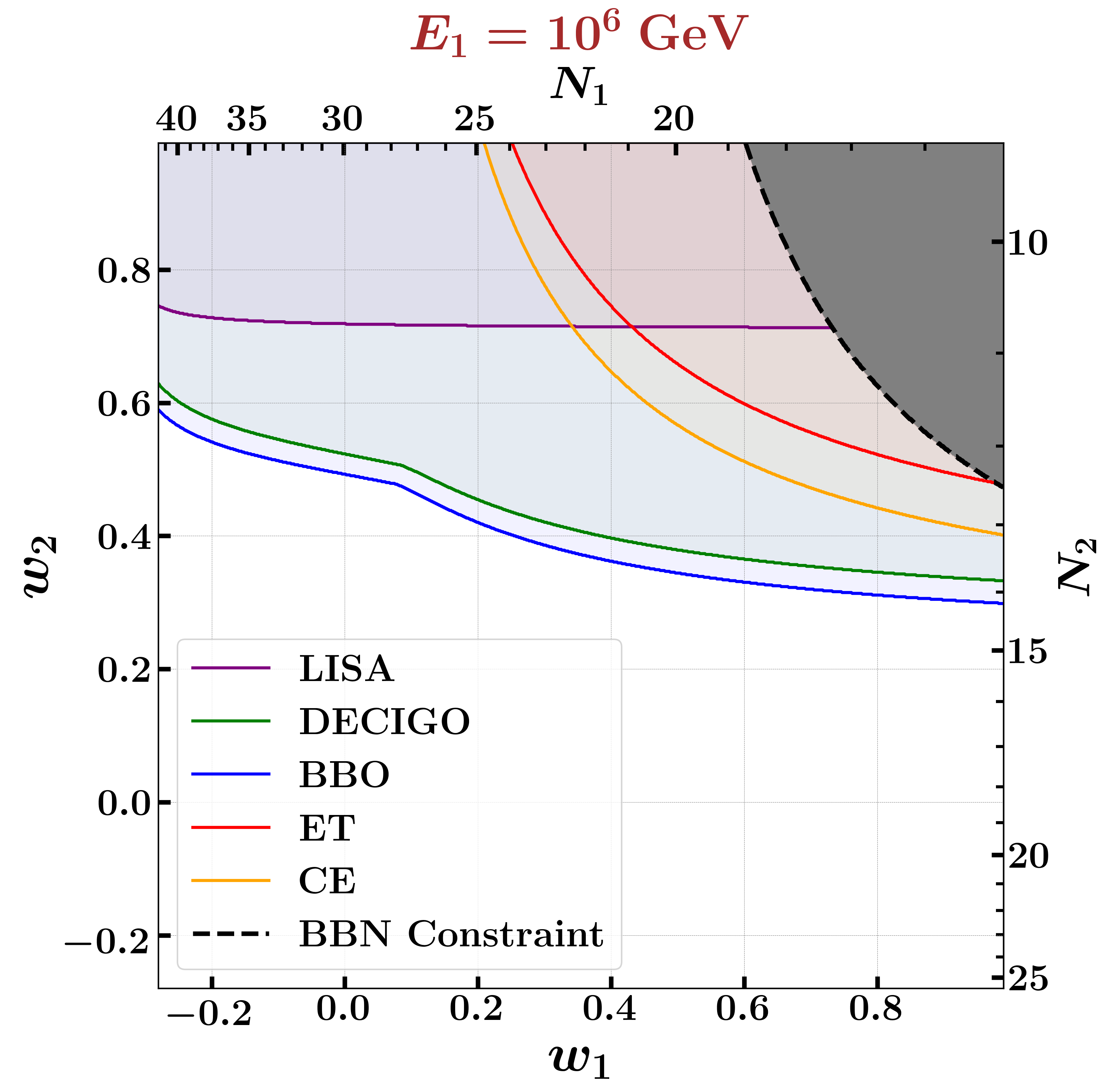}\label{subfig; two epoch 10^6}}
  \subfigure[]{\includegraphics[width=0.385\textwidth]{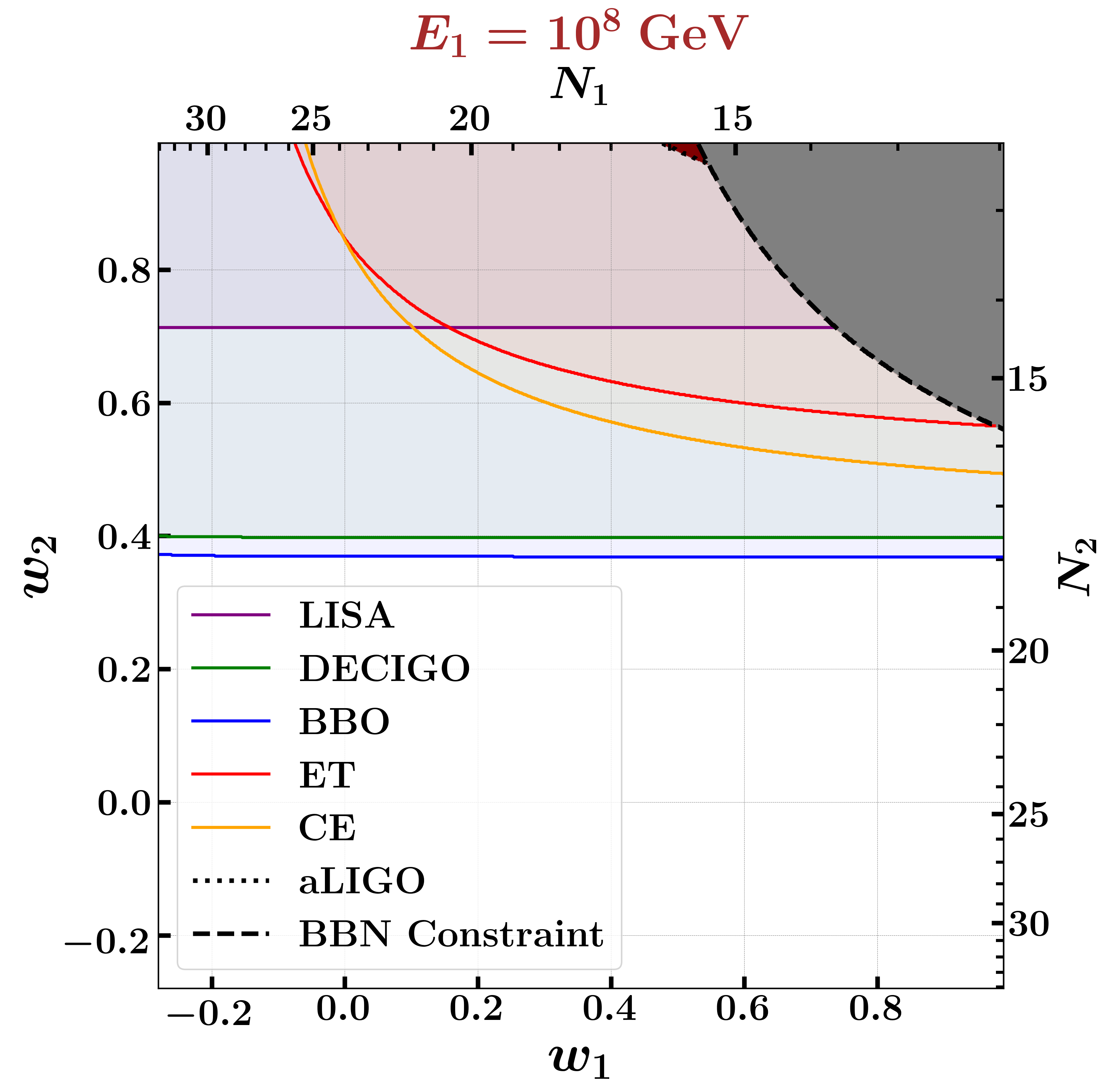}\label{subfig; two epoch 10^8}}
  \subfigure[]{\includegraphics[width=0.385\textwidth]{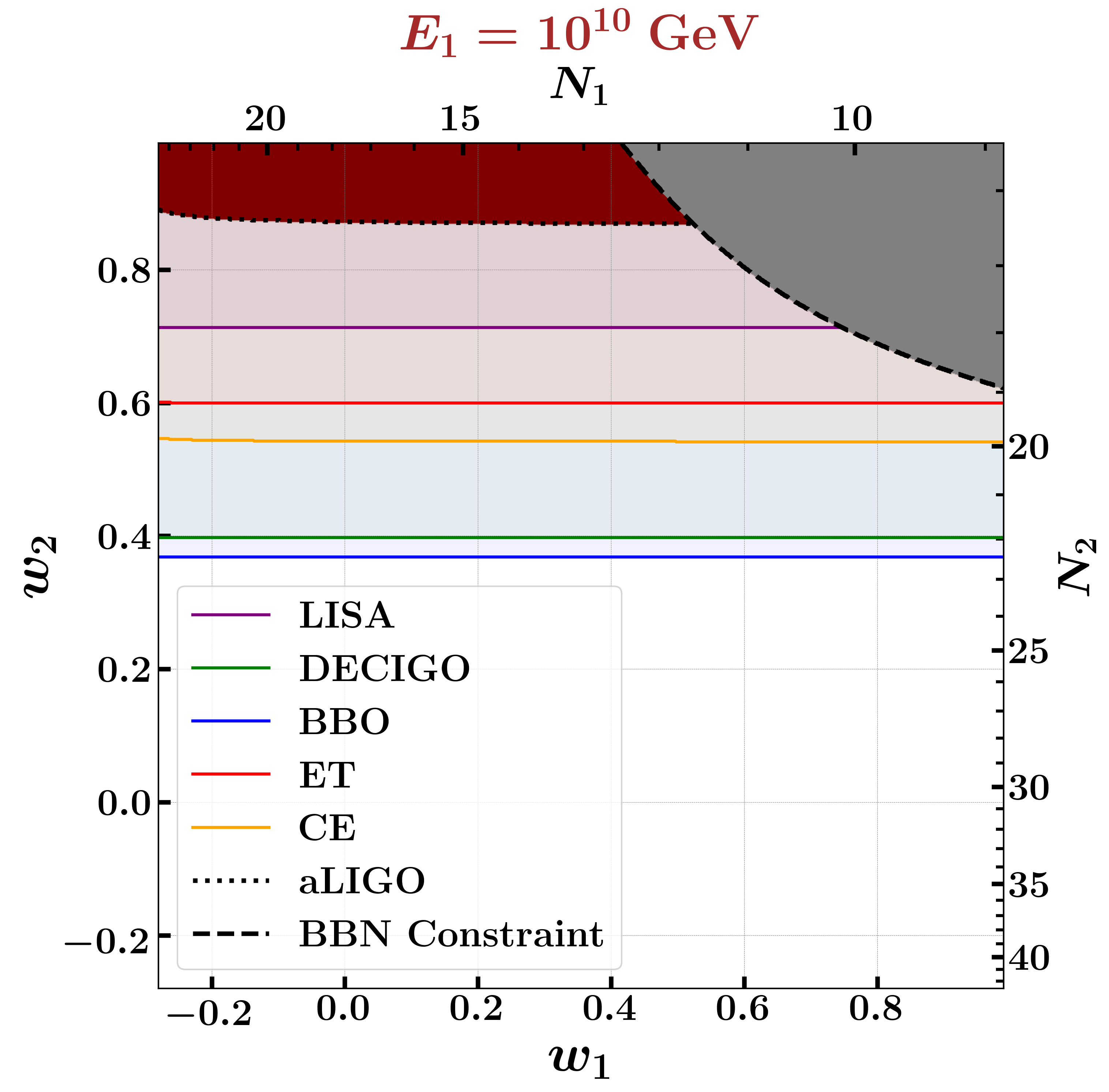}\label{subfig; two epoch 10^10}}
  \caption{The parameter space for two-epoch reheating phase with EoS parameters $w_1$ and $w_2$ that leads to a potentially detectable GW signal in the  LISA, BBO, DECIGO, CE, ET, and aLIGO detectors. The grey-shaded region in each plot corresponds to the combination of $w_1$ and $w_2$ that violate the BBN constraint, while the maroon-shaded region represents those that are ruled out by the aLIGO null detection. Energy scale of the universe at the end of the first post-inflationary epoch is labelled above in each plot. Note that, the duration of each epoch (number of $e$-folds) has been provided opposite to the corresponding EoS axis.}
  \label{fig; two epoch param space}
\end{figure}

\begin{figure}
  \centering
  \subfigure[]{\includegraphics[width=0.99\textwidth]{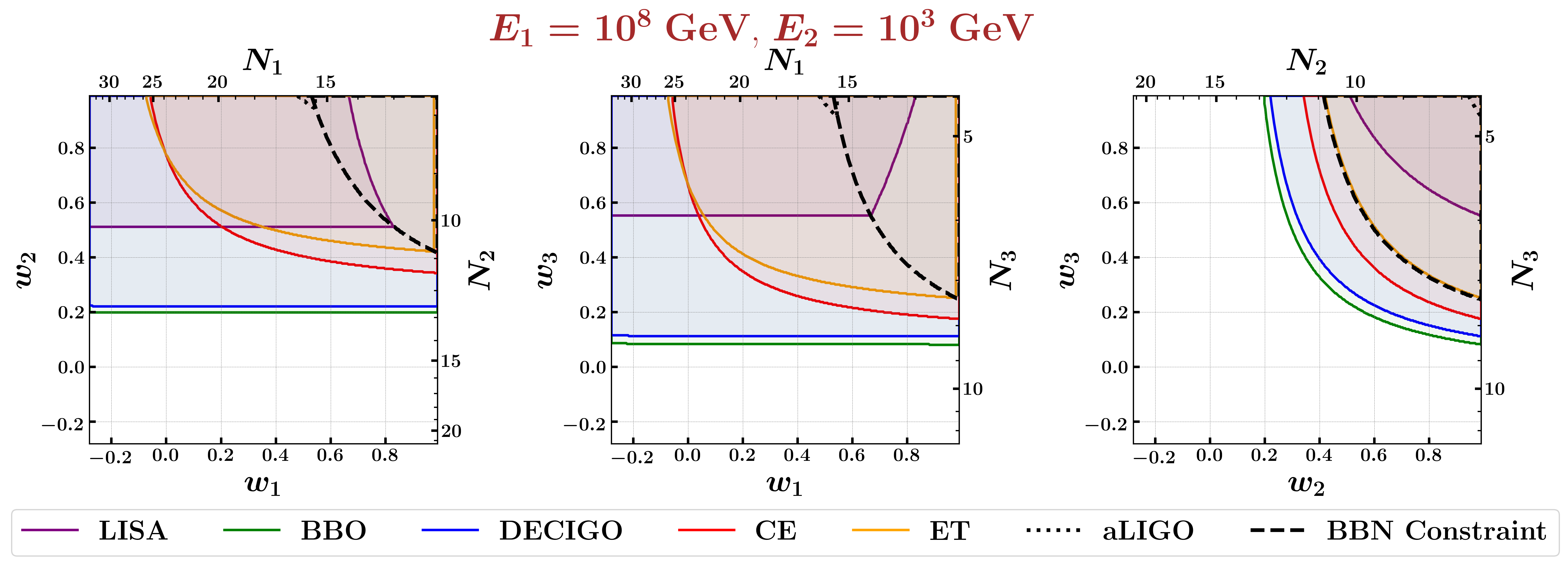}\label{subfig; three_10^8 and 10^3}}
  \subfigure[]{\includegraphics[width=0.99\textwidth]{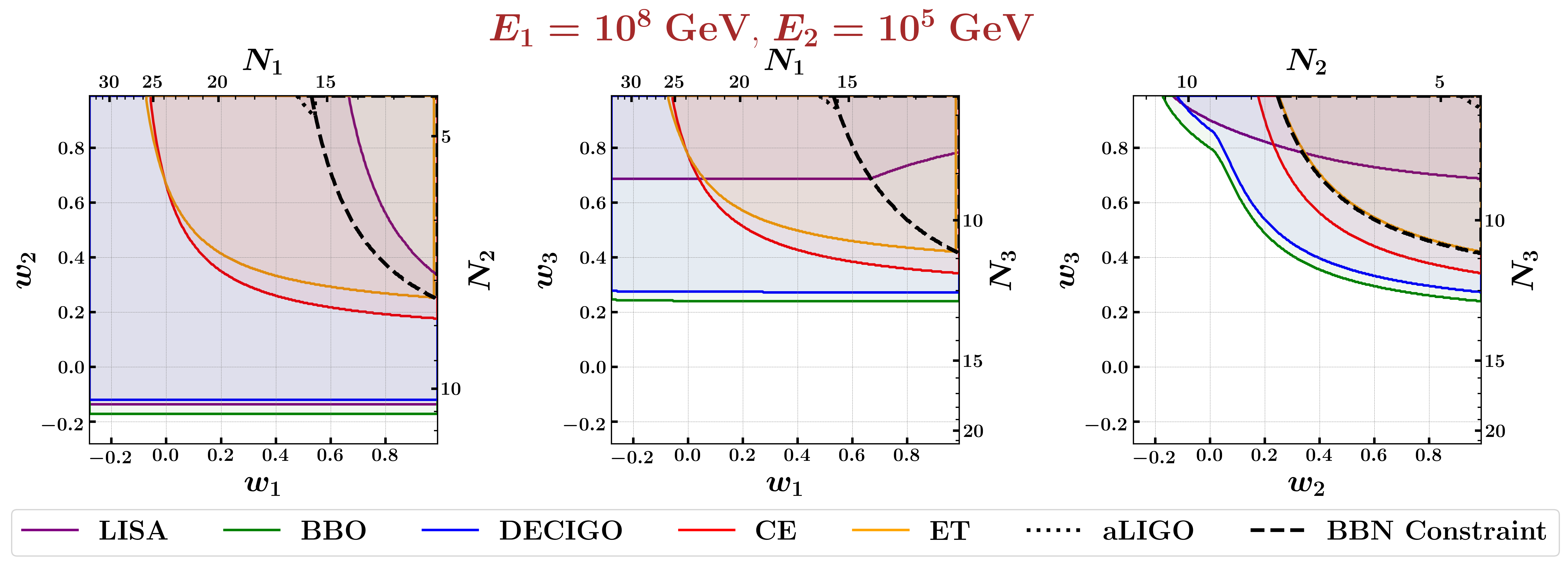}\label{subfig; three_10^8 and 10^5}}
  \subfigure[]{\includegraphics[width=0.99\textwidth]{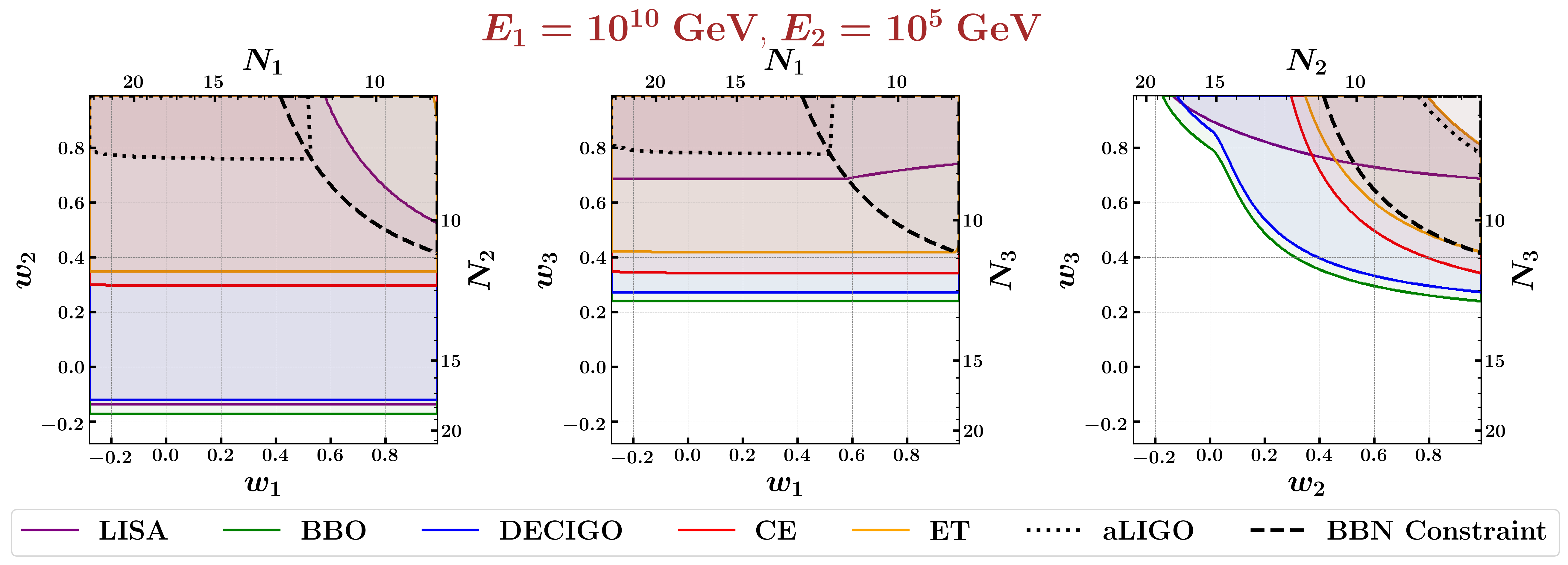}\label{subfig; three_10^10 and 10^5}}
  \caption{The parameter space for three-epoch reheating phase with EoS parameters $w_1, \, w_2$ and $w_3$ that leads to a potentially detectable GW signal in the  LISA, BBO, DECIGO, CE, ET, and aLIGO detectors. The plots are obtained by taking 2-dimensional  (integrated) projections of the full 3-dimensional subspace of  EoS parameters. Energy scales  at which the first and second post-inflationary epochs end are given as $E_1$ and $E_2$. The duration of each epoch (number of $e$-folds) has been provided opposite to the corresponding EoS axis. The projections of the BBN-constrained and aLIGO-constrained regions are enclosed by the dashed and dotted black lines respectively.}
  \label{fig; three epoch param part_one}
\end{figure}

In this section, we determine the subspace of equation of state parameters (and their corresponding duration) during reheating, comprising of two and three epochs of piece-wise constant EoS, which leads to a detectable signal for the inflationary GWs in the upcoming GW detectors\footnote{Important additional constraints come from the bound on CMB spectral distortions~\cite{Kite:2020uix, Cyr:2023pgw} which are more relevant for relatively lower frequency GW signals in between the CMB and PTA sensitivity curves.} such as LISA, CE, ET, BBO, and DECIGO. These GW detectors can be classified into three groups based on their optimal frequency range of operation (see Fig.~\ref{fig:limit for BBN}): $10^{-4}-10^{-1} \, {\rm Hz}$ for LISA, $10^{-3}-10 \, {\rm Hz}$ for BBO and DECIGO, and $1-10^3 \, {\rm Hz}$ for aLIGO, CE and ET~\cite{Danzmann, Kawamura2006CQGra..23S.125K, Harry:2006fi, Evans:2021gyd, Grado_2023, Martynov_PhysRevD.93.112004}. As mentioned earlier, all of the following results are obtained assuming a tensor-to-scalar ratio  $r = 0.001$, and for a fixed reheating energy scale $E_{\rm r*} = 1$ GeV. Note that even though the  post-inflationary EoS  can vary between $-1/3 < w \leq 1$, we restrict ourselves to  $ -0.28 \leq w < 1 $ due to computational limitations associated with the values of EoS closer to $w = -1/3$, which lead to divergences in our numerical computation.

 While, in general one should utilize Eq.~\eqref{eq; BBN bound general approx} to derive BBN constraints for multiple epochs of reheating, it is important to state that plots in this  section have been generated by incorporating  the following  additional approximation in order to speed up the computation. Note that, for the specific  choice of parameters used in our plots, namely, $r=0.001$, $E_{\rm r*} =  1 \, {\rm GeV}$ and $10^2 \, {\rm GeV} \leq E_i \leq 10^{10} \, {\rm GeV} $ (where $E_i$'s are the energy scale of the universe at the time of transition from $i^{\rm th}$ epoch to the next during reheating),  it is easy to see that the BBN constraint is violated only for  stiff-enough values of $w_1$, especially, since we restrict $E_1 \leq 10^{10} \, {\rm GeV} $. 

To illustrate this, let us stick to  the case $w_1 \leq 1/3$, for which  the left-hand-side of the BBN constraint  Eq.~\eqref{eq; BBN bound general approx} is maximum when $w_1 = 1/3$ and  $w_2 \simeq 1$, as shown by the solid line in Fig.~\ref{fig:limit for BBN}. In fact, for the specific  aforementioned value of parameters ($E_1$, $w_2$, $E_{\rm{r}*}$), Eq.~\eqref{eq; BBN bound general approx} for the solid line yields $1.86 \t 10^{-7} < 1.13 \t 10^{-6}$, demonstrating that the BBN constraint is easily satisfied. In fact, we numerically find that the BBN constraint is satisfied for $w_1 \leq 0.416$ for these specific choice of parameters. Since we keep the other parameters fixed in our plots,   we only need to check cases with $w_1 \geq 0.417$ for plausible violation of BBN constraint.
% in order to determine whether the BBN constraint is satisfied, we only needed to check whether  $w_1 < 0.417$.
Therefore, the exercise reduces to looking for the simple case of single-epoch reheating with a stiff equation of state ($w_1 \geq 0.417$), with the peak of the spectrum being located at the UV-cutoff frequency $f=f_e$.  In this regime,  following  Ref.~\cite{Figueroa:2019paj}, a weaker constraint on the parameters  can be obtained by assuming  the integral in Eq.~\eqref{eq; BBN constraint integral} to be dominated by the value of $\Og$ at $f_e$, namely,
\begin{equation}\label{eq; BBN bound for numerical}
    h^2 \Og(\tau_0, \, f_e) < 2(1 - \alpha_1) \, 1.13 \t 10^{-6}, \, \quad \text{for} ~ w_1 > \f{1}{3} \, .
\end{equation}
 Note that, the above expression is not valid when $w_1 \to (1/3)^+$ (approaching from above). In such a situation the integral in Eq.~\eqref{eq; BBN constraint integral} will naturally generate a logarithm term of  $\ln \l(f_e/f_1 \r)$  for the first epoch. Therefore, the BBN constraint becomes
\begin{equation}
     h^2 \Og(\tau_0, \, f_e)  < \frac{1.13 \t 10^{-6}}{\ln{\l(f_e/f_1\r)}}  \, \label{eq; BBN bound numerical limiting case} \, ,
\end{equation}
 where $f_1$ is the present-day frequency of tensor modes that entered the Hubble radius at the time of transition from the first to the second epoch during reheating.
Let us proceed to determine a critical value of $ w_{1, \rm min}$  so that as long as $w_1 > w_{1,min} >1/3$,   Eq.~\eqref{eq; BBN bound for numerical} remains to be a good approximation of the BBN constraint Eq.~\eqref{eq; BBN constraint integral}. $w_{1, \rm min}$ can be obtained by imposing
\begin{equation}
    \frac{1}{\ln{\l(f_e/f_1 \r)}} = 2 \l(1 -\frac{2}{1+3 w_{1, \rm min} } \r) \implies w_{1, \rm min} = \frac{1}{3}  \l[\frac{\ln \l( f_e/f_1\r)^4}{\ln \l( f_e/f_1\r)^2 -1} - 1 \r] \label{eq; w_1_min} \,.
\end{equation}
Therefore, we suggest the following simplified prescription for checking whether the BBN constraint is satisfied in this specific scenario (choice of $r$, $E_1$, $E_{\rm r *}$)\,--\,
\begin{equation}\label{eq; BBN bound for numerical split}
    h^2 \Og(\tau_0, \, f_e) <
    \begin{cases}
        \l( \frac{3w_1 -1}{3w_1 +1} \r) \, 2.26 \t 10^{-6}, & w_1 > w_{1, \rm min} \, ,\\
        \l( \frac{3w_{1, \rm min} -1}{3w_{1, \rm min} +1} \r) \, 2.26 \t 10^{-6}, & 1/3 < w_1 \; \leq \; w_{1, \rm min} \, .
    \end{cases}
\end{equation}

 We again emphasise that the weaker BBN constraint, as given by Eq.~\eqref{eq; BBN bound for numerical} or \eqref{eq; BBN bound for numerical split}, has been used purely to avoid computational expenses, and it is applicable  for the choice of parameters used in our plots. Therefore, we caution the reader that if $E_1 > 10^{10} \, {\rm GeV}$ (or $E_{\rm r*} < 1\, {\rm GeV}$, or both) and if $w_2$ is stiff-enough, which correspond to a very long stiff dominated epoch with EoS $w_2$ then BBN constraints can be violated, even when $w_1 < 1/3$. In this case, one must use  Eq.~\eqref{eq; BBN bound general approx} in order to derive accurate constraints.

The EoS parameter space, $\lbrace w_1\, ,~ w_2 \rbrace$, for reheating with two epochs is shown in Fig.~\ref{fig; two epoch param space}, with $E_1$ being the energy scale of the universe at the transition between the two epochs. In order for the signal to possess a greater likelihood of detection with the upcoming GW detectors (without violating BBN and aLIGO bounds), a higher value of $w_2$ is preferred, as evident from Fig.~\ref{fig; two epoch param space}. 
However, having higher values of both $w_1$ and $w_2$ would violate  the BBN constraint given in Eq.~\eqref{eq; BBN bound for numerical split} as indicated by the grey shaded region.  Similarly, the dark maroon-shaded region  represents the subspace outside the BBN constraint that are ruled out by the null detection of stochastic GWs by aLIGO. From Fig.~\ref{fig; two epoch param space}, we also notice that with an increase in the value of $E_1$, the range of values of $w_2$ that leads to a detection in the future detectors actually decreases; while that of $w_1$ becomes relatively unconstrained. This is because the physical frequency of GWs corresponding to the Hubble-entry at the end of the first epoch during reheating becomes higher than the frequency ranges of the aforementioned GW detectors. As a result, these GW detectors fails to put any constraint on $w_1$; as evident from Figs.~\ref{subfig; two epoch 10^6}, \ref{subfig; two epoch 10^8} and \ref{subfig; two epoch 10^10}. Similar things  happen for each detector, albeit at different values of $E_1$. Specifically, the detectable values of $w_2$ for LISA reaches its minimum around $E_1 \geq 10^6$ GeV. For BBO and DECIGO, this happens around $E_1 \geq 10^8$ GeV, while for CE and ET, operating at higher frequencies, the same takes place around $E_1 \geq 10^{10}$ GeV. 

Furthermore, we can also constrain the duration of each epoch, in terms of the number of $e$-folds ($N$), corresponding to an EoS ($w$), given the energy scales at the beginning ($E_i$) and end ($E_f$) of that epoch. Since the energy density  scales as $\rho \simeq E^4 \propto a^{-3(1+w)}$, the relation between $N$ and $w$ is given by
\begin{equation}\label{eq; Number of e-folds vs energy scale}
    N = \ln(\frac{a_f}{a_i}) = \frac{4}{3(1+w)} \, \ln(\frac{E_i}{E_f}) \, .
\end{equation}
Accordingly, we have also provided the corresponding number of $e$-folds in the plots appearing in Fig.~\ref{fig; two epoch param space} for each value of $w$. 

Following the above procedure, one can also obtain the parameter space for a multiple-epoch reheating phase with the EoS parameters $\lbrace w_1, \, w_2, \, w_3, ..., w_n \rbrace$. For example, we have used said procedure to determine the parameter space for a three-epoch reheating space, and have illustrated the results here with  (integrated) 2-dimensional projections along the axes $w_1, \, w_2 \, {\rm and} \, w_3$. The projected two-dimensional plots for different values of $E_1$ and $E_2$ (energy scales of the universe at the end of the first and the second epochs, respectively) are shown in Fig.~\ref{fig; three epoch param part_one}.

However, interpreting conclusively from  the projected plots for three-epoch reheating phase is quite difficult, and at times very ambiguous. Hence, we provide a link to the GitHub repository \href{https://github.com/athul104/Spectral_Energy_Density_FO_GWs}{\faGithub} which contains a \texttt{Python} code to directly plot the spectral energy density of first-order GWs. The reader can use this tool to conclusively determine whether a particular combination of EoS parameters would result in a potentially detectable GW signal (which is consistent with BBN and aLIGO constraints) for multiple transitions during reheating.

As mentioned in the case of two-epoch reheating, higher values of $E_1$ implies a shorter duration of the first epoch during reheating following the end of inflation. As a result, the aforementioned  GW detectors will  be able to impose weaker constraints on $w_1$ for large values of  $E_1$, as evident from Fig.~\ref{fig; three epoch param part_one}. This is because if the first post-inflationary epoch is short, then the detectability essentially relies on the subsequent epochs, specifically on the final epoch before BBN.  Similarly,  higher values of  $E_2$ widen the range of detectable values of $w_2$ further, with the range of $w_3$ being shortened. To conclude, our results in this section indicate that higher values of $w_2$ and $w_3$ would result in detectable GW signal, as expected.

\section{Effect of varying reheating temperature on GW spectral energy density}
\label{sec:effect_GW_Trh}
In the previous sections, we had fixed the reheating temperature to be $T_{\rm r*}=0.45\,{\rm GeV}$, or equivalently, the energy scale at the end of reheating to be $E_{\rm r*}=1\,{\rm GeV}$, for the purpose of illustration. However, in general, the reheating energy scale can range from the end of inflation all the way down to the onset of BBN. Hence, in this section, we discuss the impact of varying the reheating temperature on the spectral energy density of inflationary GWs.

Additionally, we had fixed the tensor-to-scalar ratio during inflation to be $r=0.001$, or equivalently, the energy scale at the end of inflation to be $E_{\rm inf}=5.76\times10^{15}\,{\rm GeV}$, keeping in mind that the upcoming CMB detectors will be able to probe $r$ up to ${\cal O}\l(10^{-3}\r)$. Since inflation does not make a  prediction for $r$, it is possible that $r \ll 10^{-3}$. Therefore, we proceed to first discuss the impact of small $r$ on the detectability of inflationary GWs \textit{via}  the upcoming GW observatories, before moving on to determine the  detectable parameter space of equation of states for fixed $r=0.001$, but different values of the reheating energy scales $E_{\rm r*}$. 

\subsection{Effect of varying tensor-to-scalar ratio on GW signal}
\label{app_sub; r_vs_T_r* }

\begin{figure}[htb]
    \centering
    \includegraphics[width=0.9\linewidth]{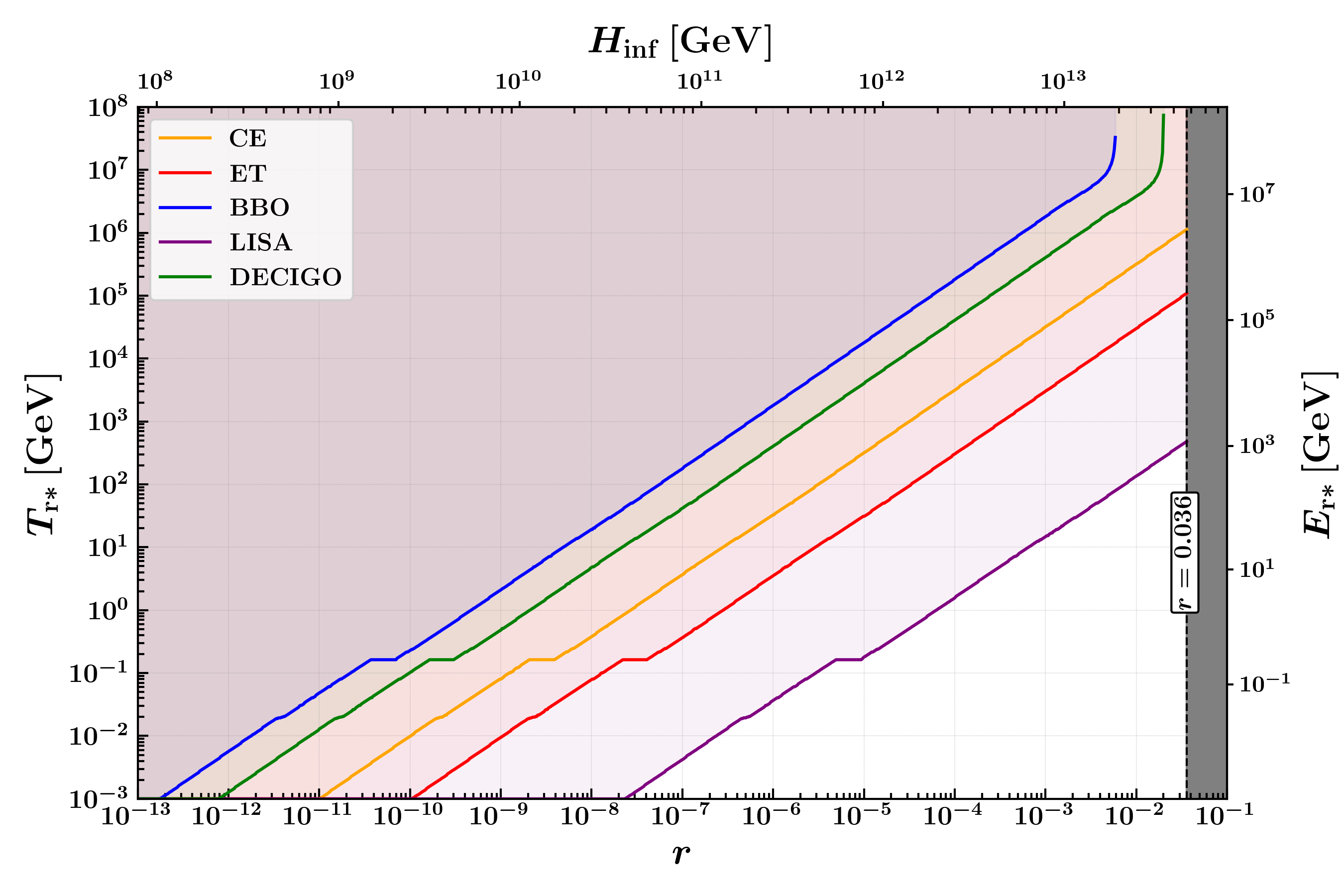}
    \caption{The figure illustrates the effect of varying tensor-to-scalar ratio ($r$), and  reheating temperature ($T_{\rm r*}$) on the detectability of inflationary GWs in various GW detectors.  The coloured shaded regions indicate the parameter space of the $r$, and $T_{\rm r*}$, that can never yield a potential detection of inflationary GW signal in  CE, ET, BBO, LISA and DECIGO detectors. The grey-shaded region is ruled out by the CMB observations, namely, $r \leq 0.036$. The Hubble parameter during (at the end of) inflation ($H_{\rm inf}$) corresponding to each value of $r$ and the energy scale corresponding to each value of $T_{\rm r*}$ are provided in the respective opposite axes.}
    \label{fig:r_vs_T_r*}
\end{figure}

A quantity of central importance is the minimum value of reheating temperature corresponding to a given value of tensor-to-scalar ratio which leads to a detectable GW signal in various GW detectors. In fact, for a given GW detector, and for a fixed value of  $r$, we find that upon increasing the minimum value of the reheating temperature, there is a critical value of $T_{\rm r*}$ (or $E_{\rm r*}$) above which the inflationary GW signal will lie outside the sensitivity region of that detector,  even if the EoS during reheating is that of a  stiff-matter, \textit{i.e.} $w\simeq 1$. 

In Fig.~\ref{fig:r_vs_T_r*}, we show the (shaded) region in the parameter space of $r$ and $T_{\rm r*}$ (or $E_{\rm r*}$) for different GW detectors that will never lead to a detectable inflationary GW signal. This does not necessarily mean  that the unshaded region corresponding to each detector will guarantee a detection in that detector, which of course depends upon the number of epochs and their EoS and duration during reheating, as well as on whether the GW spectrum satisfies the current observational constraints on BBN and aLIGO. In short, the unshaded regions have the feasibility to yield detection as well as may include regions that violate current observational constraints.

The solid lines in Fig. \ref{fig:r_vs_T_r*} are obtained by assuming the reheating to consist of only single epoch with equation of state $w \simeq 1$, which is an extreme scenario of having a maximum tilt for $\Og$ curve in the reheating regime. All other possible scenarios of multiple epochs will make the $\Og$ curve in the reheating regime lower than the single epoch case with $w \simeq 1$ for a fixed $r$ and $T_{r*}$. This gives the maximum value $T_{r*}$ above which no detection is possible no matter whatever the EoS during reheating for a fixed $r$.

For large enough values of $r \gtrsim {\cal O}\l( 10^{-2}\r)$, height of flat (nearly scale-invariant) region of $\Og(f)$, corresponding to modes making their Hubble-entry during the thermal RD epoch, increases enough to  reach the sensitivity curves of BBO and DECIGO, as long as the reheating temperature is also large enough. In such cases, at large reheating temperatures, these detectors will detect a GW signal with certainty. That is why we see nearly vertical curves for BBO and DECIGO at high values of $r$ in Fig.~\ref{fig:r_vs_T_r*} (blue and green curves). 

However, the same is not true for other GW detectors, because their sensitivity curves are above the highest allowed height of $\Og$, corresponding to $r=0.036$, for modes that entered in the RD epoch. Therefore, if $T_{\rm r*}$ is high enough, then the GW signal will be absent in these detectors, independent of the EoS during reheating.

Note that the small kinks seen in the solid curves are due to the sudden drop in the relativistic degrees of freedom in the early Universe, within the Standard Model of particle physics. In particular, the prominent horizontal shift in all curves around $T_{\rm r*} \simeq 150$ MeV is due to a  large drop in relativistic degrees of freedom closer to QCD phase transition. 

\begin{figure}[htb]
  \centering
  \subfigure[]{\includegraphics[width=0.4\textwidth]{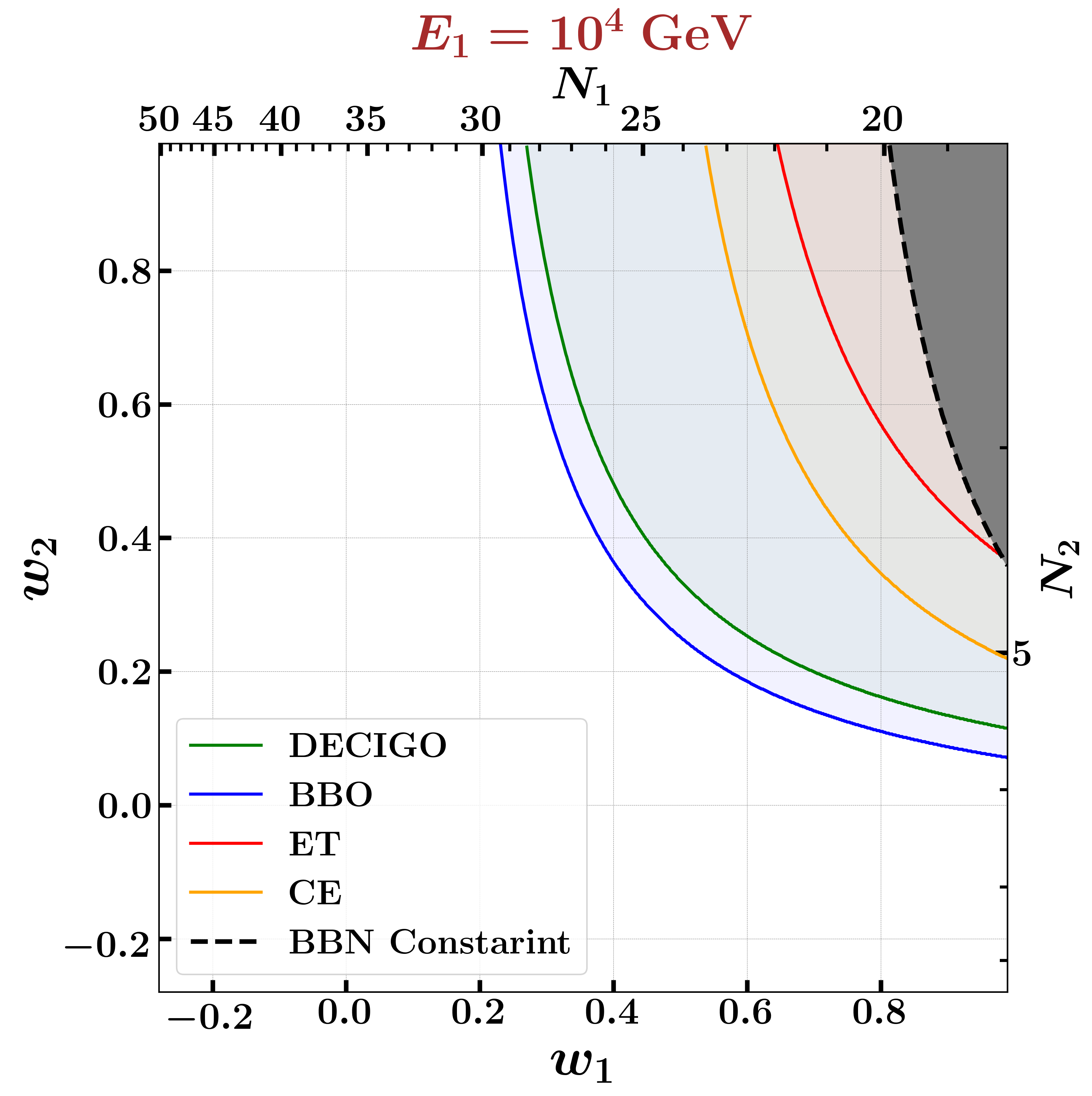}}
  \subfigure[]{\includegraphics[width=0.4\textwidth]{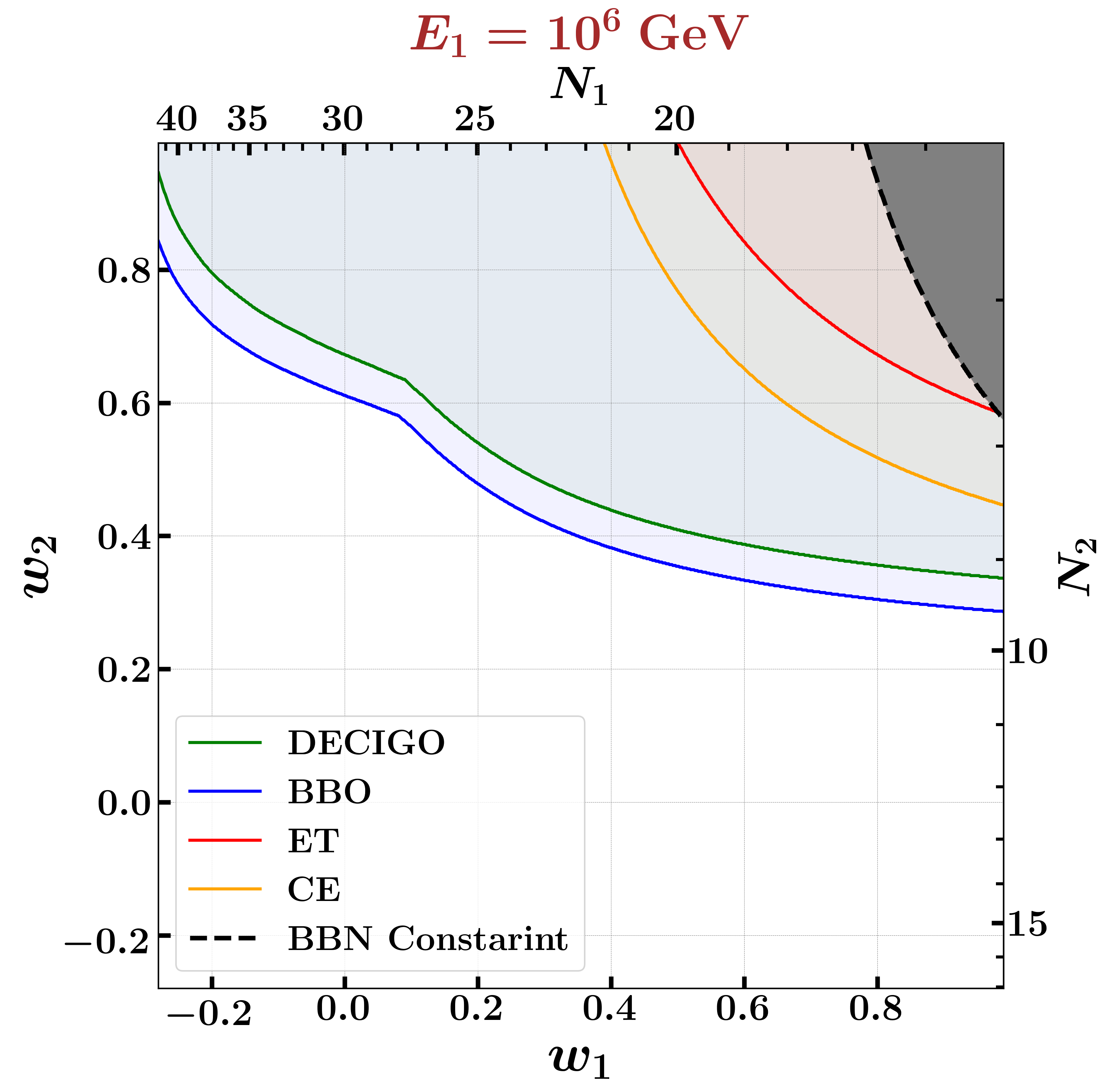}}
  \subfigure[]{\includegraphics[width=0.4\textwidth]{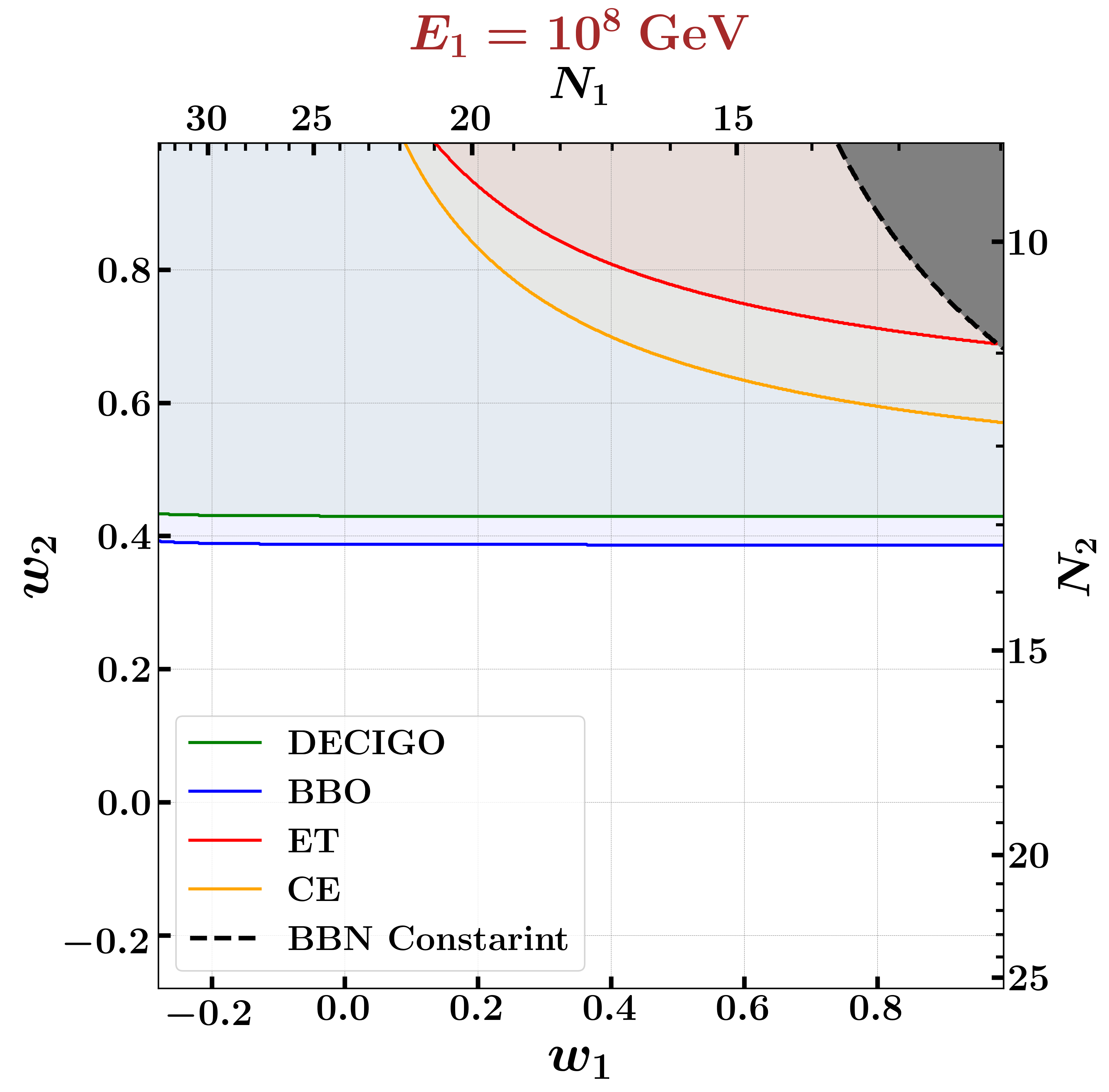}}
  \subfigure[]{\includegraphics[width=0.4\textwidth]{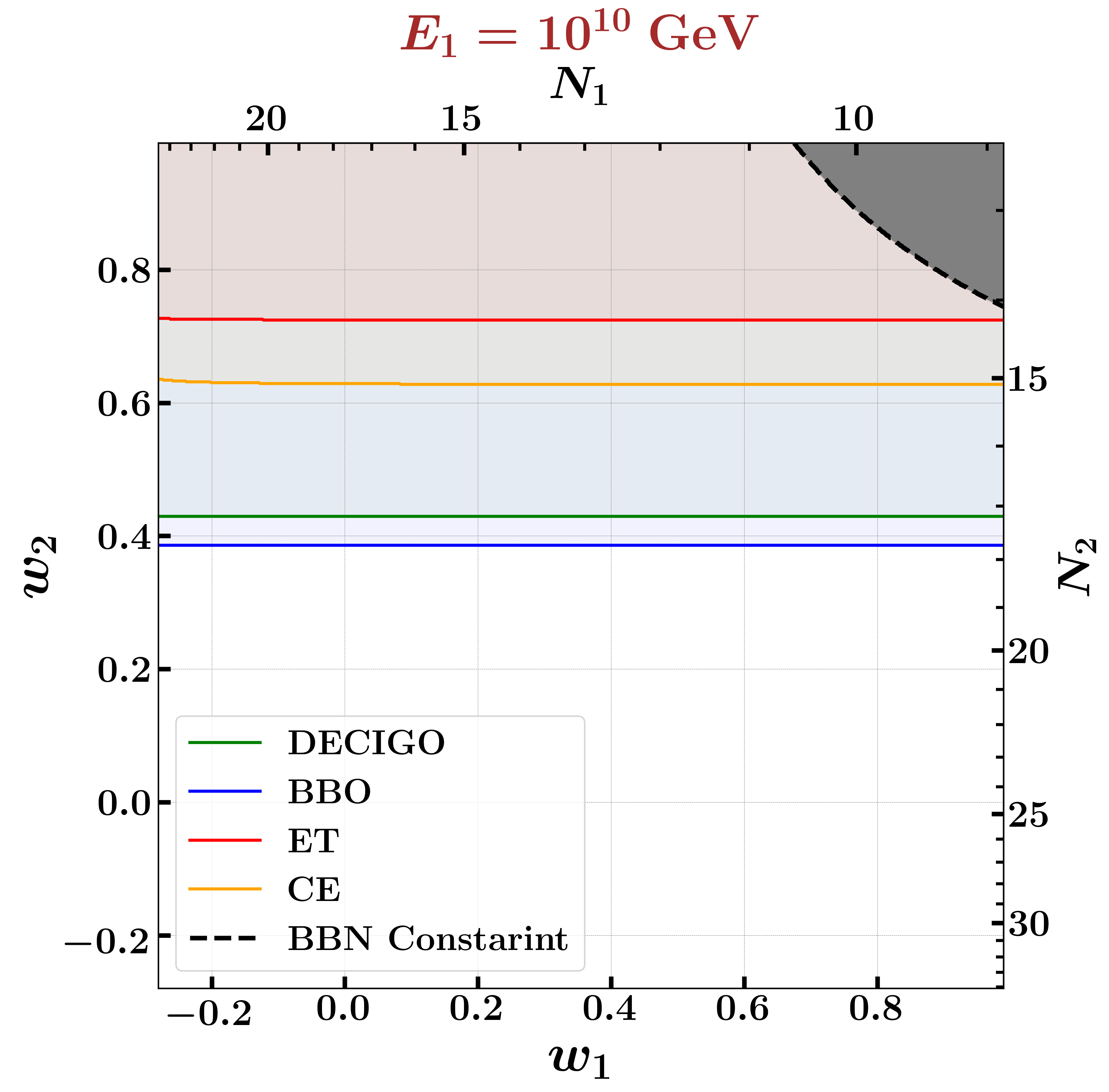}}
  \caption{The parameter space for two-epoch reheating phase with EoS parameters $w_1$ and $w_2$ that leads to a potentially detectable GW signal in the  LISA, BBO, DECIGO, CE, and ET detectors for $E_{\rm r*} = 100$ GeV and $r = 0.001$. The grey-shaded region in each plot corresponds to the combination of $w_1$ and $w_2$ that  results in the $\Og$ curve intersecting the horizontal BBN constraint curve at  $ 1.13 \t 10^{-6}$, while the maroon-shaded region represents those that are ruled out by the aLIGO null detection. The energy scale of the universe at the end of the first post-inflationary epoch is labelled above in each plot. The duration of each epoch (number of $e$-folds) has been provided opposite to the corresponding EoS axis.}
  \label{fig;two epoch E_r* = 100 GeV}
\end{figure}

\subsection{Two-epoch reheating with different reheating temperature}\label{app_sub; two epoch param space}

In Sec.~\ref{sec:probing_primodial_EOS}, we  determined  the parameter space of EoS $w_1$ and $w_2$, that results in a potentially detectable  inflationary GW signal in future GW detectors,  for fixed values $E_{\rm r*} = 1 \, {\rm GeV}$ and $r = 0.001 \, (E_{\rm inf} = 5.76 \t 10^{15} \, {\rm GeV})$. As stressed before,  the aforementioned values of $r$ and reheating temperature were used purely for the purpose of illustrating how to to use our calculations of $\Og$ to check whether a particular set of parameters leads to a detectable signal. 

\begin{figure} 
  \centering
  \subfigure[]{\includegraphics[width=0.35\textwidth]{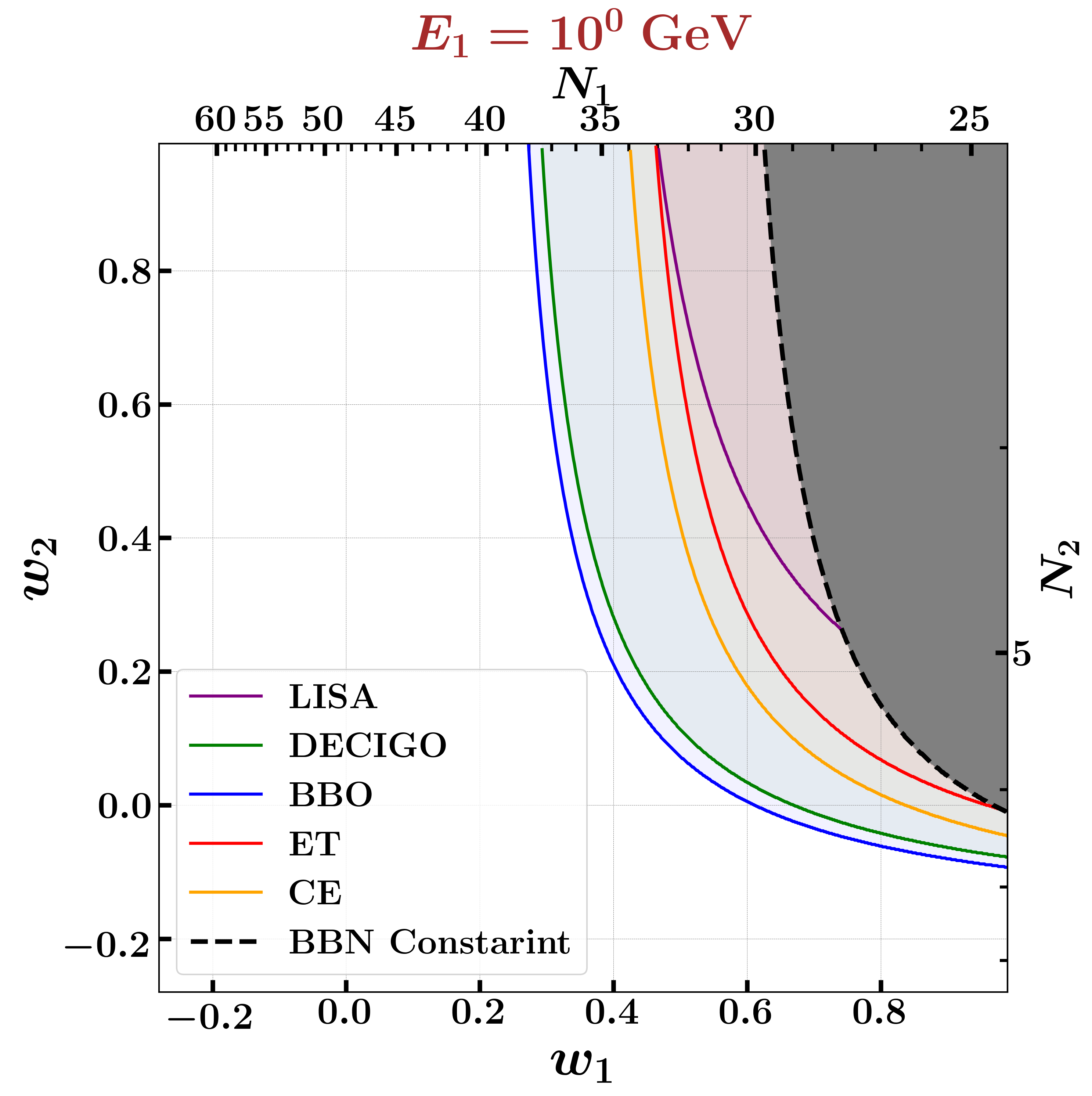}}
  \subfigure[]{\includegraphics[width=0.35\textwidth]{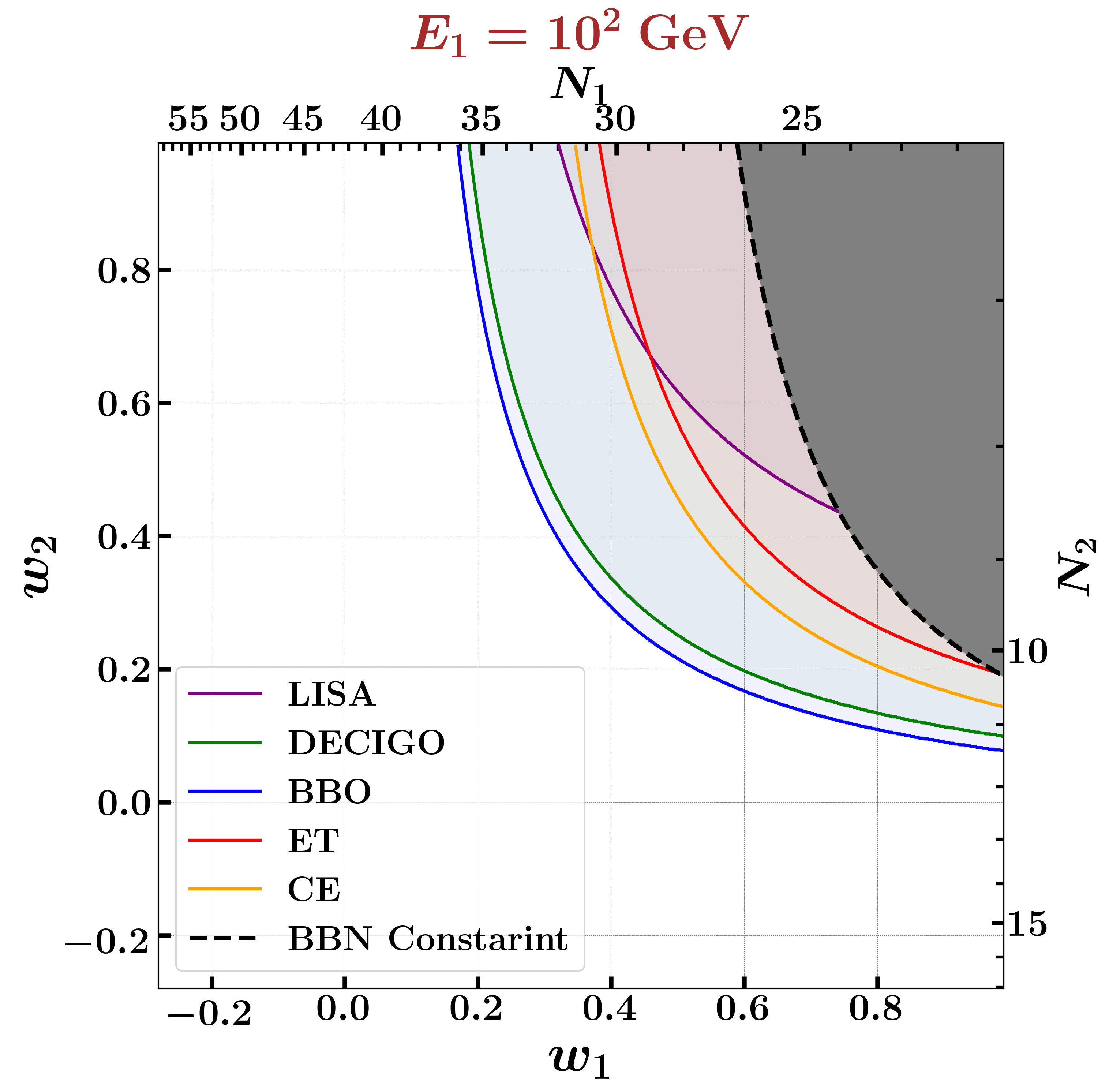}}
  \subfigure[]{\includegraphics[width=0.35\textwidth]{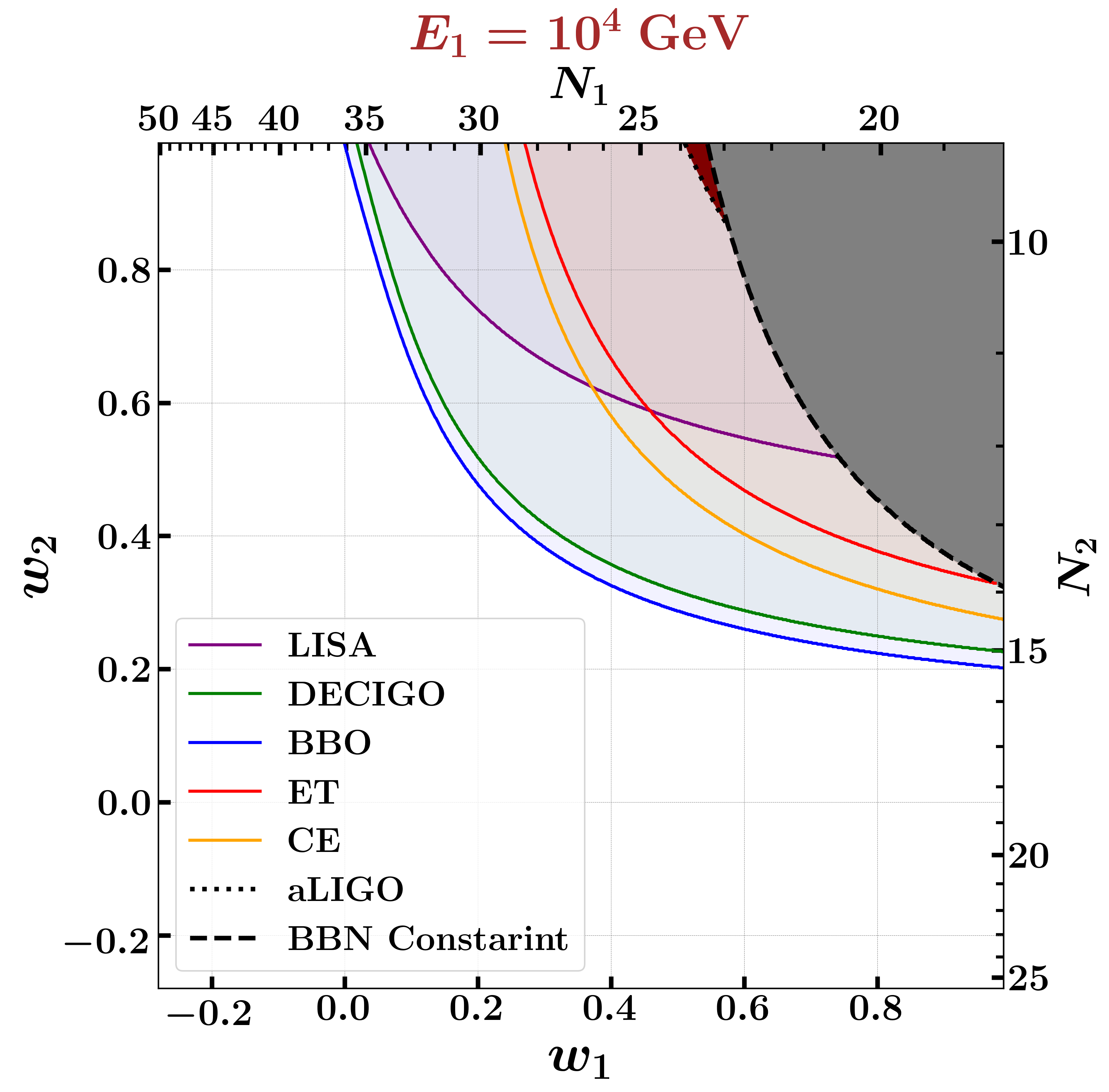}}
  \subfigure[]{\includegraphics[width=0.35\textwidth]{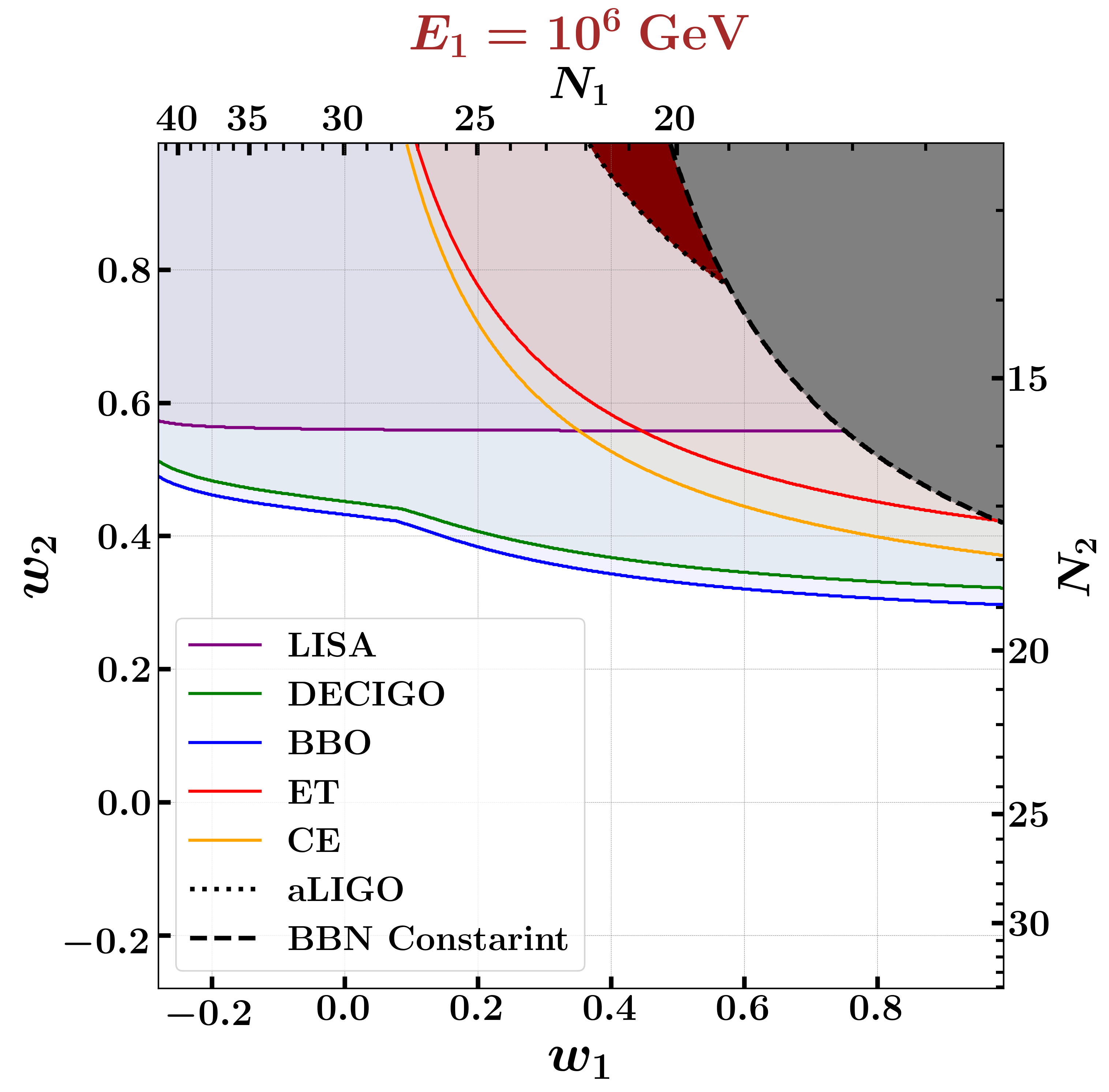}}
  \subfigure[]{\includegraphics[width=0.35\textwidth]{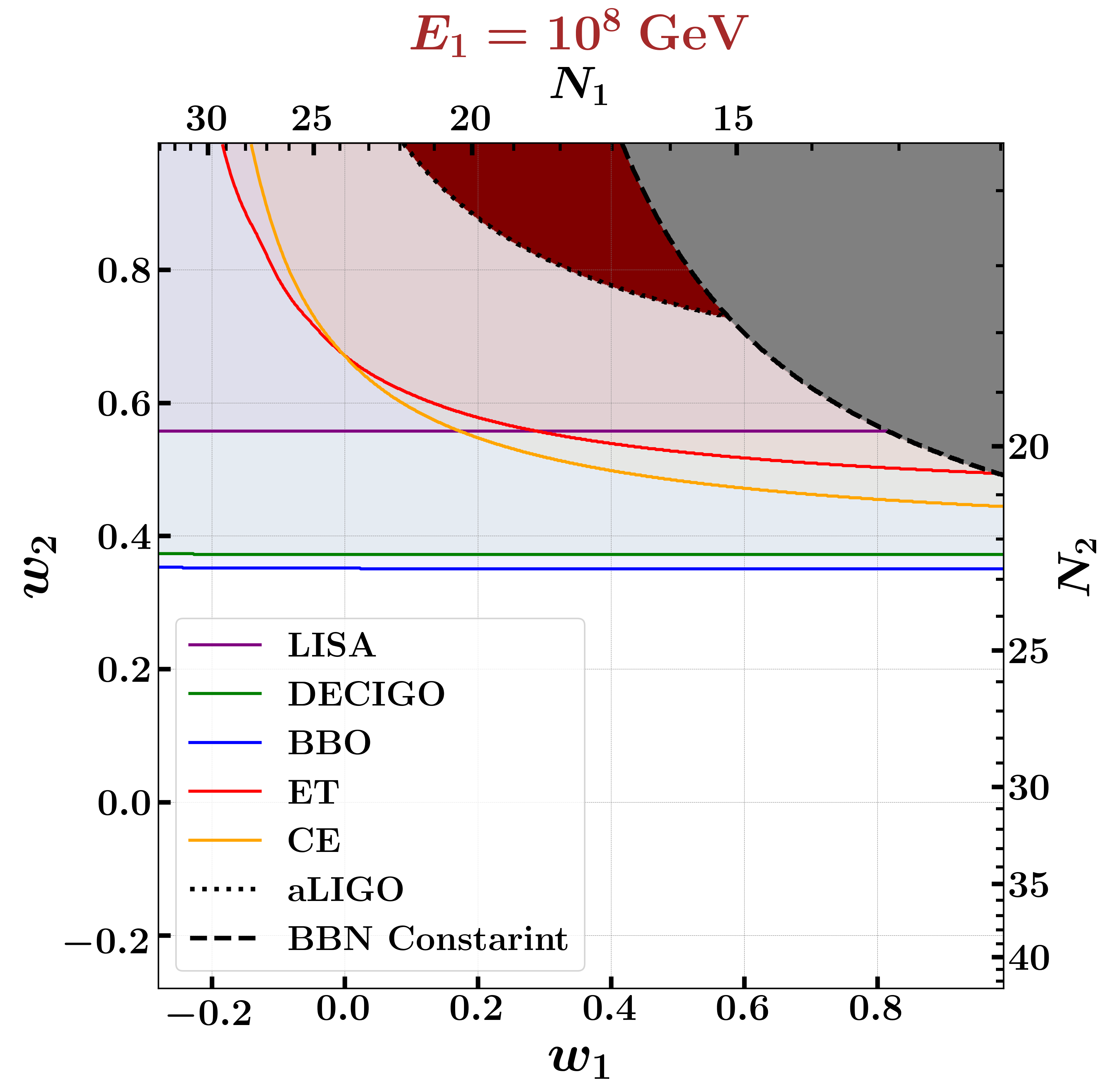}}
  \subfigure[]{\includegraphics[width=0.35\textwidth]{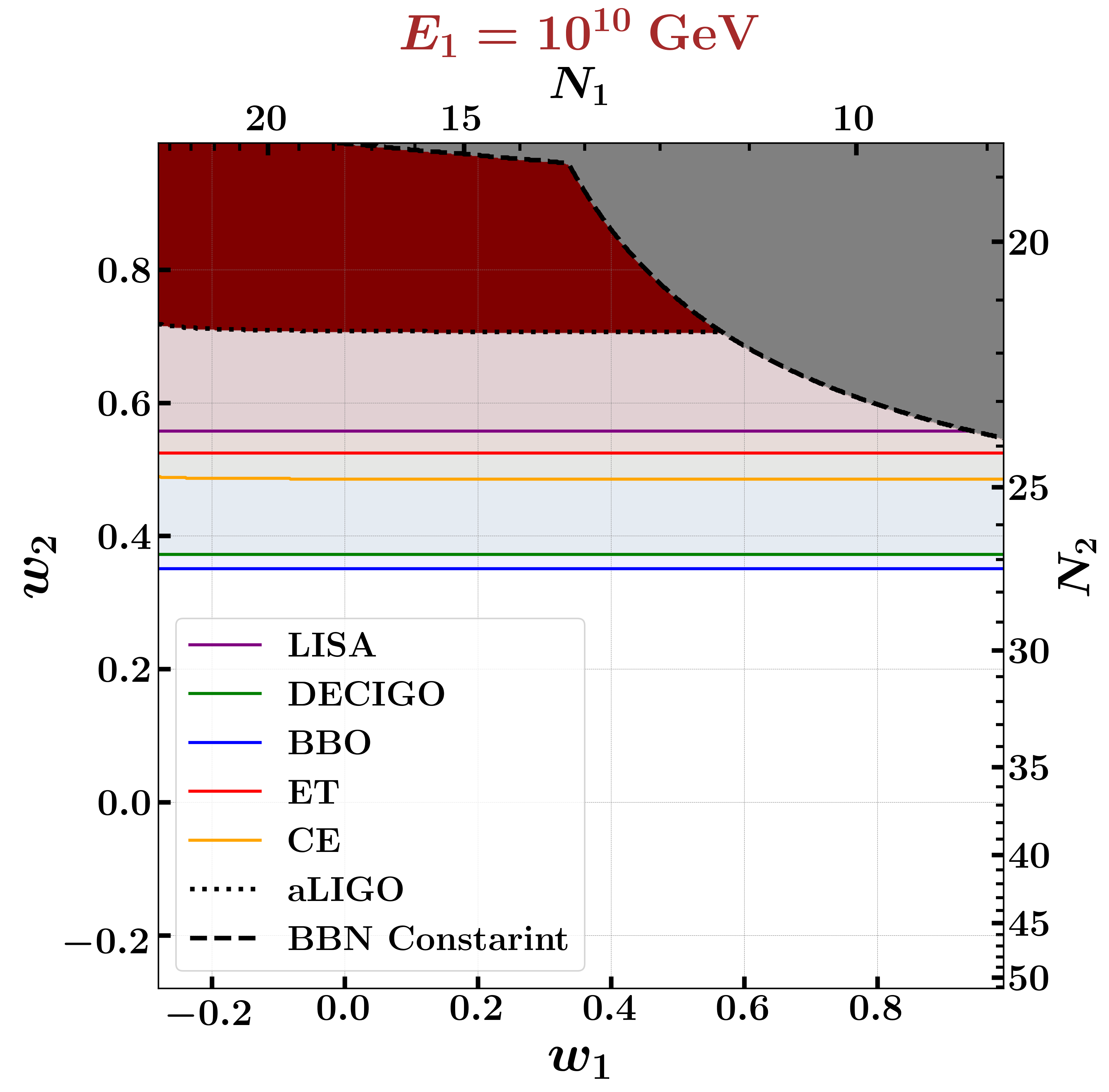}}
  \caption{The parameter space for two-epoch reheating phase with EoS parameters $w_1$ and $w_2$ that leads to a potentially detectable GW signal in the  LISA, BBO, DECIGO, CE, and ET detectors for $E_{\rm r*} = 10$ MeV and $r = 0.001$. The grey-shaded region in each plot corresponds to the combination of $w_1$ and $w_2$ that  results in the $\Og$ curve intersecting the horizontal BBN constraint curve at  $ 1.13 \t 10^{-6}$, while the maroon-shaded region represents those that are ruled out by the aLIGO null detection. Energy scale of the universe at the end of the first post-inflationary epoch is labelled above in each plot. The duration of each epoch (number of $e$-folds) has been provided opposite to the corresponding EoS axis.}
  \label{fig; two epoch E_r* = 10 MeV}
\end{figure}

However, for a given value of $r$,  reheating temperature can be very different, ranging in between $1\,{\rm MeV} \leq T_{\rm r*} \lesssim E_{\rm inf}$. Therefore, in this section, we include the parameter space plots of $w_1$ and $w_2$ corresponding to a  two-epoch reheating scenario (similar to Fig.~\ref{fig; two epoch param space}), specifically for  $E_{\rm r*} = 100 \, {\rm GeV}$ and $10 \, {\rm MeV}$ in Figs.~\ref{fig;two epoch E_r* = 100 GeV} and \ref{fig; two epoch E_r* = 10 MeV}, respectively. We fix the value of $E_{\rm inf}$ to be the  same as before. Note that the BBN constraint that we have imposed  in this section  is not the same as the one quoted in Eq.~\eqref{eq; BBN bound for numerical split}. Here, we  check whether the $\Og$ curve intersects the horizontal BBN constraint curve at $ 1.13 \t 10^{-6}$.

\section{Post-inflationary evolution in a scenario inspired from String Theory}
\label{sec:String_phenomenology}

\begin{figure}[htb]
    \centering
    \includegraphics[width = 1\textwidth]{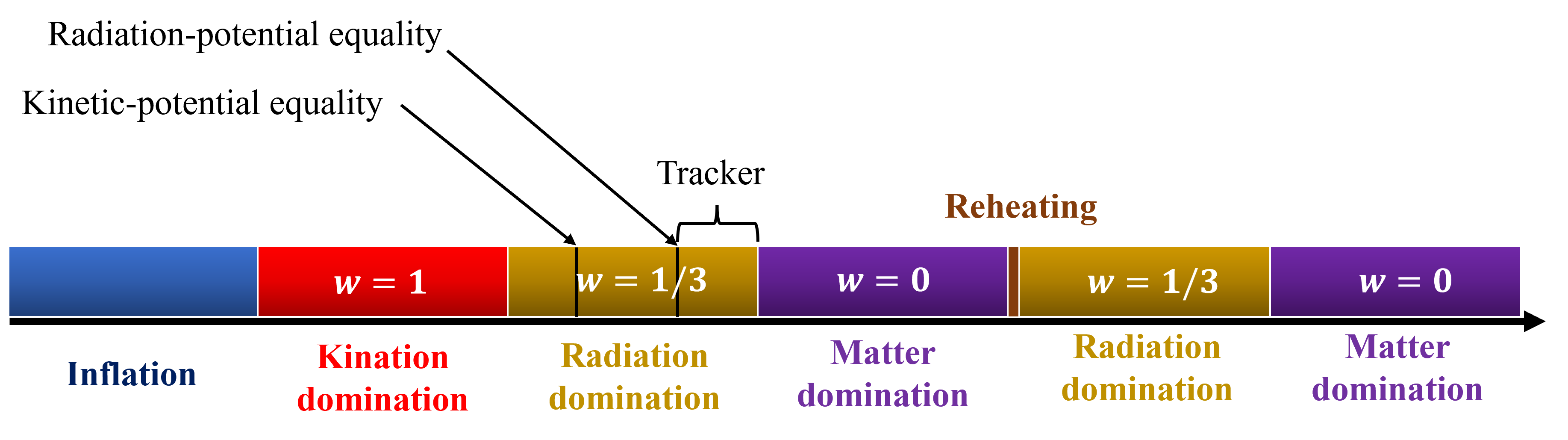}
    \caption{A schematic depiction  of the chronology of post-inflationary epochs in the String-inspired model considered in Ref.~\cite{Apers:2024ffe}, where the final epoch of reheating is assumed to be instantaneous. Note that the horizontal length is not an accurate representation of the  duration of each epoch.}
    \label{fig: timeline_string}
\end{figure}

In our analysis of the spectral energy density of GWs in Sec.~\ref{sec:probing_primodial_EOS}, we remained agnostic about the exact values of the post-inflationary EoS parameters and focused on determining the sub-space of the post-inflationary parameter space that results in a potentially detectable signal in the upcoming GW detectors, without violating the existing constraints. In this section, we focus on a particular post-inflationary scenario considered in a recent study~\cite{Apers:2024ffe}, which is phenomenologically inspired from String Theory~\cite{Zwiebach:2004tj, Polchinski:1998rq, Tong:2009np, Cicoli:2023opf, Conlon:2024ene}.

In type-\Romannum{2}B String Theory within the large volume scenario (LVS)~\cite{Cicoli:2023opf}, a resultant \textit{moduli field} $\chi$ (which is an artifact of string compactification) may persist throughout the  universe, potentially long enough until the onset of BBN. During inflation, the  moduli field is sub-dominant and remains nearly frozen in its potential,  while after the end of inflation, the volume modulus comes to dominate the universe and commences rolling swiftly down its steep exponential potential. The moduli field eventually reaches the minimum of its potential around which it oscillates and later decays, thereby reheating the universe to the hot Big Bang phase. Thus, the dynamics of the universe from the end of inflation until the onset of BBN is primarily dictated by the moduli field in this scenario~\cite{Apers:2024ffe}.

The succeeding epoch after the end of inflation is  therefore a {\em  kination epoch}\footnote{See Refs.~\cite{Gouttenoire:2021jhk, Gouttenoire:2021wzu} for discussions on kination cosmology.}  during which the moduli exhibits a fast-roll phase, with $\dot{\chi}^2 \gg V(\chi)$, down a steep region of its potential of the form
\begin{equation}
    V(\chi) \simeq V_0 \, e^{-\lambda \, \f{\chi}{ m_p}} \, ,
\label{eq; generic moduli potential} 
\end{equation}
where $\lambda = \sqrt{27/2}$ and $V_0 = m_p^4$ for LVS. 
The EoS during this kination epoch is $w = \frac{\dot{\chi}^2 - 2 V(\chi)}{\dot{\chi}^2 + 2 V(\chi)} \simeq 1$.
While the  universe, at this epoch, is also filled with a small fraction of radiative degrees of freedom that was generated from the decay of the inflaton at the end of inflation. Due to a sharp decline of the moduli energy density during kination, the universe  enters into a radiation-dominated epoch, following the kination phase.

The moduli field, rolling down its steep potential in the radiation dominated epoch, then enters into a tracker regime~\cite{Copeland:1997et,Zlatev:1998tr,Sahni:1999qe,Mishra:2017ehw,Bag:2017vjp}, as  illustrated in Fig.~\ref{fig: timeline_string},  where its energy density scales proportionally to the energy density of the background universe. Eventually, the field leaves the steep region of its potential and rolls down to its stable (quadratic) local minimum about which it begins to oscillate. The oscillating moduli field then dominates the energy budget of the universe again, leading to a matter-dominated universe. The oscillating field then decays, potentially non-perturbatively (hence efficiently), thereby reheating the universe to give rise to the hot Big Bang phase.

\begin{figure}[hbt]
    \centering
    \includegraphics[width=0.85\linewidth]{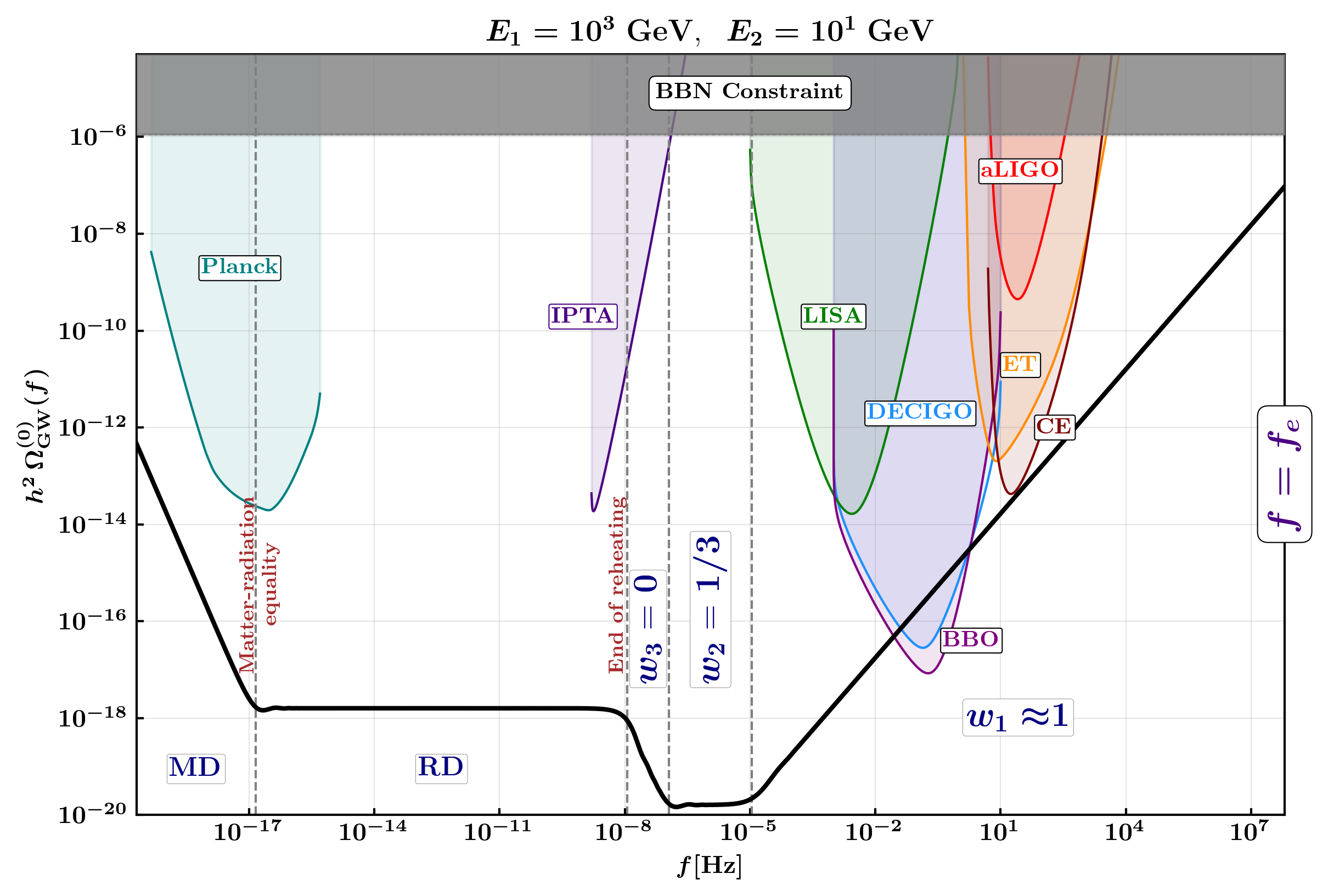}
    \caption{The spectral energy density ($\Og$) of inflationary GWs for the String-inspired scenario as illustrated in Fig.~\ref{fig: timeline_string}, where we have an early kination epoch ($w \simeq 1$) following the end of inflation, which was later followed by a radiation-dominated epoch ($w = 1/3$) and subsequently by a matter-dominated epoch ($w = 0$). The final epoch of reheating is assumed  to be instantaneous at an energy scale $E_{\rm r*} = 1$ GeV. The energy scales $E_i$ (with $i \in \mathbb{Z}^+$) corresponding to the end of each epoch, are given in the top label of the plot. The energy scale at the end of inflation is the same as in our previous plots, namely, $E_{\rm inf} = 5.76 \t 10^{15}$ GeV, so as to have the tensor-to-scalar ratio $r = 0.001$. The maximum frequency considered here (the rightmost point in the plot) is the UV cut-off frequency, $f_e = 6.37 \t 10^7$ Hz, which corresponds to $k_e = 4.05 \t 10^{22} \, {\rm Mpc}^{-1}$ for $r = 0.001$.}
    \label{fig:spectral_energy_density_string}
\end{figure}

In short, the specific chronology of post-inflationary epochs in this scenario involves an early kination epoch $(w \approx 1)$ following the end of inflation, followed by a radiation-dominated epoch $(w = 1/3)$, and  subsequently, a matter-dominated epoch $(w = 0)$, before reheating the universe prior to the commencement of the BBN. Note that this final epoch of reheating is assumed to be instantaneous. Fig.~\ref{fig:spectral_energy_density_string} depicts a typical plot of spectral energy density against the current frequency of GWs for this particular model.

\begin{figure}[htb]
    \centering
    \includegraphics[width = 0.85\textwidth]{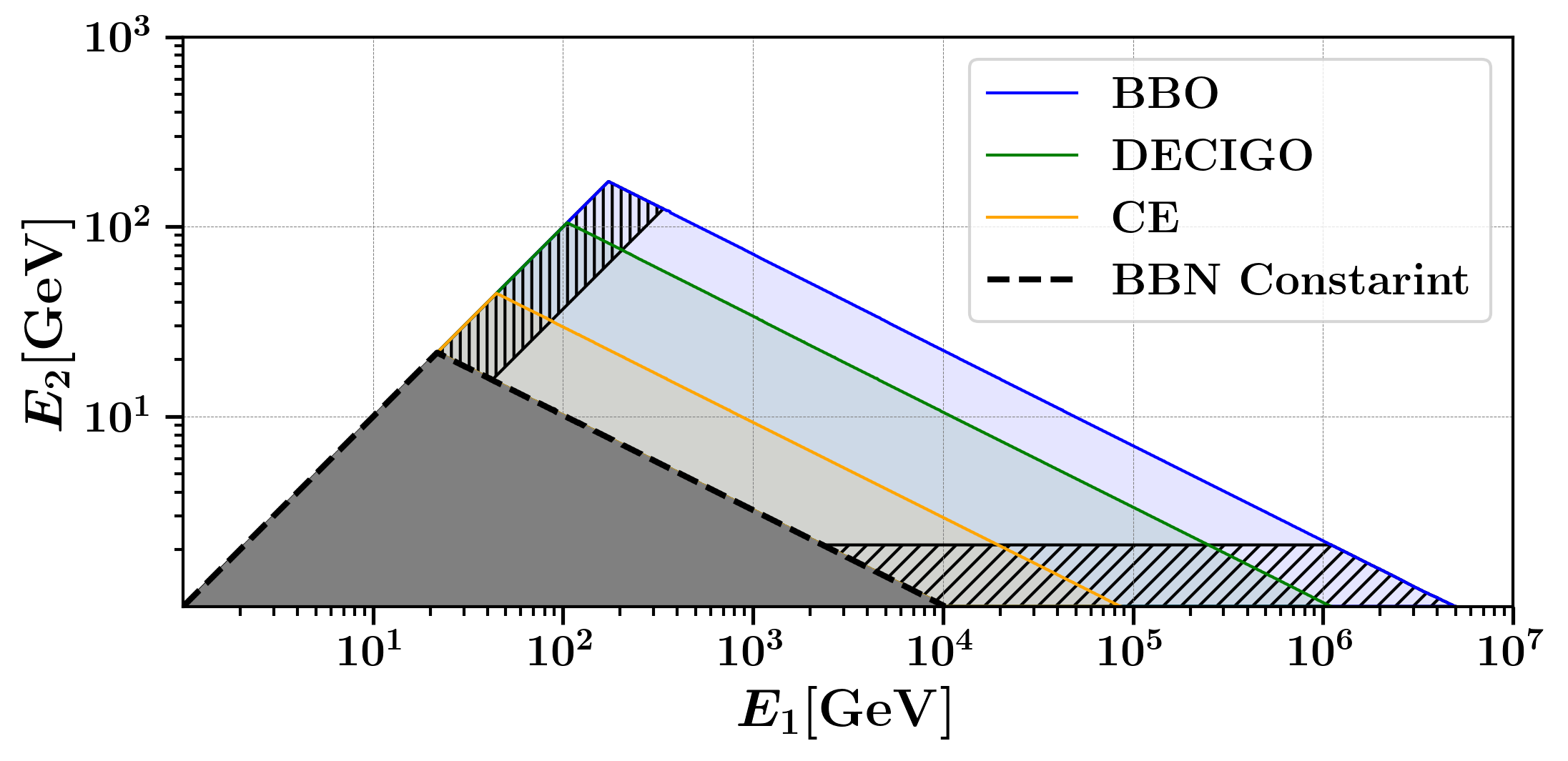}
    \caption{The parameter-space plot for the energy scale at the end of the kination epoch ($E_1$), and that at the beginning of the early matter domination ($E_2$), which leads to a detectable GW signal in the future detectors (assuming the final epoch of reheating to be instantaneous) for the case of a String-inspired scenario considered in Ref.~\cite{Apers:2024ffe} and illustrated in Fig.~\ref{fig: timeline_string}. The grey-shaded region violates the BBN constraint. The hatched shading regions corresponds to scenarios in which the number of $e$-folds are less than one. The diagonal hatching corresponds to the early matter-dominated epoch being less than one $e$-folds and the vertical hatching corresponds to that of early raditaion-dominated epoch.}
    \label{fig:instant rehating}
\end{figure}

Following the methodology as spelled out systematically  in Sec.~\ref{sec:probing_primodial_EOS}, we determine the parameter space of the energy scales at the end of each of the aforementioned epochs  which gives a signal in GW detectors. To be consistent with our previous analyses, let us assume the energy scale at the end of inflation to be $E_{\rm inf} = 5.76 \t 10^{15}$ GeV, and that at the end of (the instantaneous period of)  reheating to be $E_{\rm r*} = 1$ GeV. Specifically, we determine the range of energy scale at the end of the kination epoch ($E_1$) and that at the end of (the early) radiation domination epoch ($E_2$), which results in detectable GW signals.
The corresponding parameter-space plot is shown in Fig.~\ref{fig:instant rehating}. In our computation, we have taken one $e$-fold as the minimum duration of epochs and the regions corresponding to less than one $e$-fold are hatch shaded.

From Fig.~\ref{fig:instant rehating}, we observe that this particular scenario results in a GW signal which will be detectable by BBO, DECIGO, and CE (for our choice of $E_{\rm inf}$ and $E_{\rm r*}$) for a large range of $E_1$ and $E_2$. However, the combination of $E_1$ and $E_2$, that leads to  a detectable signal in LISA and ET, actually violates the BBN constraint given in Eq.~\eqref{eq; BBN bound for numerical}. Furthermore,  the spectrum of only those tensor modes that entered the  Hubble radius during the kination epoch are high enough to intersect the sensitivity curves of the detectors, since the kination epoch results in a highly blue-tilted GW spectrum. The plot also demonstrates that a detection of primordial GWs is possible in this scenario only if  $E_2 \lesssim {\cal O}(10^2)$ GeV, or equivalently, if the duration of the early matter-dominated epoch is roughly less than 7 $e$-folds, \textit{i.e.,} $N_{\rm matter} \lesssim 7$.

\section{Discussion and conclusions}
\label{sec; discussion}

One of the key predictions of the inflationary paradigm is the generation of tensor fluctuations which later constitute a stochastic background of gravitational waves (GWs) upon their Hubble-entry after the end of inflation~\cite{Starobinsky:1979ty}. These primordial GWs encode  crucial information about the state of the universe, both at the time of their production, and the subsequent epochs through which they propagate before reaching the terrestrial and celestial detectors. Below the Planck scale, the GWs interact minimally with matter and energy, making them the most ideal candidate for  probing  the dynamics of the very early universe~\cite{Starobinsky:1979ty,Giovannini:1999hx, Giovannini:2019oii, Guzzetti:2016mkm, Wang:2016tbj, Haque:2021dha}. Numerous studies have recently examined the feasibility of detecting these GWs using both the current and upcoming GW detectors~\cite{Figueroa:2019paj, Liu:2015psa, Boyle:2005se, Bernal:2019lpc,Bernal:2020ywq, Caprini:2024ofd, Wang:2016tbj, Campeti:2020xwn,Braglia:2024kpo}. 

In this paper, we focused on the possibility of probing the unknown post-inflationary history of the universe via primordial GWs. Since the dynamics of reheating involves a number of complex phases during which the universe transits from a highly  non-thermal state at the end of inflation to the thermal hot Big Bang phase~\cite{Kofman:1994rk, Shtanov:1994ce, lozanov2019lectures}, it is important to describe reheating  by a series of epochs with distinct EoS parameters~\cite{Ng:1993pv,Antusch:2021aiw}. Hence, it is phenomenologically important to determine the possible  post-inflationary scenarios that would result in a detectable stochastic GW signal in the upcoming GW detectors, without violating the existing constraints. Therefore, we have dedicated this paper in computing the spectral energy density of  (first-order) inflationary GWs at the present epoch by considering the post-inflationary universe to feature  multiple successive phases, each with a (nearly) constant EoS parameter. We  determined the subspace  of the EoS parameters of different epochs, as well as the corresponding  duration of each epoch, during reheating that leads to a potentially  detectable GW signal. 

We focused on studying the evolution of inflationary (first-order) tensor fluctuations in the post-inflationary universe in Sec.~\ref{sec:GWs_theory}. We  began by  solving the evolution equation for the  tensor mode functions, where we obtained the general solution for a given post-inflationary epoch described by a constant EoS parameter in terms of the Bessel functions. By assuming the  transitions between successive epochs of constant EoS  to be instantaneous, we obtained  expressions for the unknown coefficients appearing in  the general solutions for the tensor mode functions; be it for the  epoch immediately following the end of inflation, as given by Eq.~\eqref{eq; coeff A_k and B_k for first epoch}, or any other epoch succeeding that, as given by Eqs.~\eqref{eq; coeff A_k,n} and \eqref{eq; coeff B_k,n}.  We computed the spectral energy density $\Og(f)$ of GWs using Eq.~\eqref{eq; exp for Omega_GW as fun of freq}, and checked whether their resulting spectrum will be detectable by the upcoming GW observatories, such as LISA, ET, CE, DECIGO and BBO; for a variety of reheating scenarios as demonstrated in Fig.~\ref{fig: EDS two, three, four epoch}. In the case of a reheating scenario consisting of two successive phases, the EoS parameter space which results in a detectable GW signal is plotted in Fig.~\ref{fig; two epoch param space}. Similarly for a three-epoch reheating phase, the projections of the three-dimensional EoS parameter space are shown in Fig.~\ref{fig; three epoch param part_one}.

Our results primarily indicate that if inflationary  GWs are detected by the aforementioned observatories in the near future, then it is highly likely that the temperature at the end of reheating was low enough, and that the universe  evolved through  one or more epochs of stiff-matter ($w > 1/3$) dominated phase towards the end of reheating. More importantly, our analysis demonstrates that a predicted GW signal, that could have violated the BBN and/or  aLIGO constraints when modelled by a single  stiff EoS, $w_{\rm stiff}  > 1/3$,  may actually be consistent with the aforementioned constraints if the post-inflationary history is comprised of additional phases with softer EoS parameters, $w \leq 1/3$, along with the stiff-matter phase. Therefore, our work   highlights the drawback of using a single average EoS to describe the post-inflationary expansion, and opens up a broad range of parameter space which generates robust forecasts for the upcoming GW observatories.  However, it is important to investigate the possible degeneracies between the GW spectra coming from the parameter space of multiple EoS, and durations of individual epochs. Consequently, a more careful analysis would require parameter estimation via Bayesian inference technique, \textit{e.g.}, see Ref.~\cite{Duval:2024jsg}, which we plan to carry out in a future work.

Finally, in Sec.~\ref{sec:String_phenomenology} we applied our techniques to a particular post-inflationary scenario described in Ref.~\cite{Apers:2024ffe}, which is phenomenologically inspired from String Theory, in which a scalar moduli field plays a cosmologically important role, and dictates  the evolution of  the universe in between the end of inflation and the onset of BBN. In this scenario,   the post-inflationary universe transits through a series of chronological epochs, namely, a kination-dominated epoch, an intermediate radiation-dominated epoch, and an early matter-dominated epoch (see Fig.~\ref{fig: timeline_string}), prior to the standard radiative hot Big Bang phase. We explored the possibility of observing the imprints of this scenario in the inflationary GW spectrum via upcoming GW observatories. Representing the  duration of each epoch in terms of the energy scale at the end of that epoch, we obtained  the corresponding parameter space that results in a detectable signal, as shown in Fig.~\ref{fig:instant rehating}; with the assumption that the final stage reheating is instantaneous. We found  that  GWs, corresponding to  tensor modes that made their Hubble-entry during the kination phase, will be detected by  BBO, DECIGO and CE, and not by LISA and ET (due to violation of BBN constraints). Furthermore, we obtain a necessary condition for this scenario to result in a detectable GW signal to be that  the early matter-dominated epoch prior to the end of  reheating must be shorter in duration, lasting no more than $ 7 \, e$-folds of expansion. 

Before concluding, we emphasize that the specific analysis carried out in this work comes with a number of caveats, which we list in the following.

\begin{itemize}

\item Throughout this paper, we have mostly fixed the value of the tensor-to-scalar ratio at $r = 0.001$ which is a standard target for the upcoming CMB missions~\cite{CMB-S4:2020lpa, LiteBIRD:2024twk}. In this sense, our analysis was focused on the question:  `\textit{if low-frequency primordial GWs get detected by their imprints on the CMB B-mode polarization in the upcoming decade, then what will the implications be for the upcoming GW observatories functioning at higher frequencies?}'. However, one can easily consider lower values of tensor-to-scalar ratio and incorporate our computational scheme to generate updated forecast for GWs. Nevertheless, we provide a detailed discussion on the effect of changing reheating temperature on the GW spectral energy density in Sec.~\ref{sec:effect_GW_Trh}.

\item Similarly, the energy scale of the universe at the end of reheating was fixed to be $E_{\rm r*} = 1$ GeV in this paper in order to demonstrate our approach and to focus on the EoS parameter and duration of each post-inflationary (pre-BBN) epoch. However, the hot Big bang phase might have commenced at a much lower temperature (as low as ${\cal O}(1)\,{\rm MeV}$) or even at higher temperatures (as high as ${\cal O}(10^{15})\,{\rm GeV}$). The lower reheating temperature is of greater phenomenological interest, and one can utilize our scheme to generate the corresponding GW forecast. However, much higher reheating temperature usually results in a low amplitude of GWs which may be of phenomenological interest for the high-frequency resonant cavity GW detectors~\cite{Gatti:2024mde,Kanno:2023whr,Ito:2022rxn,Aggarwal:2020olq}.

\item More importantly, while realistic transitions (no matter how sharp they are) in the early universe are expected to be smooth~\cite{Felder:1998vq,Figueroa:2019paj}, in this work we modelled the transitions between different post-inflationary epochs to be instantaneous, primarily for two reasons: \textit{(i)} for the convenience of solving the tensor mode equations analytically in terms of Bessel functions, and \textit{(ii)} to remain agnostic about the exact functional form of the EoS $w(\tau)$. However, it is possible to carry out a fully numerical approach by either using a (phenomenologically) smoothed EoS parameter, or by considering a specific model where the exact evolution of $w(\tau)$ is known. It has been shown that a smoothed reheating EoS in the perturbative regime leads to transient oscillatory features in the GW spectrum~\cite{Haque:2021dha}. Nevertheless, since we primarily focused on  determining the detectable parameter space of post-inflationary dynamics,  the implementation of instantaneous transitions suffices for our purpose. 

\item In our analysis, we did not take into account the finer effects coming from the variation in the effective number of relativistic degrees of freedom, $g_*(T)$, in the early universe~\cite{Watanabe:2006qe} as well as due to the anisotropic stress induced by freely streaming relativistic neutrinos~\cite{Weinberg:2003ur} on the spectrum of GWs.

\item  We assumed a (nearly-) flat primordial tensor power spectrum (at the end of inflation) to be the initial condition for evolving the tensor modes post inflation and for computing the present-day GW spectrum. However, towards the end of inflation, the tensor power-spectrum acquires a scale dependence which affects the GW spectrum~\cite{Haque:2021dha} closer to the highest frequency values $f_e$.  Additionally, we did not consider the UV tail of GWs~\cite{Pi:2024kpw} corresponding to tensor modes that never became super-Hubble during inflation. However, the aforementioned effects are more relevant for detecting  ultra-high frequency GWs, potentially via resonant cavity detectors, which was not the primary focus of our work.

\item Our treatment of tensor modes in this paper was based on first-order inflationary GWs. However, at second-order in cosmological perturbation theory, GWs can be induced by the presence of potentially large scalar perturbations in the primordial universe~\cite{Ananda:2006af, Baumann:2007zm, Kohri:2018awv, Domenech:2021ztg}. They  get amplified by the enhancement of scalar curvature perturbations at small scales and/or by the presence of an early matter-dominated epoch in the  post-inflationary universe~\cite{Assadullahi:2009nf, Kohri:2018awv}, which might lead to a larger background of second-order GWs in comparison to their first-order counterparts. The CMB observations can  probe scalar fluctuations with comoving wavenumber only in the range around $k = 0.05 \,  {\rm Mpc}^{-1}$, hence they are insensitive to small scale primordial fluctuations. Large enough fluctuations at  smaller cosmological scales may also lead to the  formation of copious amounts of primordial black holes (PBHs), a plausible candidate for dark matter. Hence, a potential  detection of these induced GWs would allow us to put strong constraints on the small-scale primordial density perturbations~\cite{Assadullahi:2009jc} and on the abundance of PBHs~\cite{Saito:2008jc, Saito_2010, Bartolo:2018rku}. We will revert to second-order GWs in a future project.

\item Finally, it is plausible that non-linear solitonic and non-solitonic configurations can contribute considerably towards the energy budget during the early stages of the post-inflationary evolution~\cite{Lozanov:2019ylm,Shafi:2024jig,Piani:2023aof}. They further induce second and higher-order GWs that is expected to be detected by the planned GW detectors. For example, high-frequency (GHz-scale) stochastic GWs can result from oscillon formation (and inflaton fragmentation in general) during reheating~\cite{Lozanov:2019ylm, Hiramatsu:2020obh}. However, such signals are beyond the sensitivities of the current and proposed detectors, and are relevant for the resonant cavity detectors~\cite{Herman:2022fau}.  Another example would be the stochastic GW associated with cosmic strings which can be picked up by LISA, DECIGO and BBO~\cite{Gouttenoire:2019kij,Gouttenoire:2019rtn,Blanco-Pillado:2024aca, Schmitz:2024hxw}. Thus, it is important to classify the possible GW signal resulting from such differing scenarios, which is outside the scope of this paper. 

\end{itemize}

\newpage

\section{Acknowledgements}
SSM is supported by the STFC Consolidated Grant [ST/T000732/1] at the University of Nottingham. AKS was supported by the INSPIRE scholarship of the Department of Science and Technology (DST), Govt. of India, during his Master's thesis work. AKS thanks IISER TVM for the institutional support and acknowledges SISSA for the PhD funding during the final revision.  SSM is grateful to IUCAA for the hospitality during the earliest stage of this work. SSM thanks Varun Sahni for insightful discussions over the years which led to the inception of this project. SSM is also thankful to Ed Copeland,  Oliver Gould, Sanjit Mitra and Ranjeev Misra for helpful discussions at various stages of this project.  Numerical computations were performed on the \texttt{Padmanabha HPC Cluster} at IISER TVM. 

\medskip

This paper is {\bf dedicated to the  cherished memory of Prof.~P.~P.~Divakaran} (1936\,--\,2025), whom SSM is thankful for many stimulating discussions.

\bigskip

For the purpose of open access, the authors have applied a CC BY public copyright license to any Author Accepted Manuscript version arising. 

\medskip

\noindent {\bf Data Availability Statement:} This work is entirely theoretical and has no associated data.  A \texttt{Python} code to generate the plots of  spectral energy density of first-order inflationary GWs can be found in the GitHub repository \href{https://github.com/athul104/Spectral_Energy_Density_FO_GWs}{\faGithub}.

\appendix
\section*{Appendices}
\section{Inflationary dynamics}
\label{sec:Inf_dyn}

We work in the simplest scenario where inflation  is sourced by a single canonical scalar field (inflaton) $\varphi(t, \, \xvec)$, whose homogeneous part is denoted as $\phi(t)$, \textit{i.e.}, $\varphi(t, \, \xvec) =  \phi(t) + \delta \varphi(t, \, \xvec)$.
The action governing the dynamics of the scalar field $\varphi$ is given by~\cite{Linde:1990flp, Baumann_TASI, Mishra:2024axb}
\begin{equation}
    S[g_{\mu\nu},\, \varphi] = \int \,  \d^4x \,  \sqrt{-g} \,  \l( \half \, m_p^2 \,  R - \f{1}{2}  \, g^{\mu\nu}\,  \partial_{\mu}\varphi \,  \partial_{\nu}\varphi  \, -V(\varphi)\r)\, , \label{eq:Action_phi_g_munu}
\end{equation}
where $g_{\mu \nu}$ is the metric tensor, $R$ is the Ricci scalar curvature and $V(\varphi)$ is the inflaton potential. The background spacetime is described by the flat FLRW line element
\begin{equation}
    \d s^2 = -\d t^2 + a^2(t) \l[ \d x^2 + \d y^2 + \d z^2 \r]\, ,
    \label{eq:FLRW_bg_metric}
\end{equation}
where $a(t)$ is the scale factor whose time evolution is governed by the Friedmann equation
\begin{equation}
    H^2 = \frac{1}{3 \, m_p^2} \, \rho_{\phi} \equiv  \frac{1}{3 \, m_p^2} \, \l[ \half \, \dot{\phi}^2 + V(\phi) \r] \, ,  
    \label{eq:Friedmann_eqn}
\end{equation}
where $H = \dot{a}/a$ is the Hubble parameter. The homogeneous inflaton condensate satisfies
\begin{equation}\label{eq; EOM of inflation}
    \ddot{\phi} + 3 \,  H  \, \dot{\phi} + \frac{\d V}{\d \phi} = 0 \, .
\end{equation}
Departure from a pure de Sitter (exponential) expansion  during inflation is characterised by the first  slow-roll parameter
\begin{equation}
    \epsilon_{_H} =  -\f{\d  \ln H}{\d N} = -\frac{\dot{H}}{H^2} = \frac{1}{2 \, m_p^2} \, \frac{\dot{\phi}^2}{H^2} \, ,
    \label{eq:epsilon_H}
\end{equation}
while a small enough value of the  second  slow-roll parameter, defined by 
\begin{equation}
    \eta_{_H} = -\frac{\ddot{\phi}}{H \, \dot{\phi}}=\epsilon_{_H}  - \frac{1}{2} \, \frac{\d \ln \epsilon_{_H}}{\d N} \, ,
\label{eq:eta_H}
\end{equation}
ensures that inflation lasts for long enough. Note that $ N(a) = \ln(a/a_i) $ marks the number of $e$-folds of expansion  during inflation, with $a_i$ being the scale factor at an arbitrary initial time during inflation~\cite{Baumann_TASI}. Slow-roll inflation is defined by
\beq
\epsilon_{_H} \ll 1 \, , \quad {\rm and} \quad |\eta_{_H}| \ll 1 \, .
\label{eq:SR_conditions}
\eeq

\subsection{Quantum fluctuations during Inflation}
\label{sec:inf_dyn_QF}
\begin{figure}[htb]
    \centering
    \includegraphics[width =0.8\textwidth]{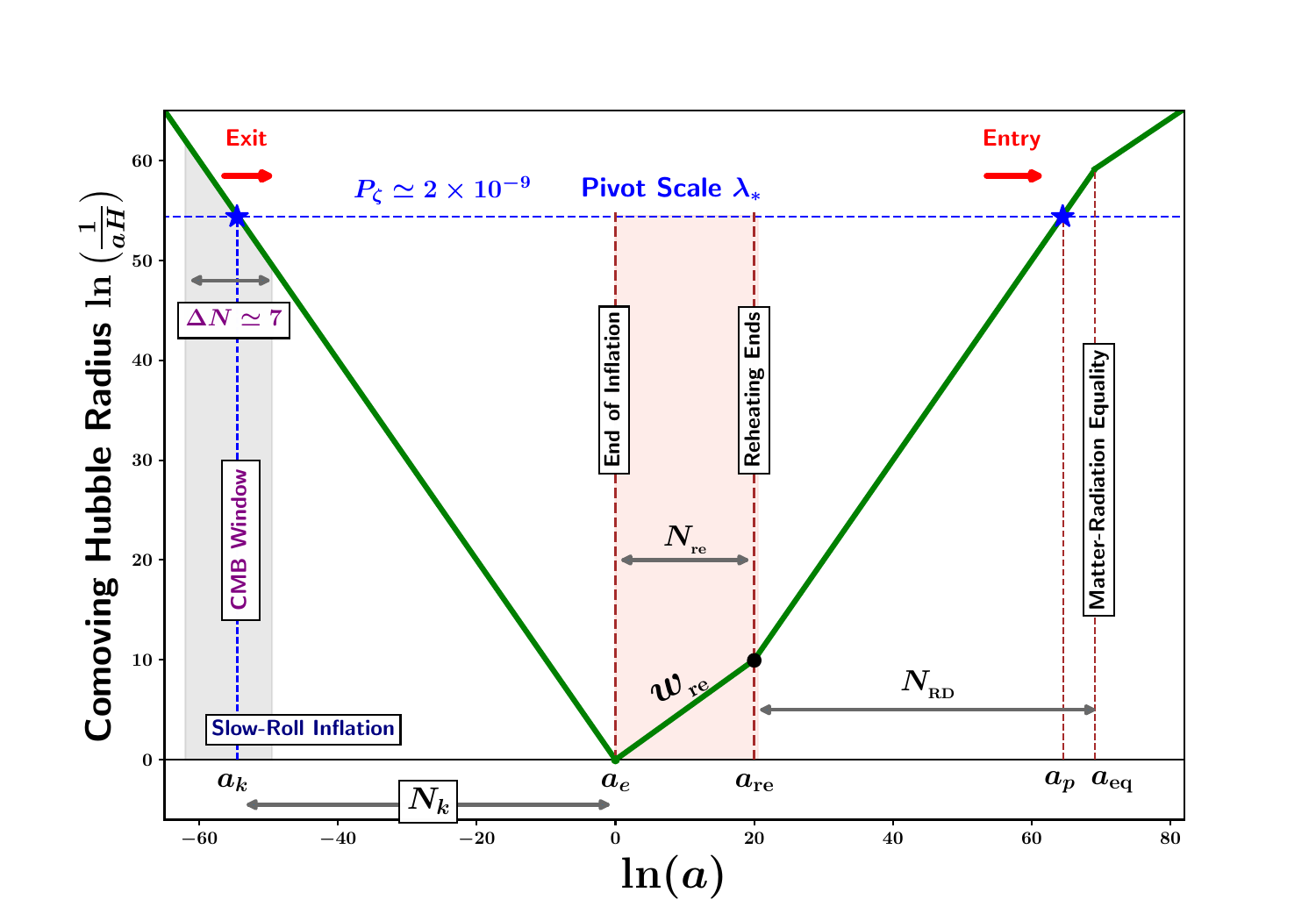}
    \caption{The figure illustrates the variation of the comoving Hubble radius $(aH)^{-1}$ with the scale factor $a(t)$ (both in the logarithmic scale).  During inflation, the comoving Hubble radius decreases, causing the fluctuations to exit the comoving Hubble radius. On super-Hubble scales, the amplitude of the fluctuations remains frozen. After the end of inflation, these fluctuations enter the comoving Hubble radius, which begins to increase due to the decelerating expansion of space. Modes with larger wavelengths exit the comoving Hubble radius earlier during inflation, as compared to the ones with smaller wavelengths. The CMB pivot scale (corresponding to a wavenumber $k_* = 0.05 \, {\rm Mpc}^{-1}$), shown in blue dashed line, makes its Hubble-exit about $N_k$ $e$-folds before the end of inflation, while it enters the comoving Hubble radius prior to the matter-radiation equality.  The black dot represents the end of reheating. The figure has been plotted  assuming a matter-like  reheating  equation of state, \textit{i.e.} $w_{_{\rm re}} = 0$, for the purpose of illustration.}
    \label{fig:Comoving Hub rad vs scale factor}
\end{figure}

The rapid accelerated expansion during inflation stretches the small-scale (sub-Hubble) vacuum fluctuations to large super-Hubble scales, which then remain frozen until their subsequent Hubble entry in the post-inflationary epochs, as illustrated in Fig.~\ref{fig:Comoving Hub rad vs scale factor}. In fact, the simplest scenario  of single field slow-roll inflation generates both scalar- and tensor-type fluctuations, see Refs.~\cite{Baumann_TASI, Mishra:2024axb}; the former induce temperature and density fluctuations in the primordial plasma, while the latter lead to a stochastic GW background, as discussed in  Sec.~\ref{sec:GWs_theory}.  Since CMB observations suggest that the primordial fluctuations are small, their dynamics is well described by the framework of  linear  (cosmological) perturbation theory, where the scalar and the tensor perturbations are decoupled (SVT decomposition theorem). Thus, they can be treated independently of each other~\cite{Baumann_TASI}.

\subsubsection{Scalar fluctuations during inflation}
\label{sec:inf_dyn_scalar_QF}
Scalar fluctuations during inflation are described  by the gauge-invariant comoving curvature perturbation $\zeta(\tau, \, \xvec)$, whose action, in the comoving gauge ($\delta \varphi = 0$), takes the form~\cite{Maldacena:2002vr, Baumann_TASI} 
\begin{equation}
    S[\zeta(\tau, \vec{x})] = \f{1}{2} \int \d\tau \, \d^3 \vec{x} \,  \l(\frac{a \, \phi'}{\cal H} \r)^2 \l[    (\zeta')^2 - (\grad \zeta)^2 \r] \, ,
    \label{eq:action_zeta}
\end{equation}
where ${\cal H} = a'/a$, is the conformal Hubble rate. To transform the action into its Minkowski equivalent, we define the Mukhanov-Sasaki (MS) variable $v = (a \, \phi' / {\cal H}) \, \zeta$.  For simplification, we denote $z = a \, \phi' / {\cal H}  = \sqrt{2 \, \epsilon_{_H}} \, a \, m_p$, where Eq.~\eqref{eq:epsilon_H} has been used to get the second equality. The equation of motion in terms of the MS variable in Fourier space is given by
\begin{equation}
    v''_k \, + \, \l( \, k^2 - \f{z''}{z} \,  \r) v_k = 0 \, ,
    \label{eq:MS_eqn}
\end{equation}
where
\begin{equation}
    \frac{z''}{z} = {\cal H}^2 \l[ 2 + 2 \epsilon_{_H} - 3\eta_{_H} + 2 \epsilon_{_H}^2 +  \eta_{_H}^2  - 3 \epsilon_{_H} \eta_{_H} - \frac{\eta_{_H}'}{\cal H}  \,  \r] \, .
    \label{eq:Meff_MS}
\end{equation}
To solve the MS equation analytically, we define a new parameter $\nu$ such that $z''/z = {\cal H}^2 (\nu^2 - 1/4)$.  For slow-roll inflation ($\epsilon_{_H}, \, |\eta_{_H}|, \, |\eta'_{H}| \ll 1 )$, we can treat $\nu$  to be (approximately) a constant. The MS Eq.~\eqref{eq:MS_eqn} can be written in terms of $\nu$ as
\begin{equation}
    v''_k \, + \, \l( \, k^2 - \f{\nu^2 -1/4}{\tau^2} \,  \r) v_k = 0 \, ,
    \label{eq; MS_eqn_nu}
\end{equation}
where we have substituted ${\cal H} = - \tau^{-1}$  due to  the quasi-de Sitter expansion during inflation. Eq.~\eqref{eq; MS_eqn_nu} is similar to the dynamic equation of a simple harmonic oscillator with a time-dependent frequency. Imposing the Bunch-Davies initial condition~\cite{baumann2012tasi, Kundu_2012}, $ v_k \xrightarrow[]{-k \tau \gg 1} e^{-i k\tau}/\sqrt{2k}$, results in a unique solution~\cite{Mishra:2023lhe}
\begin{equation}\label{eq; soln to MS eqn}
    v_k(\tau) = \frac{1}{\sqrt{2k}} \sqrt{\frac{\pi}{2}} \, e^{i(\nu+\frac{1}{2})\frac{\pi}{2}} \sqrt{-k \tau} \, H_{\nu}^{(1)}(- k\tau) \, ,
\end{equation}
where $ H_{\nu}^{(1)}$ is the Hankel function of first kind~\cite{NIST:DLMF}. The derivation of Eq.~\eqref{eq; soln to MS eqn} is given in App.~\ref{appendix; soln to MS eqn}. For slow-roll inflation, $\nu$ can be approximated as $\nu \approx  3/2 $. Substituting this into Eq.~\eqref{eq; soln to MS eqn} leads to the following expression for the mode function
\begin{align}\label{eq; MS variable v_k for SR}
    v_k(\tau) =  \f{1}{\sqrt{2k}} \l(1-\f{i}{k\tau}\r) e^{-ik\tau} \, .
\end{align}
The power spectrum for scalar fluctuations  is defined as
\begin{equation}
    {\cal P}_{\zeta} (k) = \f{k^3}{2\pi^2} \l \vert \zeta_k \r \vert^2  = \f{k^3}{2\pi^2} \l \vert \frac{v_k}{z} \r \vert^2 \, .
    \label{eq:scalar_power}
\end{equation}
Substituting  $z=am_p\sqrt{2\epsilon_{_H}}$,  $a = -1/(H \tau)$ and taking the super-Hubble limit $-k\tau \to 0$, we obtain the scalar power spectrum during inflation to be 
\begin{equation}
    {\cal P}_{\zeta} (k) = \frac{1}{8 \pi^2 \epsilon_{_H}} \l(\frac{H}{m_p}\r)^2  \, .
    \label{eq:inflationary scalar PS}
\end{equation}

\subsubsection{Tensor fluctuations during inflation}
\label{sec:inf_dyn_tensor_QF}

The action for the  gauge-invariant, transverse and traceless tensor fluctuations $h_{ij}$  is given by~\cite{Maldacena:2002vr,Baumann:2018muz}
\begin{equation}
    S[h_{ij}] = \frac{m_p^2}{8} \int \d\tau~\d^3\vec{x}~a^2 \left[ (h_{ij}')^2 - (\grad h_{ij})^2\right] \, .
    \label{eq:action_tensor}
\end{equation}
Using the rotational symmetry of the background metric, we align the z-axis along the direction of the momentum of the tensor mode, \textit{i.e.}, $\kvec = (0, \, 0, \, k)$, and upon decomposing the tensor perturbation into its two polarization components, we obtain
\begin{align}
    h_{ij} &= \frac{1}{\sqrt{2}} 
    \begin{pmatrix} 
        h^+ & ~~ h^{\times} &~~ 0 \\
        h^{\times} & -h^+ & ~~0 \\
        0~ & ~0 & ~~0 
    \end{pmatrix}  = \epsilon^+_{ij} \,  h^+ + \, \epsilon^{\times}_{ij} \,  h^{\times} \, , \label{eqn; h_ij lin comb of two polarisarion}
\end{align}
where  $ +, \, \t $ stands for the two polarisation states of GWs and
\begin{equation}
    \epsilon^+_{ij} = \frac{1}{\sqrt{2}} 
    \begin{pmatrix} 
        1 & ~~ 0 &~~ 0 \\
        0 & -1 & ~~0 \\
        0 & ~~0 & ~~0 
    \end{pmatrix};  \quad  \epsilon^{\t}_{ij} = \frac{1}{\sqrt{2}} 
    \begin{pmatrix} 
        0 & ~~1 &~~ 0 \\
        1 & ~~0 &~~ 0 \\
        0 & ~~0 &~~ 0 
    \end{pmatrix}  \, .
\end{equation}
Hence, the action can be rewritten as~\cite{Mishra:2024axb}
\begin{equation}\label{eq;action in terms of h_lambda}
    S[h^+,\, h^\times] = \frac{1}{2} \int \d\tau~\d^3\vec{x} \, \l(\frac{a \, m_p}{2} \r)^2 \, \sum_{\lambda=+,\times} \, \left[ \l(h^{' \lambda} \r)^2 - \l(\grad h^\lambda \r)^2 \right] \, ,
\end{equation}
which is analogous to the action of two massless scalar fields~\cite{Mishra:2024axb} as given in Eq.~\eqref{eq:action_zeta}. For each polarisation mode of the tensor fluctuations, we define a corresponding Mukhanov-Sasaki variable $v^\lambda = \left(a \, m_p/2 \right) h^\lambda$. The full tensor power spectrum  during inflation, consisting of both the aforementioned polarisation modes,  is given by

\begin{align}
    {\cal P}_T^{\rm inf}(k,\tau) &= \frac{k^3}{2 \pi^2} \left(\abs{h_k^+}^2 + \abs{h_k^{\times}}^2\right) \, ,
    \label{eq:tensor_power}
\end{align}
where each tensor mode can be expressed as following after substituting Eq.~\eqref{eq; MS variable v_k for SR} as
\begin{equation}
    h_k^{\lambda} (\tau) = \frac{2}{a \, m_p} \, \frac{1}{\sqrt{2k^3} \, \tau } \l( k \tau  - i\r) e^{-i k \tau} \, . \label{eq; h_k with MS soln}
\end{equation}
Substituting $\tau = -1/(aH)$ and taking the super-Hubble limit ($-k\tau \rightarrow 0 $), we end up with the expression
\begin{equation}
    h_{k, \, {\rm inf}}^{\lambda} (\tau) = i \sqrt{\frac{2}{k^3}} \, \frac{H}{m_p} \, . \label{eq; h_k, inf}
\end{equation}
Now substituting Eq.~\eqref{eq; h_k, inf} in Eq.~\eqref{eq:tensor_power}, the inflationary tensor power spectrum in the super-Hubble limit $k \ll aH$ becomes
\begin{equation}
    {\cal P}_T^{\rm inf}(k) = \frac{2}{\pi^2} \left(\frac{H}{m_p}\right)^2 \, ,
    \label{eq:inflationary tensor PS}
\end{equation}
 which is time-independent (frozen)\footnote{ Note that $H=H_{\rm inf}$ is the (nearly-constant) Hubble parameter during inflation.}. To facilitate the comparison  between the inflationary predictions and the observational data, it is instructive to express the  frozen inflationary scalar and tensor power spectra  on super-Hubble scales as power laws around a pivot scale $k_*$, such as
\begin{align}
    {\cal P}_T^{\rm inf}(k) &= A_{_T} \l( \frac{k}{k_*} \r)^{n_{_T}} \, ; &  
    {\cal P}_{\zeta}^{\rm inf}(k) &= A_{_S} \l( \frac{k}{k_*} \r)^{n_{_S} - 1} \, , \label{eq; PS_power_law}
\end{align}
where $A_{_S}$ and $A_{_T}$ are the  amplitudes of the scalar and tensor power spectra, respectively, at the CMB pivot scale $k_* = 0.05 \, \rm Mpc^{-1} $; while   $n_{_S}-1$ and  $ n_{_T} $ are the scalar and tensor spectral indices, given by
\beq
n_{_S} -1 = 2\, \eta_{_H} - 4 \, \epsilon_{_H} \, ; \quad n_{_T} = -2 \, \epsilon_{_H} \, .
\label{eq:SR_nS_nT}
\eeq
The tensor-to-scalar ratio is defined as,
\begin{equation}
\label{eq:tensor_scalar_ratio}
r = \frac{A_{_T}}{A_{_S}} \, .
\end{equation}
From Eqs.~\eqref{eq:inflationary scalar PS} and \eqref{eq:inflationary tensor PS}, the tensor-to-scalar ratio can be related to the Hubble slow-roll parameter as
\begin{equation}
\label{eq; r and epsilon_H}
r = 16 \, \epsilon_{_H} \, ,
\end{equation}
which, using Eq.~\eqref{eq:SR_nS_nT}, leads to the {\em single field consistency relation}
\begin{equation} \label{eq; r and n_T}
r= -8 \, n_{_T} \, .
\end{equation}

\subsection{Implications of the observational constraints}
\label{sec:inf_dyn_obs}

Latest observations of the anisotropies of CMB temperature and polarisation fluctuations, combined with an absence of detection of the primordial tensor fluctuations on large cosmological scales,  allow us to put stringent constraints on the inflationary  power spectra in the single field slow-roll paradigm. In particular,  the  primordial scalar fluctuations are observed to be predominantly adiabatic, highly Gaussian, and nearly scale-invariant on CMB scales, which is in accordance with the predictions of the standard single field slow-roll models~\cite{Planck:2019kim}. The amplitude of the scalar power spectrum at CMB scales is obtained from the Planck mission to be  $A_{_S} \simeq 2.1 \t 10^{-9}$~\cite{Planck:2018vyg, Mishra:2022ijb}. 
The scalar spectral index is constrained to be $n_{_S} = 0.9649 \pm 0.0012$ at $68\%$ confidence~\cite{Planck_inflation}. The improved constraints from the BICEP/Keck Array~\cite{2021PhRvL.127o1301A}, combined with the Planck 2018 observations, lead to an upper bound on the tensor-to-scalar ratio $r < 0.036$ at the CMB pivot scale $k_* = 0.05 \,  \mpc$. This further constrains the amplitude of the tensor fluctuations  to be  $A_{_T} < 7.56 \times 10^{-11}$ on CMB scales.

 Let us derive a relation between the tensor-to-scalar ratio $r$ and the energy scale of inflation $E_{\rm inf}$. Note that using Eq.~(\ref{eq:tensor_scalar_ratio}), we can write
$$r = \f{A_{_T}}{A_{_S}} ~ \Rightarrow ~ A_{_T} = A_{_S} \times r \, ,$$
which, using Eqs.~(\ref{eq:inflationary tensor PS})~and~(\ref{eq; PS_power_law}), along with $A_{_S}=2.1 \times 10^{-9}$, can be written as
\beq
\l(\f{H_{\rm inf}}{m_p}\r)^2 = 1.05 \, \pi^2 \times 10^{-12} \times \l(\f{r}{0.001}\r) \, .
\label{eq:H_r_SR_rel}
\eeq
Finally,  since $H_{\rm inf}^2 = E_{\inf}^4/(3m_p^2)$, we derive the expression for the energy scale of inflation to be
\beq
E_{\rm inf} =  \l( 3.15 \, \pi^2\r)^{1/4}  \times 10^{-3} \times \l(\f{r}{0.001}\r)^{1/4} \, m_p \simeq  5.76 \times 10^{15} \times  \l(\f{r}{0.001}\r)^{1/4} \, {\rm GeV} \, .
\label{eq:E_r_SR_rel}
\eeq

The corresponding  upper bounds on the expansion rate and energy scale during inflation can be found using Eq.~\eqref{eq:inflationary tensor PS}. The Hubble expansion rate during (slow-roll) inflation is bounded by $H_{\rm inf} <  4.64 \times 10^{13} \ \rm{GeV}$. Using the Friedmann equation $H^2_{\rm inf} \simeq \rho_{\rm inf}/{3 m_p^2}$, the bound on $E_{\rm inf}$ is given by
\begin{equation}
    \label{eq:inf_Energy_scale_inflation}
    E_{\rm inf} = (\rho_{\rm inf})^{1/4} \implies E_{\rm inf} < 1.39 \times 10^{16} \ \rm{GeV} \, ,
\end{equation}
which is orders of magnitude below the Planck scale ($\sim 10^{18} \, {\rm GeV}$). For slow-roll inflation, the constraint on the tensor-to-scalar ratio, along with Eq.~\eqref{eq; r and epsilon_H} and Eq.~\eqref{eq; r and n_T} leads to $\epsilon_{_H} < 2.25 \t 10^{-3}, \, |\eta_{_H}| \simeq 0.02 $ and $|n_{_T}| < 4.5 \t 10^{-3}$. These constraints further reinforce  the near-scale invariance of the inflationary power spectra; in particular, the  tensor power spectrum has a negligible scale dependence~\cite{Mishra:2022ijb}.

\section{Analytical solution to the  Mukhanov-Sasaki equation}\label{appendix; soln to MS eqn}

The Mukhanov-Sasaki (MS) Eq.~\eqref{eq:MS_eqn} can be written as
\begin{equation}\label{eq; MS eqn in app}
    v_k'' + \l(k^2 - \frac{\nu^2 - \frac{1}{4}}{\tau^2}\r)v_k = 0 \, ,
\end{equation}
where $\frac{z''}{z} = (aH)^2 (\nu^2 - 1/4)$ with $z = a m_p \sqrt{2 \epsilon_{_H}}$. For slow-roll inflation, at early times all relevant scales were sub-Hubble, $(\tau \xrightarrow{} -\infty \;\; \text{or} \;\; |k \tau| \gg 1)$. In this limit, Eq.~\eqref{eq; MS eqn in app} reduces to the form
\begin{equation}\label{eq; MS eqn_sub-Hubble limit}
    v_k'' + k^2 v_k = 0 \, ,
\end{equation}
which is similar to the equation for a simple harmonic oscillator with a time-independent frequency. A unique solution to Eq.~\eqref{eq; MS eqn_sub-Hubble limit} is obtained by fixing the vacuum in the sub-Hubble regime  to be the Bunch-Davies vacuum~\cite{baumann2012tasi, Kundu_2012}, namely,
\begin{equation}\label{BD-IC}
    \lim_{k \tau \rightarrow -\infty} v_k \simeq \frac{1}{\sqrt{2k} } \, e^{-ik\tau} \, .
\end{equation}
In terms of the  variable to be $T = -k \tau$, the MS equation becomes
\begin{equation}\label{eq;pre_bessel}
    \frac{\d^2 v_k }{\d T^2}+ \l(1- \frac{\nu^2 - \frac{1}{4}}{T^2}\r)v_k = 0 \, .
\end{equation}
Furthermore, we also introduce the variable $v_k = F \sqrt{T}$, in order to transform  Eq.~\eqref{eq;pre_bessel} to the Bessel equation
\begin{equation}\label{eq; Bessel eqn in apeendix}
    \frac{\d^2 F}{dT^2} + \frac{1}{T} \frac{\d F}{\d T} + \l(1- \frac{\nu^2}{T^2}\r)F = 0 \, ,
\end{equation}
whose general solution can be written in terms of Hankel functions as
\begin{equation}
    F(T) = C_1 \, H_{\nu}^{(1)}(T) + C_2 \, H_{\nu}^{(2)}(T) \, . \label{eq; soln of Bessel eqn in appendix}
\end{equation}
Hence, the general solution of the MS equation can be written as
\begin{equation}
    v_k(T) = \sqrt{T}\l[C_1 \, H_{\nu}^{(1)}(T) + C_2 \,  H_{\nu}^{(2)}(T) \r] \, \label{eq; soln to MS variable v} . 
\end{equation}
In the sub-Hubble limit, $T\gg 1$, the Hankel functions become
\begin{align}
     H_{\nu}^{(1)}(T)\Big|_{T \rightarrow \infty} &\simeq
     \sqrt{\frac{2}{\pi T}} \, e^{iT} \, e^{-i(\nu+\frac{1}{2})\frac{\pi}{2}} \, , \label{eq; Hankel_1 sub-Hubble limit; appendix} \\
     H_{\nu}^{(2)}(T)\Big|_{T \rightarrow \infty} &\simeq
     \sqrt{\frac{2}{\pi T}} \, e^{-iT} \, e^{i(\nu+\frac{1}{2})\frac{\pi}{2}} \, . \label{eq; Hankel_2 sub-Hubble limit; appendix} 
\end{align}
From the Bunch-Davies initial condition in Eq.~\eqref{BD-IC}, we have
\begin{equation}
     v_k(T)\Big|_{T \rightarrow \infty}= \frac{1}{\sqrt{2k} } e^{iT} = \sqrt{T} \, C_1 \, H_{\nu}^{(1)}(T) \Big|_{T \rightarrow \infty} \, , \label{eq; MS varible v sub-Hubble limit}
\end{equation}
which implies that the coefficients in Eq.~\eqref{eq; soln to MS variable v} are
\begin{equation}
    C_1 = \frac{1}{\sqrt{2k}} \, \sqrt{\frac{\pi}{2}} \, e^{i(\nu+\frac{1}{2})\frac{\pi}{2}} \quad , \quad C_2 = 0 \, . \label{eq; Coeff after applying BD}
\end{equation}
Ultimately, the expression for the mode function becomes
\begin{equation}
    v_k(T) = \frac{1}{\sqrt{2k}} \, \sqrt{\frac{\pi}{2}} \, e^{i(\nu+\frac{1}{2})\frac{\pi}{2}} \, \sqrt{T} \, H_{\nu}^{(1)}(T)  \, . \label{eq; final expr. for MS var. v}
\end{equation}

\section{Expression for the conformal Hubble parameter}\label{appendix; aH}
Consider an epoch with EoS, $w$. Let `$i$' denote the beginning of that epoch, then the energy density evolves as
\begin{equation}
    \rho(\tau) = \rho_i \l(\frac{a(\tau)}{a_i}\r)^{-3(1+w)} \label{eq; rho vs scale factor} \, .
\end{equation}
Hence, we can write the Hubble rate during the epoch to be, $ H^2 =  H_i^2 \, \l(a/a_i \r)^{-3(1+w)}$. Integrating the above expression from $a_i$ to an arbitrary $a$ during this epoch, we obtain
\begin{align}
    \int_{a_i}^{a} \d a \l(\frac{a^{(3(1+w)/2) -1 } }{a_i ^{3(1+w)/2}}\r) &= \int_{\tau_i}^{\tau} a_i H_i  \d \tau  \, , \label{eq; integral over scale factor and time}
\end{align}
which leads to the expression for the scale factor $a(\tau)$ during that epoch as
\begin{equation}
    a(\tau)  = a_i \l[1+ \frac{ a_i H_i  (\tau-\tau_i)}{\alpha}\r]^{\alpha}  \,  \label{eq; scale factor},
\end{equation}
where $\alpha = 2/(1 + 3w)$. Hence,  the conformal Hubble parameter ${\cal H}$ becomes
\begin{align*}
    {\cal H} &= a H = \frac{\d \ln a}{\d \tau}  \, , \\
    \ln a  & = \ln a_i + \alpha \, \ln{\l[1+ \frac{ a_i H_i  (\tau-\tau_i)}{\alpha}\r]}  \, , \\
     \Rightarrow a H&= \frac{ a_i H_i  }{\l[1+ \frac{ a_i H_i  (\tau-\tau_i)}{\alpha}\r]} \num \label{eq; conformal hubble eqn ; appendix} \, .
\end{align*}

\section{Spectral energy density of GWs}
\label{App:Om_GWs}
\subsection{Energy-momentum tensor of GWs}\label{appendix; energy momentum tensor}

The quadratic action for the tensor fluctuations $h_{ij}$ is given by
\begin{equation}
    S_2 [h_{ij}] =\frac{\mpl^2}{8} \int \d^4x \;\sqrt{-\bar{g}}\l[-\bar{g}^{\mu \nu}(\p_{\mu}h_{ij}) (\p_{\nu} h^{ij})\r]  \, ,
\end{equation}
where $\bar{g}_{\mu \nu}$ is the background metric and $\bar{g} \equiv {\rm det} \, \bar{g}_{\mu \nu}$. The energy-momentum tensor is obtained from the variation of this action with respect to the background (inverse) metric~\cite{Giovannini:2023itq}
\begin{equation}
    \delta S = -\half \int \d^4x \sqrt{-\bar{g}} \,  T_{\mu \nu} \, \delta \bar{g}^{\mu \nu} \,, \label{eq; variation of action}
\end{equation}
leading to
\begin{align*}
    T_{\alpha \beta} &= \frac{\mpl^2}{4} \l[-\frac{1}{2} \,  \bar{g}_{\alpha \beta} \, \bar{g}^{\mu \nu} \, (\p_{\mu}h_{ij}) \, (\p_{\nu} h^{ij}) +  (\p_{\alpha}h_{ij}) \,(\p_{\beta} h^{ij})\r]  \, ,\\
    \tensor{T}{^\mu_\nu} &=\frac{\mpl^2}{4} \, \l[-\frac{1}{2} \,  \tensor{\delta}{^\mu_\nu} \, \bar{g}^{\alpha \beta} (\p_{\alpha}h_{ij}) \, (\p_{\beta} h^{ij}) +  (\p^{\mu}h_{ij}) \, (\p_{\nu} h^{ij})\r] \num  \, . \label{eq; T^u_v}
\end{align*}
The time-time component of the energy-momentum tensor is the energy density of GWs,  given by
\begin{equation}
    \rho_{_{\rm GW}}(\tau,\vec{x}) = -\tensor{T}{^0_0} = \frac{\mpl^2}{8a^2 (\tau)} \l[(h'_{ij}(\tau,\vec{x}))^2 + (\grad h_{ij}(\tau,\vec{x}))^2\r] \label{eq; rho_GW final form} \, . 
\end{equation}

\subsection{Derivation of the present-day GW spectral energy density}
\label{appendix; Omega_GW}

The derivation is done for the general case of reheating consisting of multiple equation of state parameters. For $\tau = \tau_0$ in Eq.~\eqref{eq; equation of Omega_GW }
\begin{equation}
    \Omega_{_{\text{GW}}}(\tau_0, k)=\frac{k^2 }{12\, a^2 (\tau_0) H^2 (\tau_0)} \; {\cal P}_h (\tau_0,k)  \, . \label{eq; Omega_GW ; appendix}
\end{equation}
Substituting $k = y_{\rm eq} \, a_{\rm eq} \, H_{\rm eq}$ and  $ {\cal P}_h (\tau_0,k) = (k^3 / \pi^2)\, \overline{|h_{k, \, {\rm MD}}^{\lambda} (y \gg 1)|^2}$ in Eq.~\eqref{eq; Omega_GW ; appendix}, we obtain
\begin{align}
    \Omega_{_{\rm GW}}(\tau_0, k) &= \frac{1}{12 \,  a_0^2  H_0^2 } \, \l(y_{\rm eq} \, a_{\rm eq} \, H_{\rm eq} \r)^2 \,   \frac{k^3}{\pi^2} \,   \frac{1}{(2 \, y_0)^{4}} \, \frac{1}{\pi} \l[|A_{k, \, {\rm MD}}|^2 +  |B_{k, \, {\rm MD}}|^2\r] \label{eq; Omega_GW_deriv_pre_form}  \, ,
\end{align}
where  $\overline{|h_{k, \, {\rm MD}}^{\lambda} (y \gg 1)|^2}$ was substituted from Eq.~\eqref{eq; h_k in MD sq_avg value}. We can rewrite $y_0$ as
\begin{align*}
    y_0 &= \frac{k}{a_0 H_0} = y_{\rm eq} \, \frac{a_{\rm eq} H_{\rm eq}}{a(\tau_0) H(\tau_0)}  = y_{\rm eq} \, \l[1+ \frac{ a_{\rm eq} H_{\rm eq}  (\tau_0 -\tau_{\rm eq})}{2}\r]  \, ,
\end{align*}
\beq 
\Rightarrow y_0  = y_{\rm eq} \,  \l( \frac{a(\tau_0)}{a_{\rm eq}}\r)^{1/2}
\label{eq; y_0 vs y_eq relation} \,. 
\eeq
Incorporating Eq.~(\ref{eq; y_0 vs y_eq relation}) into Eq.~(\ref{eq; Omega_GW_deriv_pre_form}), we obtain
\begin{align*}
    \Omega_{_{\rm GW}}(\tau_0, k)&= \frac{1}{192} \frac{a^4_{\rm eq} H^2_{\rm eq}}{a_0^4  H_0^2 } \,  \frac{k^3}{\pi^3}   \frac{1}{y_{\rm eq}^2} \l[\tilde{A}^2_{k, \, {\rm MD}} +  \tilde{B}^2_{k, \, {\rm MD}}\r] \, \l| h_{k, \, {\rm inf}}^{\lambda} \r|^2 \, ,\\
    &=\frac{1}{96 \pi^3} \frac{a^4_{\rm r*} H^2_{\rm r*}}{a_0^4  H_0^2 } \,    \frac{1}{y_{\rm eq}^2} \l[\tilde{A}^2_{k, \, {\rm MD}} +  \tilde{B}^2_{k, \, {\rm MD}}\r]  \,\frac{H^2_{\rm inf}}{\mpl^2} \l(\frac{k}{k_*}\r)^{n_{_T}} \label{eq; Omega_GW second last expression} \num  \, ,
\end{align*}
 where $\tilde{A}_{k, \, {\rm MD}} = \frac{A_{k, \, {\rm MD}}}{h_{k, {\rm inf}}^\lambda}$ and similarly, $\tilde{B}_{k, \, {\rm MD}} = \frac{B_{k, \, {\rm MD}}}{h_{k, {\rm inf}}^\lambda}$. Furthermore,  we have made use of the following relations:
\begin{equation}
    {\cal P}_{T}^{{\rm inf}} (k) = \frac{k^3}{\pi^2} \, \l| h_{k, \, {\rm inf}}^{\lambda} \r|^2 = \frac{2}{\pi^2} \left(\frac{H_{\rm inf}}{m_p}\right)^2 \, \l( \frac{k}{k_*} \r)^{n_{_T}} \,, \label{eq; inflationary tensor power spectrum}
\end{equation}
\beq
    a^4_{\rm eq} H^2_{\rm eq} = (a_{\rm eq} H_{\rm eq})^2 a^2_{\rm eq}  = \l[ \frac{a_{\rm r*} H_{\rm r*}}{1+a_{\rm r*} H_{\rm r*} (\tau_{\rm eq} - \tau_{\rm r*})}\r]^2 a_{\rm r*}^2 [1+a_{\rm r*} H_{\rm r*} (\tau_{\rm eq} - \tau_{\rm r*})]^2 =  a_{\rm r*}^4 H_{\rm r*}^2 \, . \label{eq; relation of a^4 H^2 for 'eq' and 're'}
\eeq
where  $H_{{\rm inf}}$ corresponds to the Hubble scale  during  inflation 
 at the Hubble-exit time of the CMB pivot scale, \textit{i.e.}, $H_{{\rm inf}} = H_*$. The Hubble parameter can be written  as
\beq
    H^2(\tau) = \frac{\rho_{\rm rad}}{3 \, \mpl^2 \, \Omega_{\rm rad}}  = \frac{\rho_{\rm rad} \, \Omega_{\rm rad, \, 0} \, H_0^2}{\rho_{\rm rad ,\, 0} \, \Omega_{\rm rad}}  \,  . 
\label{eq; rel bw Hubble H and rho}
\eeq
Radiation energy density is given by
\begin{equation}\label{eq; energy density of realtivistic species}
    \rho_{\rm rad}(T) = \frac{\pi^2}{30} \,  g_*(T) \, T^4  \, , 
\end{equation}
while the  conservation of entropy implies
\begin{equation}\label{eq; conservation of entropy}
    T \propto g_{s}^{-1/3} a^{-1}  \, .
\end{equation}
Substituting Eq.~\eqref{eq; energy density of realtivistic species} and  Eq.~\eqref{eq; conservation of entropy} in Eq.~\eqref{eq; rel bw Hubble H and rho}, we obtain
\begin{equation}
    H^2(T)= \frac{g_*(T)\; g_{s}^{-4/3}(T)\; a^{-4} (T) \;\Omega_{\rm rad, \, 0}\; H_0^2}{g_{*, \, 0}\; g_{s, \, 0}^{-4/3} a_0^{-4} \Omega_{\rm rad}(T)}  \,  , \label{eq; H^2_numerical computation form}
\end{equation}
which can be rewritten as
\begin{equation}\label{eq; rel bw Hubble, scale factor and g*}
    \frac{a_k^4 H_k^2}{a_0^4 H_0^2} = \frac{g_{*,k}}{g_{*, \, 0}} \l(\frac{g_{s, \, 0}}{g_{s,k}}\r)^{4/3} \frac{\Omega_{\rm rad, \, 0}}{\Omega_{\rm rad}(\tau_k)} \, ,
\end{equation}
where $g_{*, \, k}, \, g_{s, \, k}$ are the number of relativistic degrees of freedom  in energy density and entropy density, respectively, at the time when the tensor mode with comoving wavenumber $k$ makes its Hubble-entry, while $g_{*, \, 0}, \, g_{s, \, 0}$ correspond to the same  at the present epoch.
When the tensor mode of interest enters during the RD epoch, $\Omega_{\rm rad} \approx 1$. So, \eqref{eq; rel bw Hubble, scale factor and g*} becomes
\begin{equation}\label{eq; rel bw Hubble, scale factor and g* for RD}
     \frac{a_k^4 H_k^2}{a_0^4 H_0^2} \approx \frac{g_{*,k}}{g_{*, \, 0}} \l(\frac{g_{s, \, 0}}{g_{s,k}}\r)^{4/3} \, \Omega_{\rm rad, \, 0} \, ,
\end{equation}
where  the label `$k$' refers to any mode that entered  the Hubble radius during the RD epoch.
Using Eq.~\eqref{eq; rel bw Hubble, scale factor and g* for RD} in Eq.~\eqref{eq; Omega_GW second last expression} at $\tau = \tau_{\rm r*}$ (beginning of RD epoch), we arrive at the final expression for the  spectral energy density of GWs
\begin{gather}
    \Omega_{_{\rm GW}}(\tau_0, k) = \frac{1}{96 \pi^3} \, \frac{g_{*, \, {\rm r*}}}{g_{*, \,0}} \l(\frac{g_{s, \, 0}}{g_{s, \, {\rm r*}}}\r)^{4/3} \, \Omega_{\rm rad, \, 0} \,    \frac{1}{y_{\rm eq}^2} \l[\tilde{A}^2_{k, \, {\rm MD}} +  \tilde{B}^2_{k, \, {\rm MD}}\r]  \,\frac{H^2_{\rm inf}}{\mpl^2} \l(\frac{k}{k_*}\r)^{n_{_T}} \, . \label{eq; Omega_GW final form; appendix}
\end{gather}

\subsection{Spectral tilt of inflationary GWs during the post-inflationary epoch}
\label{App:GW_tilt}
 The tensor mode functions during the epoch of reheating,  characterised by a single constant (effective) EoS, $\wre$, is given in Eq.~(\ref{eq; h_k for reheating}) to be 
    $$
         h_{k, \, {\rm re}} ^{\lambda} (y) = \frac{1}{(\alpha_{\rm re} \, y)^{\alpha_{\rm re} - \half}} \l[A_{k, \, {\rm re}} \, J_{ \l(\alpha_{\rm re} - \half \r)} (\alpha_{\rm re} \, y) +  B_{k, \, {\rm re}} \, J_{- \l(\alpha_{\rm re} - \half \r)} (\alpha_{\rm re} \, y) \r] \, ,
    $$
    where  $$y = \f{k}{aH} \, ; ~~ {\rm and} ~~\alpha_{\rm re} = \f{2} {1+ 3 \, w_{\rm re}} \, .$$
The spectral energy density of (sub-Hubble) GWs, as defined in Eq.~(\ref{eq; equation of Omega_GW }), is given by 
$$
    \Omega_{_{\text{GW}}}(\tau, k)=\frac{k^2 }{12\, a^2 (\tau) H^2 (\tau)} \times \f{k^3}{2\pi^2} \Bigl[\overline{ \l| h_{k, \, {\rm re}} ^{+} (y) \r|^2 + \l| h_{k, \, {\rm re}} ^{\times} (y) \r|^2}\Bigr] = \frac{k^5 }{24\pi^2\, (aH)^2} \times 2 \, \overline{\l| h_{k, \, {\rm re}} ^{\lambda} (y) \r|^2} \, ,
$$
which can be written as
\beq
\Omega_{_{\text{GW}}}(\tau, k) = \frac{k^5 }{12\pi^2\, (aH)^2} \times \overline{\l| h_{k, \, {\rm re}} ^{\lambda} (y) \r|^2} \, .
\label{eq:Om_GW_reheating_1}
\eeq
Using the expression for $h_{k, \, {\rm re}} ^{\lambda}$, and using junction matching conditions on super-Hubble scales -- which yields Eq.~(\ref{eq; coeff A_k and B_k for reheating}), we obtain
\beq
 \overline{\l| h_{k, \, {\rm re}} ^{\lambda} (y) \r|^2}(\tau, k) = \l(\f{aH}{\alpha_{\rm re} \, k}\r)^{\l( 2 \alpha_{\rm re} -1 \r)} \times \l| A_{k, \, {\rm re}} \r|^2 \times \overline{\l| J_{ \l(\alpha_{\rm re} - \half \r)} (\alpha_{\rm re} \, y) \r|^2} \, .
\label{eq:hk_GW_reheating_2} 
\eeq

\begin{figure}
    \centering
    \subfigure[]{\includegraphics[width=0.48\textwidth]{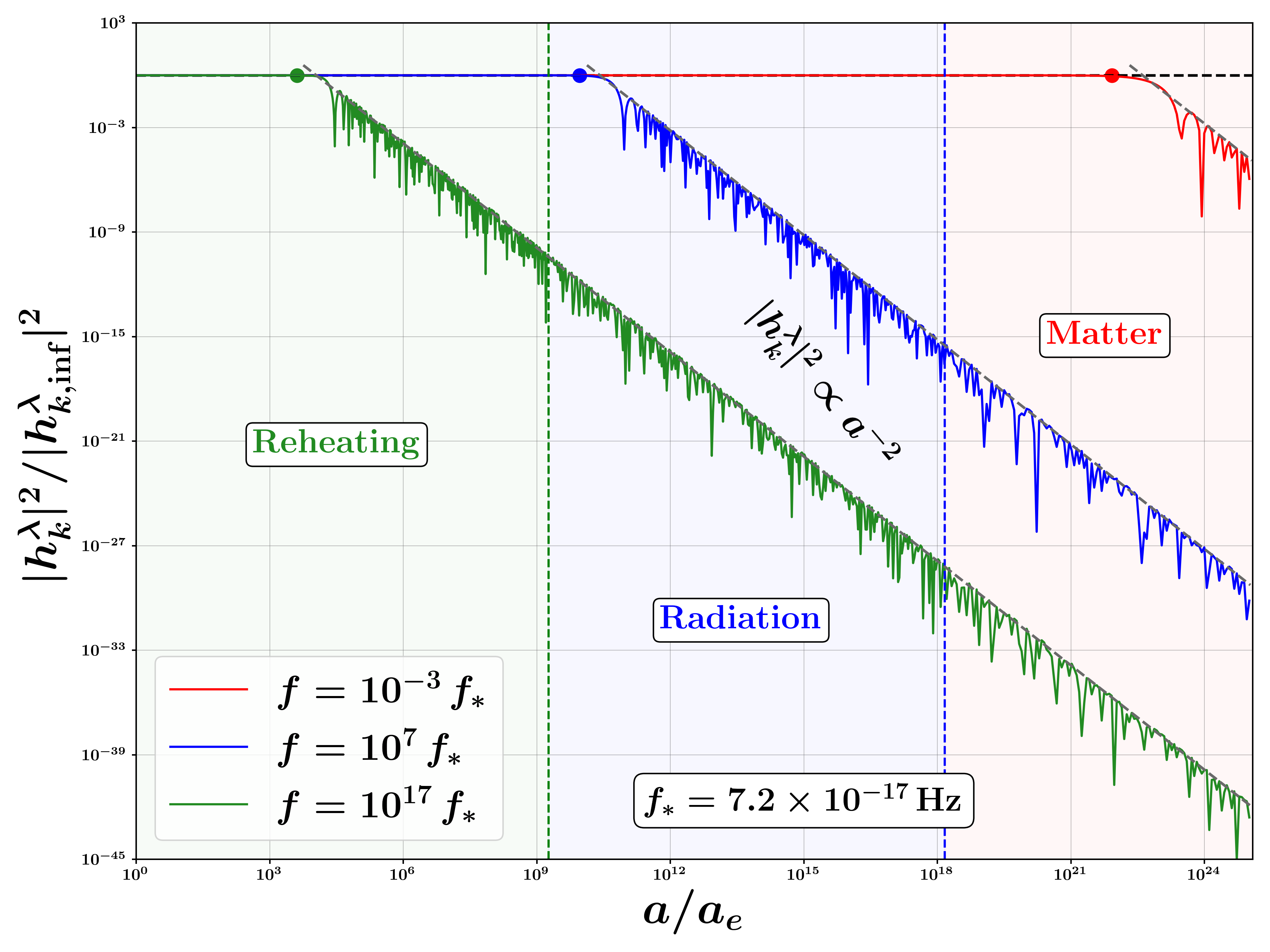}}
    \subfigure[]{\includegraphics[width=0.48\textwidth]{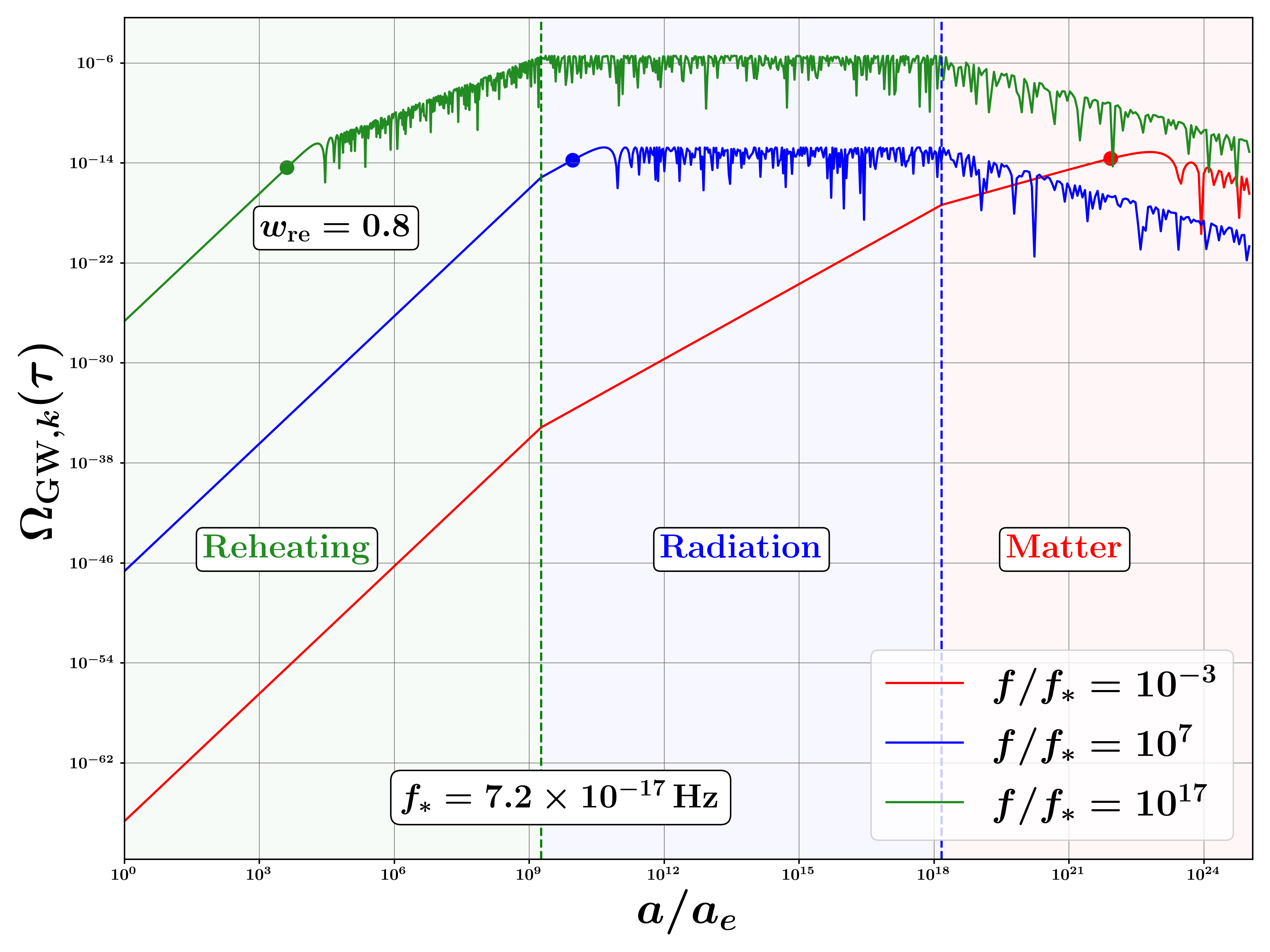}}
    \caption{ The sub-Hubble evolution of inflationary tensor modes (\textbf{left panel}) and their corresponding energy density spectrum (\textbf{right panel}). Hubble-entry epoch of a mode has been marked by a coloured circle. Observe that, in accordance with Eq.~(\ref{eq:hk_GW_reheating_3}), the amplitude of a tensor mode (after Hubble-entry) scales as $|h_k|^2 \propto a^{-2}$, which is independent of $w_{\rm re}$ (while at a given epoch, $|h_k|^2 \propto k^{-\l(7 + 9\, w_{\rm re} \r)/\l(1+3 \, w_{\rm re}\r)}$). The  spectral energy density of GWs for a given $k$ (after Hubble-entry), from Eq.~\eqref{eq:Omega_GW_reh_2a}, scales as $\Omega_{_{\text{GW}}}(\tau, k) \propto  a^{3\,w_{\rm re} -1 }$, which shows that $\Omega_{_{\text{GW}}}$ is constant during radiation domination (blue curve in the right panel), while it grows during stiff matter domination (green curve) and decays during matter domination (red curve). Similarly, at any given epoch, $\Omega_{_{\text{GW}}}(\tau, k) \propto k^{2\left(\frac{3\,w_{\rm re} - 1}{3\,w_{\rm re} + 1}\right)}$,  implying the spectrum is flat for $w_{\rm re} = 1/3$, blue-tilted for $w_{\rm re} > 1/3$ and red-tilted for $w_{\rm re} < 1/3$.}
    \label{fig:hk_Omegak_sub_Hubble} 
\end{figure}
From Eq.~(\ref{eq; coeff A_k and B_k for reheating}), we have
$$\l| A_{k, \, {\rm re}} \r|^2 = 2^{\l(2\alpha_{\rm re} - 1 \r)} \,  \Bigl( \Gamma \l(\alpha_{\rm re} + \f{1}{2} \r) \Bigr)^2 \,  \l|h_{k, \, {\rm inf}} ^{\lambda}\r|^2 \, ,$$
where $\l|h_{k, \, {\rm inf}} ^{\lambda}\r|^2 $ can be obtained as the following
$$ {\cal P}_T^{\rm inf} (k) = A_{_T} \l(\f{k}{k_*}\r)^{n_{_T}} = 2 \times \f{k^3}{2\pi^2} \l|h_{k, \, {\rm inf}} ^{\lambda}  \r|^2 \Rightarrow \l|h_{k, \, {\rm inf}} ^{\lambda}  \r|^2 = \l( \f{\pi^2 \, r \, A_{_S}}{k^3} \r) \l(\f{k}{k_*}\r)^{n_{_T}} \, .$$
Furthermore, noting that 
$$J_{\alpha_{\rm re} - \half} \l(\alpha_{\rm re}y\r) \simeq  \sqrt{\f{2}{\pi \alpha_{\rm re} y}} \, \cos{\l[ \alpha_{\rm re} y -  \l( \alpha_{\rm re} - \half\r) \frac{\pi}{2} - \frac{\pi}{4} \r]} = \sqrt{\f{2}{\pi \alpha_{\rm re}}} \, \l(\f{k}{aH} \r)^{-1/2} \, \cos{\l[ \alpha_{\rm re} \l( \f{k}{aH} - \f{\pi}{2}\r) \r]} \, ,$$
leads to
$$\overline{\l| J_{ \l(\alpha_{\rm re} - \half \r)} (\alpha_{\rm re} \, y) \r|^2} \simeq \frac{1}{\pi} \, \l(\f{\alpha_{\rm re}k} {aH}\r)^{-1} \, .$$
Therefore from Eq.~(\ref{eq:hk_GW_reheating_2}), we have
\beq
 \overline{\l| h_{k, \, {\rm re}} ^{\lambda}\r|^2}(a, k) ~~ \propto ~~ \f{k^{n_{_T}} \, k^{-\l(3+2\, \alpha_{\rm re} \r)}}{\l(aH \r)^{-2\,\alpha_{\rm re}}} ~~ \propto ~~ \f{k^{n_{_T}} \, k^{-\l(\frac{7 + 9\, w_{\rm re} }{1+ 3 \, w_{\rm re}}\r)}}{a^2} \, ,
\label{eq:hk_GW_reheating_3} 
\eeq
and consequently, Eq.~(\ref{eq:Om_GW_reheating_1}), leads to
\begin{equation}
    \label{eq:Omega_GW_reh_2a}
    \Omega_{_{\text{GW}}}(\tau, k) ~~ \propto ~~ k^{n_{_T}} \, \l( \f{k}{aH} \r)^{2(1-\alpha_{\rm re})} ~~ \propto ~~ k^{n_{_T}} \, k^{2\left(\frac{w_{\rm re} - 1/3}{w_{\rm re} + 1/3}\right)} \, a^{3\,w_{\rm re} -1 }\, .
\end{equation}
We obtain the final expression for spectral energy density of GWs to be
\begin{equation}
    \label{eq:Omega_GW_reh_3}
    \Omega_{_{\text{GW}}}(\tau, k) = \frac{ rA_{_S}}{24 \pi} \, (aH)^{2 (\alpha_{\rm re} - 1)} \left(\frac{2}{\alpha_{\rm re}}\right)^{2\alpha_{\rm re}}\Bigl(\Gamma \l(\alpha_{\rm re} + \f{1}{2} \r)\Bigr)^2   \left(\frac{k}{k_*}\right)^{n_{_T}} \, k^{2(1 - \alpha_{\rm re})} \, ,
\end{equation}
 which can be written as
\begin{equation}
    \label{eq:Omega_GW_reh_4}
    \Omega_{_{\text{GW}}}(\tau, k) = {\cal C}\l(\tau, \alpha_{\rm re}\r) \left(\frac{k}{k_*}\right)^{n_{_T}} \times k^{2(1 - \alpha_{\rm re})} \, ,
\end{equation}
where the $k$-independent coefficient is given by
\beq
{\cal C}\l(\tau, \alpha_{\rm re}\r) = \frac{rA_{_S}}{24 \pi}  \, (aH)^{2(\alpha_{\rm re} - 1)} \left(\frac{2}{\alpha_{\rm re}}\right)^{2\alpha_{\rm re}} \Bigl(\Gamma \l(\alpha_{\rm re} + \f{1}{2} \r)\Bigr)^2  \, .
\label{eq:Coeff_Om_GW}
\eeq
Collecting the powers of $k$ in Eq.~(\ref{eq:Omega_GW_reh_4}), the spectral tilt of inflationary GWs is given by  
\begin{align}
    \label{eq:nGW_1}
    n_{_{\rm GW}} = n_{_T} + 2(1 - \alpha_{\rm re}) = n_{_T} + 2\left(1 - \frac{2}{1 + 3w_{\rm re}}\right) =  n_{_T} + 2\left(\frac{w_{\rm re} - 1/3}{w_{\rm re} + 1/3}\right) \, . 
\end{align}
Ignoring the contribution from the tiny red-tilt $n_{_T} \simeq -r/8$, we arrive at the expression in Eq.~\eqref{eq; spectral tilt},
\begin{equation}
    \label{eq:nGW}
    n_{_{\rm GW}} \simeq  2\left(\frac{w_{\rm re} - 1/3}{w_{\rm re} + 1/3}\right)
\end{equation}

\section{Energy density  of  photons and  neutrinos}\label{app; energy denisty param of CMB photons and neutrinos}
From the present-day CMB temperature, $T_0 = 2.73 \, {\rm K}$~\cite{Fixsen:2009ug}, we obtain the energy density of CMB photons
\begin{equation}
    \rho_{\gamma, \, 0}  = \frac{\pi^2}{30} \t 2 \t T_0^4 = 2.00 \t 10^{-51} \, {\rm GeV}^4 \,, \label{eq; energy den of CMB photon}
\end{equation}
where the factor `2' accounts for the internal degrees of freedom of photons. The present-day critical density of the universe is 
\begin{equation}
    \rho_{c, \, 0} = 3 \, H_0^2 \, m_p^2 = 8.12 \t 10^{-47} \, h^2\, {\rm GeV}^4\,, \label{eq; present critical density}
\end{equation}
where $H_0 = 100 \, h \, {\rm km} \, {\rm s}^{-1} \, {\rm Mpc}^{-1} =  2.132 \t 10^{-42} \, h \, {\rm GeV} $~\cite{Planck:2018vyg}. Hence, the density parameter of CMB photons is given by
\begin{equation}
    h^{2} \,\Omega_{\gamma, \, 0} = h^{2} \, \frac{\rho_{\gamma, \, 0}}{\rho_{c, \, 0}} = 2.46 \t 10^{-5} \,. \label{eq; density param of CMB photons}
\end{equation}
Post electron-positron annihilation, the neutrino temperature is given by~\cite{Baumann:2022mni}
\begin{equation}
    T_{\nu} = \l(\frac{4}{11} \r)^{1/3} \, T_{\gamma} \,, \label{eq; rel bw temp of photons and neutrino}
\end{equation}
from which, we can derive a relation between the energy density of SM neutrinos and CMB photons to be
\begin{equation}
    \rho_{\nu} = \frac{7}{8} \, N_{\rm eff} \, \l(\frac{4}{11} \r)^{4/3} \, \rho_{\gamma}\,, \label{eq; rel bw rho of photons and neutrinos}
\end{equation}
where $N_{\rm eff} = 3.046$~\cite{Mangano:2005cc} is the effective number of neutrino species in the SM. Using Eq.~\eqref{eq; rel bw rho of photons and neutrinos}, the present-day density parameter for neutrinos becomes
\begin{equation}
    h^{2} \,\Omega_{\nu, \, 0} = \frac{7}{8} \, N_{\rm eff} \, \l(\frac{4}{11} \r)^{4/3} \, h^2 \, \Omega_{\gamma, \, 0} = 1.70 \t 10^{-5}\,. \label{eq; present den param of neutrinos}
\end{equation}
Hence, the present-day density parameter of  radiation becomes 
\begin{equation}
    h^{2} \, \Omega_{{\rm rad}, \, 0} = h^{2} \, (\Omega_{\gamma, \, 0} + \,\Omega_{\nu, \, 0}  )= 4.16 \t 10^{-5}\,. \label{eq; present den param of rad}
\end{equation}

\clearpage

% From MS:
% If PRD returns the mail regarding improper citation, kindly activate the following (and deactivate '\printbibliography'):

\printbibliography
% \input{Reference.bbl}

% If the fix doesn't work, message me (it won't work here, but PRD should render it properly), I'll fix it.
\end{document}